\documentclass[twocolumn,useAMS,usenatbib]{mn2e}
\usepackage{graphicx,verbatim,amsfonts,amsmath,natbib}
\usepackage{bm}

\newcommand{\hinvk}{$h^{-1}$kpc}
\newcommand{\hinvm}{$h^{-1}$Mpc}
\newcommand{\meanz}{\ensuremath{\langle z\rangle}}
\newcommand{\rmd}{{\rm d}}
\newcommand{\rme}{{\rm e}}
\newcommand{\rmi}{{\rm i}}
\newcommand{\sigmacrit}{\ensuremath{\Sigma_c}}
\newcommand{\scinv}{\ensuremath{\Sigma_c^{-1}}}
\newcommand{\avgscinv}{\ensuremath{\langle\Sigma_c^{-1}\rangle}}
\newcommand{\scinvtwo}{\ensuremath{\Sigma_c^{-2}}}
\newcommand{\dsr}{\ensuremath{\Delta\Sigma(r)}}

\newcommand{\ds}{\ensuremath{\Delta\Sigma}}
\newcommand{\mr}{\ensuremath{M_{r}}}
\newcommand{\meand}{\ensuremath{\langle\delta\rangle}}
\newcommand{\sigmad}{\ensuremath{\sigma_{\delta}}}
\newcommand{\cl}[3]{\ensuremath{#1^{+#2}_{-#3}}}

\title[Systematic errors in weak lensing]{Systematic errors in weak lensing: 
application to SDSS galaxy-galaxy weak lensing} 
\author[Mandelbaum et al.]
 {Rachel Mandelbaum$^1$\thanks{Electronic address:
    {\tt rmandelb@princeton.edu}},
  Christopher M. Hirata$^1$,
  Uro\v{s} Seljak$^{1,2}$,
\newauthor
  Jacek Guzik$^{3,4}$,
  Nikhil Padmanabhan$^1$,
  Cullen Blake$^5$,
  Michael R. Blanton$^6$,
\newauthor
  Robert Lupton$^7$,
  Jonathan Brinkmann$^8$
\\$^1$Department of Physics, Jadwin Hall, Princeton University,
      Princeton NJ 08544, USA
\\$^2$International Centre for Theoretical Physics, Strada Costiera 11,
      34014 Trieste, Italy
\\$^3$Astronomical Observatory, Jagiellonian University, Orla 171, 30-244
      Krak\'{o}w, Poland
\\$^4$Department of Physics and Astronomy, University of Pennsylvania,
Philadelphia PA 19104, USA 
\\$^5$Harvard-Smithsonian Centre for Astrophysics, MS-10, 60 Garden
      Street, Cambridge MA 02138, USA
\\$^6$Centre for Cosmology and Particle Physics, Dept. of Physics,
      New York University, 4 Washington Pl, New York NY 10003, USA
\\$^7$Princeton University Observatory, Princeton NJ 08544, USA
\\$^8$Apache Point Observatory, 2001 Apache Point Road, Sunspot NM
88349-0059, USA 
}

\date{\today}

\begin{document}
\maketitle

\begin{abstract}
Weak lensing is emerging as a powerful observational tool to constrain
cosmological models, but is at present limited
by an incomplete understanding of many sources of systematic error. 
Many of these errors are multiplicative and depend on the 
population of background galaxies. 
We show how the commonly cited
geometric test, which is rather insensitive to cosmology, 
can be used as
a ratio test of systematics in the lensing signal at the 1 per cent level.
We apply this test to the galaxy-galaxy lensing analysis of 
the Sloan Digital Sky Survey (SDSS), which at present is the sample with the
highest weak lensing signal to noise and has the additional 
advantage of spectroscopic redshifts for lenses. This 
allows one to perform meaningful geometric tests of 
systematics for different subsamples of galaxies at different 
mean redshifts, such as 
brighter galaxies, 
fainter galaxies and high-redshift luminous red galaxies, both with and without 
photometric redshift estimates. We use overlapping objects between SDSS
and the DEEP2 and 2SLAQ spectroscopic surveys to 
establish accurate calibration of photometric redshifts and 
to determine the redshift distributions for SDSS. 
We use these redshift results to compute the  
projected surface density contrast
$\Delta\Sigma$ around 259~609 spectroscopic galaxies in
the SDSS; by measuring $\Delta\Sigma$ with  
different source samples we establish consistency of the results at
the 10 per cent level ($1\sigma$).  We also use the ratio test to constrain  
shear calibration biases and other systematics in the SDSS survey data
to determine the overall galaxy-galaxy weak lensing signal calibration
uncertainty. We find no evidence of any inconsistency among many 
subsamples of the data.  
\end{abstract}

\begin{keywords}
gravitational lensing -- galaxies: distances and redshifts -- galaxies: halos.
\end{keywords}

\section{Introduction}

Recent years have seen tremendous progress in the detection of
galaxy-galaxy weak lensing (\citealt{1996ApJ...466..623B}; 
\citealt{1998ApJ...503..531H}; \citealt{2000AJ....120.1198F};
\citealt{2001ApJ...551..643S};  \citealt{2001astro.ph..8013M}; 
\citealt{2002MNRAS.335..311G}; \citealt{2003MNRAS.340..609H};
\citealt{2004AJ....127.2544S}; \citealt{2004ApJ...606...67H}; 
\citealt{2005PhRvD..71d3511S}), the tangential shear distortion around
galaxies due to their dark matter halos.  Recent measurements 
of galaxy-galaxy weak lensing (\citealt{2004AJ....127.2544S};
\citealt{2004ApJ...606...67H}; \citealt{2005PhRvD..71d3511S}) demonstrate 20--30$\sigma$ detections.   
In light of the increasing statistical precision with which this
effect is measured, it is important to revisit common sources of 
systematic error, which are currently at the 10 per cent level and
therefore already dwarf the statistical error. 

While galaxy-galaxy weak lensing is potentially a very powerful tool
for studying the dark matter halo profiles around stacked foreground
galaxies, it suffers from a large number of potential calibration biases.
Because it involves measuring the projected surface density contrast,
\begin{equation}
\dsr = \gamma_{+}(r)\Sigma_c,
\end{equation}
averaged over stacked lens and source
galaxies (where $r$ is the 
transverse separation from the lens galaxy), 
 calibration biases may be introduced via both the tangential
shear and the redshifts
used to compute $\Sigma_c$.
%(Additive errors can be easily
%determined by computing the signal using a random lens catalogs with
%the true source catalog, which 
%will reveal spurious signal that must be subtracted from the
%signal computed using the true lens catalog.)  

Here we introduce some of the notation that is used in
this paper to describe the weak-lensing signal.  We 
compute the 2-dimensional projected surface density contrast of stacked foreground
galaxies as a function of transverse separation from those galaxies,
where this separation is measured in comoving coordinates, using $r =
r_{p}(1+z_l)$ (with subscript $l$ referring to the lens, $s$ to
the source), and $r_{p}= \theta_{ls} D_A(z_l)$, the product of the
observed angular lens-source separation on the sky and the angular
diameter distance at the lens redshift.  We can then measure the
surface density contrast, which is related to the projected mass density $\Sigma(r)$
and its average value inside radius $r$, $\overline{\Sigma}(<r)$, as follows:
\begin{equation}
\dsr = \overline{\Sigma}(<r) - \Sigma(r) = \gamma_{+}(r)\Sigma_c
\end{equation}
The inverse critical surface density \scinv{} in comoving coordinates is
defined by 
\begin{equation}\label{E:scinv}
\scinv = \frac{4\pi G}{c^2}\frac{D_{ls}D_l(1+z_l)^2}{D_s}
\end{equation}
in terms of angular diameter distances, and has
the following 
properties: for a given lens redshift, it is 
zero for $z_s<z_l$, then increases rapidly above $z_s>z_l$ until it
flattens out as $z_s \gg z_l$; the asymptotic value for $z_s \gg
z_l$ increases with $z_l$.  

For a full discussion
of the errors that can be introduced in the shear computation, see
\citet{2004MNRAS.353..529H}, hereinafter H04.  As shown there,
several types of error can be introduced when computing the
shear, including biases due to point-spread function (PSF) correction, noise rectification
bias, and selection
biases.  These biases are considered in detail for the linear PSF correction
method used in that paper.  In this paper, we use a different PSF
correction scheme, ``re-Gaussianization,'' which was introduced and
tested in \cite{2003MNRAS.343..459H}, and include a discussion of
the effects of that choice on systematic error in the shear.

Because systematic errors in the shear computation have been well-studied,
and statistical errors can be decreased to a fairly low level when data from 
large surveys such as the SDSS are used, systematic errors in the
redshift distribution (and consequently, in
\sigmacrit) are of much greater importance than they were previously.
For the purposes of this paper, we will assume that lens redshifts are
known to high precision via spectroscopy, so our concern is the source
redshift distribution; as shown in~\cite{2004astro.ph..4527K}, if the
lens redshifts are also 
unknown, g-g weak lensing is not nearly as powerful a tool.   Several
common methods of determining the source redshift 
distribution are inadequate for precision cosmology due to
previously unquantified systematic uncertainties that they introduce via their effects on
\scinv;
this paper includes a study of the biases that may be introduced using
these methods, and a comparison against a well-determined reference
distribution.  

In addition to biases introduced due to the
redshift distribution, we also
discuss systematics that can be introduced while
computing the signal.  These include biases due to intrinsic
alignments, selection 
effects, and several other effects.  One new effect not noted before
is a problem with the determination of the sky flux near bright objects by
the SDSS {\sc Photo} pipeline that leads to the problem in detection
of sources within about $90''$ of bright objects
(\S\ref{SSS:boostsresults}).  This problem affects 
any galaxy-galaxy or cluster-galaxy weak lensing analysis using
SDSS data.

Many of the sources of 
systematic errors discussed above
are common to both weak lensing auto-correlation 
analysis and galaxy-weak lensing correlations (galaxy-galaxy lensing). 
Weak lensing auto-correlation analysis at present is 
limited by the statistical precision (see summaries in \citealt{2003ARA&A..41..645R} and \citealt{2003IAUS..216E..11H}) and it is difficult to test 
for the presence of systematics within each data set. However, the 
size of these data sets is rapidly increasing, and in the near future 
systematic errors are likely to dominate the statistical errors. 
Understanding of weak lensing systematics is essential if one is to exploit the 
full potential of upcoming and planned surveys such as the
CFHT Legacy survey ({\slshape
  http://www.cfht.hawaii.edu/Science/CFHLS}, \citealt{mellier2001}),
Pan-Starrs ({\slshape http://pan-starrs.ifa.hawaii.edu/},
\citealt{2002SPIE.4836..154K}, \citealt{2004SPIE.5489...11K}), LSST
({\slshape http://www.lsst.org/lsst\_home.html}, \citealt{2002SPIE.4836...10T}), and 
SNAP ({\slshape http://snap.lbl.gov/}, \citealt{2004APh....20..377R}, \citealt{2004AJ....127.3089M}, \citealt{2004AJ....127.3102R}), some of which may reach statistical precision at the 0.1 per cent 
level. 

Many of our systematic tests are done using the
following method.  Several authors 
(\citealt{2003PhRvL..91n1302J}, \citealt{2004ApJ...600...17B}) have proposed
geometric tests 
of dark energy using the fact that \ds{} is an invariant of the
projected lens mass distribution, and therefore must be the same 
when measured with two different source samples at different
redshifts.  The use of this fact to test the dark energy density and
equation of 
state requires control
of systematics at  the 0.1 per cent level. 
Since systematics in weak lensing are currently only
constrained at the 10 per cent level, we turn
this test around to use \ds{} measured with reference samples to
check for systematic error, knowing that cosmology plays a negligible
roll in the comparison.
If the source samples being compared vary in quantities affecting both
the shear computation
 and the redshift distribution (e.g., if one sample is more distant,
 with lower $S/N$ shape measurement and less well-known redshift
 distribution), then the systematics in both quantities may be
different as well, and we can only use this method to test the overall
calibration of the signal rather than the shear calibration, redshift
distribution, and other effects separately.  Fortunately, the SDSS
now covers a large 
enough area that there is significant statistical power for such tests.

In \S\ref{S:tech}, we describe the data
acquisition, selection criteria, and processing.  The common redshift
distributions used for weak lensing analyses are
discussed in \S\ref{S:zdist}, including specifically  how
these methods are implemented in this paper.  Additional systematics
issues introduced in the computation of the lensing signal are
described in \S\ref{S:othersys}.  Our implementation of the
 test
for systematics is described in \S\ref{S:systest}, and the results
are given in \S\ref{S:results}.  
%An interesting application of our weak
%lensing results, the measurement
%of the mass-to-light ratio of galaxies, is described in
%\S\ref{S:app}.  
Finally, the implications of 
these tests are discussed in \S\ref{S:concl}.

Here we note the cosmological model and units used in this paper.
All computations assume a flat $\Lambda$CDM universe with
$\Omega_m=0.3$ and $\Omega_{\Lambda}=0.7$.  Distances quoted for
transverse lens-source separation are comoving (rather than physical)
\hinvk, where $H_0=100h$~km/s/Mpc.  Likewise, \ds{} is computed
using the expression for \scinv{} in comoving coordinates,
Eq.~\ref{E:scinv}.
In the units
used, $H_0$ scales out of everything, so our results are independent of
this quantity.  All confidence intervals in the text and tables are 95
per cent confidence 
level ($2\sigma$) unless explicitly noted otherwise.

\section{Technical apparatus}\label{S:tech}

In this section, we describe the data used for our computation of the
lensing signal.  The source of this data is the Sloan Digital Sky
Survey (SDSS), an ongoing survey that will eventually image
approximately one quarter of the sky (10,000 square degrees).
Imaging data is taken in drift-scan mode in 5 filters, $ugriz$,
centered at 355, 469, 617, 748, and 893 nm respectively
\citep{1996AJ....111.1748F} using a wide-field
CCD \citep{1998AJ....116.3040G}.   After the computation of an
astrometric solution \citep{2003AJ....125.1559P}, the imaging data are
processed by a 
sequence of pipelines, collectively called {\sc Photo}, that estimate
the PSF and sky brightness, identify objects, and measure their  
properties. The software pipeline and photometric quality assessment
is described in
\cite{2004astro.ph.10195I}.  Bright galaxies and other interesting
objects are 
selected for spectroscopy according to specific 
criteria (\citealt{2001AJ....122.2267E};
\citealt{2002AJ....124.1810S}; \citealt{2002AJ....123.2945R}).   The SDSS has had four major data releases: the Early Data Release or
EDR \citep{2002AJ....123..485S}, DR1 
\citep{2003AJ....126.2081A},  DR2
\citep{2004AJ....128..502A}, and  DR3
\citep{2005AJ....129.1755A}.   
While we use imaging data more up-to-date than DR3, we are limited in
area 
by the spectroscopic coverage available to us because
spectroscopy lags significantly behind photometry.

\subsection{Lens catalog}

The lens (foreground) galaxies used for this study are included in the
SDSS main galaxy spectroscopic sample (\citealt{2002AJ....124.1810S}), 
%\citealt{2000AJ....120.1579Y};
%\citealt{2003AJ....125.2276B}
as part of the NYU Value-Added Galaxy Catalog (VAGC,
\citealt{2004astro.ph.10166B}), though the version of the VAGC used
here includes more area than the public one described in
\cite{2004astro.ph.10166B}. 
The VAGC is used because of its consistent overall calibration
\citep{schlegel04a}.
The sample, after redshift and magnitude cuts to be described below,
includes 259~609 
galaxies (decreased from 314~906 after exclusion of the southern
Galactic region).
We only use lenses at
redshift $z>0.02$ because of the computational expense of computing
pairs out to 2 \hinvm{} for lenses at lower redshifts, and because
the lower redshift galaxies have low \scinv{} and therefore contribute
little weight.
Furthermore, for this study, we only use galaxies with $r$-band
Petrosian absolute magnitude $-23\le\mr\le
-17$, divided into six magnitude bins, each one magnitude wide.  The spectra
used were processed by a separate pipeline at Princeton
\citep{schlegel04b}.  The fluxes were extinction-corrected using dust
maps from~\cite{1998ApJ...500..525S}, then
$k$-corrected to $z=0.1$ using {\sc kcorrect} v1\_11 
with values given directly in the VAGC catalog.
%; the distance
%modulus computation assumes $h=1$.  
%Fig.~\ref{F:specarea} shows the
%area covered by the spectroscopic sample used for this paper.
%\begin{figure}
%\includegraphics[width=3in,angle=0]{specarea_cropped.ps}
%\caption{\label{F:specarea}The area covered by the spectroscopic
%  galaxies used for this work, shown as an Aitoff projection using J2000
%  equatorial coordinates.}
%\end{figure}

This sample is approximately flux-limited to Petrosian apparent magnitude
$r=17.77$; all absolute magnitudes for the lens sample used in this
paper are Petrosian $r$-band magnitudes.
Redshift-evolution of luminosity consistent
with~\cite{2003ApJ...592..819B} was used, so that 
the absolute magnitude used for all cuts was
\begin{equation}
\mr(\mathrm{used}) = \mr(\mathrm{measured}) + 1.6(z-0.1)
\end{equation}
The effect of this shift is to include higher luminosity lenses with
$\mr(\mathrm{measured}) < -23$ in the
brightest luminosity bin because their \meanz{} is greater than 0.1, and
to include fainter lenses in the faintest bin, because their \meanz{} is
approximately 0.03.  
\begin{comment}
Furthermore, since our absolute magnitudes are
referenced to 10 $h^{-1}$pc rather than 10 pc, the relation between
the true absolute magnitude and the one used is
\begin{equation}
\mr(\mathrm{true}) = \mr(\mathrm{used}) + 5\,\mathrm{log}_{10}h
\end{equation}
\end{comment}

\begin{figure}
\includegraphics[width=3in,angle=0]{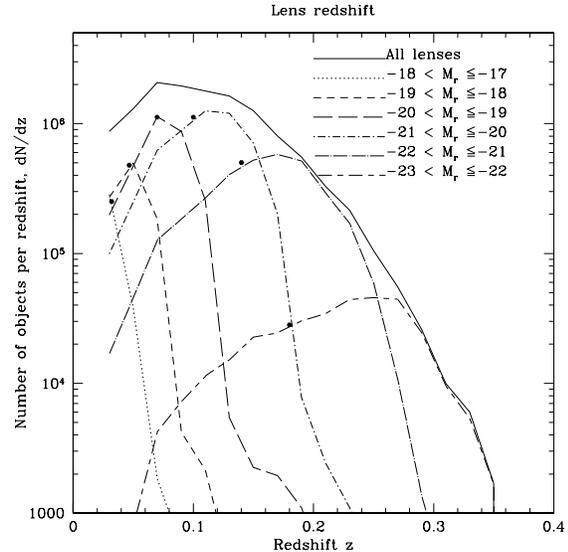}
\caption{\label{F:lens_zdist}The redshift distribution for lens
  galaxies, shown overall and for the 6 luminosity bins.  As shown,
  brighter samples peak at higher redshifts.  The cutoff at $z=0.02$
  was imposed artificially on the sample.  Solid points are used to
  indicate the weighted mean redshift of each lens sample.}
\end{figure}

%Fig.~\ref{F:magdist}  shows the absolute magnitude distribution of the lens
%galaxies (including redshift evolution), and Fig.~\ref{F:lens_zdist}
%shows their redshift distribution.  
Fig.~\ref{F:lens_zdist} shows the lens redshift distribution.   The
\mr{} limits, mean  
redshifts, and widths of the distribution for the 6 luminosity bins
used here are shown in 
table~\ref{T:lenses}, as is the mean effective redshift (taking into
account the weights used for the computation of signal) and mean
effective luminosity relative to $L_*$ (with
$M_*=-20.44$ as in~\citealt{2003ApJ...592..819B}) since they are more
relevant for the lensing signal.  Note that $z_{\rm eff}<\meanz$
because the larger number of pairs 
for lower redshift lenses (due to the larger angular scale associated
with the fixed transverse comoving scale) overcomes the fact that
\scinv{} for a given source redshift increases with lens redshift.

%\begin{figure}
%\includegraphics[width=3in,angle=0]{magdist.ps}
%\caption{\label{F:magdist}The magnitude distribution for the lens
%  sample.}
%\end{figure}

\begin{table}
\caption{\label{T:lenses} For each luminosity bin, the number of lens galaxies, \meanz{} and $\sigma(z)$ (a characteristic width,
  though the distribution is not Gaussian), 
  the mean weighted redshift $z_{\rm eff}$, and the mean weighted luminosity
  $L_{\rm eff}$ relative to $L_*$.}
\begin{tabular}{crllll}
\hline\hline
Sample, $\mr$ range$\!\!\!\!$ & $N_{gal}$  & \meanz  &
$\sigma(z)$ ~ & $z_{\rm eff}$ & $\!\!L_{\rm eff}/L_*$ \\
\hline
L1, $\![-18,-17)$  &   6 524  & 0.032$\!\!$  & 0.011$\!\!$ & 0.032$\!\!$ & 0.080$$ \\
L2, $\![-19,-18)$  &  19 192  & 0.048$\!\!$  & 0.015$\!\!$ & 0.047$\!\!$ & 0.20 \\
L3, $\![-20,-19)$  &  58 848  & 0.074$\!\!$  & 0.021$\!\!$ & 0.071$\!\!$ & 0.49 \\
L4, $\![-21,-20)$  & $\!\!$104 752  &  0.11  & 0.03 & 0.10 & 1.2 \\
L5, $\![-22,-21)$  &  63 794  &  0.16  & 0.05 & 0.14 & 2.5 \\
L6, $\![-23,-22)$  &   6 499  &  0.22  & 0.06 & 0.19 & 5.6 \\
\hline\hline
\end{tabular}
\end{table}

One important fact about the lens samples L1--L6 is that they are
flux-limited, not volume limited.  As a result, since
$\Delta\Sigma$ is averaged over lens-source pairs with a weight
proportional to $\Sigma_c^{-2}(z_l, z_s)$, the mean effective redshift
and luminosity may vary depending on the redshift distribution of the
source sample.  When comparing \ds{} for a given
lens sample but different source samples, we explicitly computed the
mean effective redshift $z_{\rm eff}$ and luminosity $L_{\rm eff}$ of
each lens sample with each 
source sample (representative values are given in
table~\ref{T:lenses}).  Variations in $1+z_{\rm eff}$ and $L_{\rm
  eff}$ for the same lens sample but different source sample were
found to be quite small, a maximum of 2 per cent;  
because this variation is so small, and because we lose significantly
in statistics by going
to a volume limited sample, we choose to keep the full flux-limited
sample and, when necessary, apply corrections to the computed $\Delta\Sigma$
when comparing between different source subsamples.  
%%Of course, this
%%procedure is not necessary when combining data computed using two
%%different source subsamples, since we simply get the signal at the
%%weighted mean redshift and luminosity of the two individual sources
%%samples, but it is necessary when comparing the signal for those
%%subsamples.  
Corrections will
be described further in \S\ref{SS:notvol}.

\subsection{Source catalog}\label{S:sourcecatalog}

\subsubsection{Constructing the catalog}\label{SSS:construct}
The source sample consists of galaxies selected from the SDSS
photometric catalog 
(\citealt{2000AJ....120.1579Y}; \citealt{2001AJ....122.2129H};
\citealt{2002AJ....123..485S}; \citealt{2002AJ....123.2121S};
\citealt{2003AJ....125.1559P}; \citealt{2003AJ....126.2081A}). The
catalog contains information about the images from the SDSS
camera \citep{1998AJ....116.3040G} processed at Princeton by the {\sc
  Photo} software (\citealt{2001adass..10..269L}; \citealt{dfink04}),
rerun 137.  Note that for the source 
catalog, we use the model magnitudes in all 5 bands rather than Petrosian
magnitudes, because of their higher signal-to-noise for fainter
galaxies.
\begin{comment}
\footnote{Unlike EDR and DR1 data, this catalog does not include the
model magnitude bug, a serious error in the determination
of the model magnitude most noticably for larger (extended) galaxies,
and consequently the model magnitude may reasonably be used rather
than the Petrosian magnitude.}
\end{comment}

The source catalog used for this work is not the same as that used in
H04.  Here we describe the catalog used, emphasizing differences 
from the previous one.  The most trivial difference is the size of the
dataset: here we use imaging data acquired from 1998 September 
19 (run 94) through 2004 June 15 (run 4682), whereas the H04 catalog
did not include imaging data acquired after 2003 March 10 (run 
3712); but there are also significant changes in the pipeline.  There
are several steps involved in the development of the catalog: 
(1) object selection based on {\sc Photo} outputs, (2) shape
measurement, and cuts on the shape measurement, (3) other cuts, and
(4) organization.

We begin by describing basic object selection starting from the {\sc
  Photo} outputs.  Star/galaxy separation was accomplished using the
  {\sc Photo} pipeline output OBJC\_TYPE\footnote{OBJC\_TYPE
  classifies objects as ``galaxies'' if the flux estimated from the
  linear combination of de  
Vaucouleurs and exponential profiles (composite model magnitude, or
  cmodel magnitude) fit to the object exceeds the flux
  estimated from the best-fit PSF by at least 0.24 magnitudes.   
This works because the profile fit will pick up more of the light from an
  extended object than the PSF fit.  At faint magnitudes ($r>21$)  
this separation scheme mistakes some stars as galaxies; see
  \S\ref{SS:sgsep}.}, and the cut on 
  resolution factor described below should further reduce
  stellar contamination.  We defer discussion of possible
stellar  contamination to \S\ref{SS:sgsep}. Unlike the catalog used for
H04, this catalog includes deblended child galaxies.
Because the deblender has
been significantly improved for DR2, the
phenomenon noted for EDR and DR1 that the deblender sometimes
``shreds'' large galaxies rarely occurs (according to \citet{2004AJ....128..502A}, inspection of several hundred deblends indicates that they are correct roughly 95 per cent of the time).  While shape measurement was
  performed for all galaxies brighter than magnitude 22.0 in $r$-band and 21.6 in
  $i$-band (no requirements on detection in $g$-band), these cuts were
  applied using model magnitudes before the extinction correction.
  Several other cuts on the {\sc Photo} flags were performed: the
  galaxy must have been detected in unbinned images in the $r$ and $i$ bands;
also, several flags indicating problems in shape measurement or problems
with the image (e.g., interpolated pixels) must not have been set.

Next, we describe the shape measurement determination. The
PSF-correction algorithm used for this work was the
``re-Gaussianization'' 
scheme described and tested in \citet{2003MNRAS.343..459H}. 
Recent SDSS lensing works, including \citet{2004AJ....127.2544S} and 
H04, have used the linear scheme described there, but as shown in
\citet{2003MNRAS.343..459H}, 
re-Gaussianization is much more successful at avoiding 
various shear calibration problems, reducing them to the several per
cent level (rather than $\sim$10 per cent) even for 
poorly-resolved galaxies.  Unlike the linear scheme, which involves
correcting the measured adaptive moments of the image by factors 
involving the adaptive moments of the PSF, re-Gaussianization involves
fitting the PSF shape to a Gaussian, and using the deviations 
of the PSF from Gaussianity in the PSF correction.  The
re-Gaussianization method was implemented by reading the atlas images
and the 
PSF maps from {\sc Photo} \citep{2002AJ....123..485S}, since it is
impossible to implement using the object catalogs alone. 

Re-Gaussianization is a perturbative PSF correction scheme based on
the observation that if the PSF $P$ and the pre-seeing galaxy 
image $f$ are Gaussians, and have covariance matrices ${\mathbfss
  M}^{(P)}$ and ${\mathbfss M}^{(f)}$, then the observed image  
$I=P\otimes f$ (here $\otimes$ represents two-dimensional convolution)
is a Gaussian of covariance ${\mathbfss M}^{(f)} + {\mathbfss  
M}^{(P)}$.  A simple PSF correction scheme is thus to find the
covariance matrices ${\mathbfss M}^{(P)}$ of the PSF and ${\mathbfss  
M}^{(I)}$ of the observed galaxy image, and estimate
\begin{equation}
{\mathbfss M}^{(f)} = {\mathbfss M}^{(I)}-{\mathbfss M}^{(P)}.
\label{eq:mfip}
\end{equation}
In practice, galaxy shapes are not perfectly Gaussian but one can fix
this by finding $A$, ${\bmath x}_I$, and ${\mathbfss M}^{(I)}$  
that minimizes
\begin{equation}
\int \left| I({\bmath x}) - A\rme^{-({\bmath x}-{\bmath x}_I)\cdot
{\mathbfss M}^{(I)\,-1}({\bmath x}-{\bmath x}_I)/2}
 \right|^2\,\rmd^2{\bmath x}.
\label{eq:adaptive}
\end{equation}
The covariance matrix ${\mathbfss M}^{(I)}$ so obtained is known as
the ``adaptive'' covariance matrix, and its trace $T^{(I)}$ is  
known as the ``adaptive'' trace.  In principle one can evaluate
Eq.~(\ref{eq:mfip}), and then 
estimate the galaxy ellipticity
\begin{equation}
(e_+^{(f)}, e_\times^{(f)}) = \frac{(M^{(f)}_{xx}-M^{(f)}_{yy},\; 2M^{(f)}_{xy})}{T^{(f)}}.
\label{eq:ef-moment}
\end{equation}

In practice, Eq.~(\ref{eq:ef-moment}) does not work very well for real
PSFs and galaxies -- both PSF and galaxy tend to have sharper 
central peaks and wider tails than Gaussians with the adaptive
covariance -- and thus two corrections are made.  The first  
correction, due to \citet{2002AJ....123..583B}, accounts for the
non-Gaussianity of the galaxy.  If the PSF is circular,  
Eq.~(\ref{eq:ef-moment}) reduces to
\begin{equation}
{\mathbfss e}^{(f)} = \frac{ {\mathbfss e}^{(I)} }{ R_2^{(I)} },
\label{eq:ef-res}
\end{equation}
where $R_2^{(I)} = 1 - T^{(P)}/T^{(I)}$
is the resolution factor.  \citet{2002AJ....123..583B} worked to first
order in $T^{(P)}$ for a non-Gaussian galaxy, and found that 
Eq.~(\ref{eq:ef-res}) still applies, but with $R_2^{(I)}$ replaced by
the non-Gaussian resolution factor 
\begin{equation}
R_2^{(I)}({\rm NG}) = 1 - \frac{ T^{(P)} }{ T^{(I)} }\,
\frac{1+a_4^{(I)}}{1-a_4^{(I)}}, 
\label{eq:r2ng}
\end{equation}
where $a_4^{(I)}$ is the dimensionless kurtosis of the galaxy, defined
to be zero for a Gaussian (see \citealt{2002AJ....123..583B}  
for a precise definition).  This equation can be generalized to an
elliptical PSF by requiring $SL(2,\mathbb{R})$ shear invariance,  
and is found to work well for Gaussian PSFs in ``toy'' simulations
\citep{2003MNRAS.343..459H}. 

The second correction required is for the non-Gaussianity of the PSF.
This correction begins by finding the Gaussian $G({\bmath x})$ 
that best fits the PSF $P({\bmath x})$ according the unweighted
least-squares method, i.e. minimizing $\int |G-P|^2\,\rmd^2{\bmath 
x}$.  It then re-normalizes $G$ to integrate to unity -- a condition
not always satisfied by the best-fit Gaussian even though $\int 
P({\bmath x})\,\rmd^2{\bmath x}=1$ -- and finds the residual
$\epsilon({\bmath x}) = P({\bmath x}) - G({\bmath x})$.  Next the 
pre-seeing image of the galaxy is approximated by a Gaussian
$f_0({\bmath x})$ whose covariance matrix ${\mathbfss M}^{(f_0)}$ is 
obtained by subtracting the adaptive covariance matrices ${\mathbfss
  M}^{(f_0)} = {\mathbfss M}^{(I)}-{\mathbfss M}^{(P)}$, and a 
``re-Gaussianized'' image $I'=I-\epsilon\otimes f_0$ is constructed,
which is supposed to approximate what would have been observed 
had the PSF been Gaussian.  The \citet{2002AJ....123..583B}
prescription is then applied to $I'$ using $R_2^{(I')}({\rm NG})$. 

The re-Gaussianization scheme is exact to first order in PSF
non-Gaussianity, however higher-order approaches have been proposed
based 
on expansions of the galaxy and PSF in orthogonal functions
(\citealt{2003MNRAS.338...35R}; \citealt{2002AJ....123..583B}).   
Suggestions include direct fitting of the convolved galaxy to the data
\citep{2002AJ....123..583B} or deconvolution regularized by a 
cutoff in the orthogonal function expansion
\citep{2003MNRAS.338...48R}.  These methods have not yet come into
general use, but are 
likely to become more widely used in the future due to the demanding
calibration requirements of cosmic shear surveys.  However, we 
note that galaxy-galaxy lensing is a promising ``testing ground'' for
these methods since the same systematics tests that we use to test 
redshift distributions in \S\ref{S:systest} could also be used to
study the relative calibrations of the various PSF correction 
methods.  Such tests could be done independent of redshift
distribution information by using the same sets of sources for a
comparison of the shear $\gamma_t$ computed from ellipticities
determined by each PSF-correction method.

Only galaxies passing certain cuts on the shape measurement were
included in the catalog.  The shape
measurements used were the average of those in the $r$ and $i$
bands; the $u$, $g$, and $z$ bands were not used because their
lower 
signal-to-noise did not justify the large computational expense of
performing the re-Gaussianization.  To eliminate galaxies that may cause large
noise-rectification bias, 
to avoid the untrustworthy results of PSF correction when the
galaxy is unresolved, and to help with star-galaxy separation, we only
include galaxies with resolution factor $R_2 >1/3$; 
$R_2$ here is defined as  
\begin{equation}
R_2 = 1 -\frac{T^{(P)}}{T^{(I')}},
\end{equation}
where $T^{(P)}$ is the trace of the moment matrix for the PSF and
$T^{(I')}$ is that quantity for the re-Gaussianized galaxy image.
Note that this cut is only applied in the 
bands used for shape measurement, and since we only attempt shape
measurement in 
$r$ and $i$, it is possible that $R_2<1/3$ in the other bands (indeed,
the object may not even be visible in $u$, $g$, or $z$).  We require
that the galaxy pass this cut in both bands, since if we only require
it in one band, then we can create a selection bias by preferentially
using the shape measurement that has $R_2>1/3$ over that with $R_2<1/3$.

Once shape measurement was complete, several other types of cuts were
applied.  To ensure a relatively uniform source sample
across the survey area and to avoid regions near the Galactic plane,
only regions in which the extinction was lower 
than 0.2 magnitudes in $r$-band were used. The extinction was
determined using dust maps from~\cite{1998ApJ...500..525S}. To ensure
a uniform sample, we also require $r<21.8$ (extinction-corrected).   

The shape error estimates $\sigma_e$ are in principle used for three
purposes: weighting, determination of the shear responsivity, and 
determination of the error bars on final quantities such  as
$\Delta\Sigma$.  We obtain the errors due to Poisson fluctuations in
the 
sky and CCD dark current via the simple formula (appropriate for
Gaussians; \citealt{2002AJ....123..583B}) 
\begin{equation}
\sigma_e({\rm sky+dark}) = \frac{\sigma^{(I)}}{ R_2F}\sqrt{4\pi n},
\label{eq:sigma-sky}
\end{equation}
where $n$ is the sky and dark current brightness in photons per pixel,
$\sigma^{(I)\,4}=\det {\mathbfss M}^{(I)}$ is the size of the  
galaxy in pixels, 
and $F$ is the flux.
This equation is crude, but for the purposes of weighting there is no
need for high accuracy, and the error bars on our results are 
computed via analytic, random catalog, and bootstrap methods that
depend on the actual dispersion of the ellipticities, including 
shape noise, rather than $\sigma_e$ itself.  (The responsivity
determination is addressed in \S\ref{SSS:shearcalibration}.)  There is 
also a contribution to the shape measurement uncertainty due to
Poisson noise from the galaxy itself; this is given by (again for  
Gaussians) 
\begin{equation}
\sigma_e({\rm gal}) = R_2^{-1}\sqrt{\frac{64}{27N_e}}
\label{eq:sigma-gal}
\end{equation}
where $N_e$ is the number of photoelectrons from the galaxy.  This is
only significant for galaxies bright compared to the sky; 
the typical sky brightness is 21.0~mag~arcsec$^{-2}$ in $r$ band and
20.3~mag~arcsec$^{-2}$ in $i$, and even a poorly resolved ($R_2=1/3$) 
galaxy usually has a full width at half maximum of $\sim 1.7$~arcsec
after seeing, so sky brightness dominates for $r\ge 20.0$ 
and $i\ge 19.3$.  Since the shape noise is dominant over measurement
noise for the brighter objects, we have not included the galaxy 
noise (Eq.~\ref{eq:sigma-gal}) in our weighting, nor has it been
included in the adaptive moment errors from {\sc Photo}. 
 
Some additional cuts were designed to eliminate regions with faulty data.
These cuts eliminated less than 1 per cent of the data total.   First,
the mean ellipticities $\langle e_1\rangle$ and $\langle 
e_2\rangle$, and their rms deviations, were computed on a run/camcol
basis.  Those few run/camcols that had $|\langle e \rangle| > 0.05$ (for either
ellipticity component, in either band) were excluded from the
analysis; visual inspection of several of those runs showed severe PSF
anisotropy (despite a reasonable PSF FWHM) for which our correction
scheme was unable to account. 
Furthermore, based on the distribution in rms ellipticities, those
run/camcols with values less than 0.38 or greater than 0.52 were
excluded (the mean value was 0.45, higher than the expected
shape noise since all galaxies, even those that had significant
measurement error, were included).  Those with rms ellipticities above
the acceptable range typically were imaged in particularly poor
seeing, which led to greater noise in the PSF-corrected
ellipticities.  Within runs that were accepted,
galaxies with total ellipticity $e^2 = e_1^2 + e_2^2 >4$ were
rejected.  Finally, those galaxies in a small
region ($\sim 4$ square degrees) that had faulty astrometry were eliminated.

Once these cuts were applied, a few final steps were
necessary to make the catalog useful.  In the case of multiple
observations of the same galaxy, only one 
observation was used, that which was taken in better seeing.  
%All other observations were not used.  
The shape measurements in the two bands  
%were compared individually to check for any systematic differences,
%but for the general analysis, they 
were combined, weighting by the
$(S/N)^2$ of the detection in each band.

Unlike in H04, photometric redshifts were assigned to all
extinction-corrected $r<21$
galaxies using a template-based program {\sc kphotoz}
v3\_2~\citep{2003AJ....125.2348B}.   The performance of these
photometric redshifts will be 
described in \S\ref{SS:zresults}.   Approximately 4 per cent of
galaxies with $r<21$ had failed photometric redshift determination, and so were
not used for any analysis requiring the use of photometric redshifts.

\begin{table}
\caption{\label{tab:sourcecat}Source catalog properties.  Note that only about 65 per cent of the sources are in regions with 
spectroscopic coverage, and hence only these are used in computing
$\Delta\Sigma$.  Also, because we require a successful measurement in
both bands, only the number measured in both $r$ and $i$
is relevant.}
\begin{tabular}{lcr}
\hline\hline
Sky coverage, $f_{sky}$ & & 0.176 \\
\hline
Successful measurements & $r$ or $i$ & 39 436 326 \\
(all galaxies) & $r$ band & 35 789 302 \\
 & $i$ band & 35 344 893 \\
 & $r$ and $i$ & 31 697 869 \\
\hline
Successful measurements & $r<21$ & 18 709 472 \\
($r$ and $i$) & $21\le r<21.8$ & 12 988 397 \\
\hline
Source density & $r$ or $i$ & 1.51 arcmin$^{-2}$ \\
(all galaxies) & $r$ and $i$ & 1.21 arcmin$^{-2}$ \\
\hline
Resolution factor $\langle R_2\rangle$ & $r$ band & $0.61\pm 0.15$ \\
(mean $\pm$ std. deviation) & $i$ band & $0.60\pm 0.16$ \\
\hline
Mean magnitude & $\langle r\rangle$ & 20.68 \\
(extinction corrected) & $\langle i\rangle$ & 20.22 \\
\hline
High-redshift LRGs, $r$ and $i$ & & \\
~~~~~successful measurements & & 2 884 242 \\
~~~~~source density & & 0.11 arcmin$^{-2}$ \\
~~~~~mean redshift & & 0.55 \\
~~~~~mean resolution factor & $\langle R_2(r)\rangle$ & 0.58 \\
& $\langle R_2(i)\rangle$ & 0.55 \\
~~~~~mean magnitude & $\langle r\rangle$ & 20.93 \\
& $\langle i\rangle$ & 20.03 \\
\hline\hline
\end{tabular}
\end{table}

\begin{figure}
\includegraphics[width=3in,angle=0]{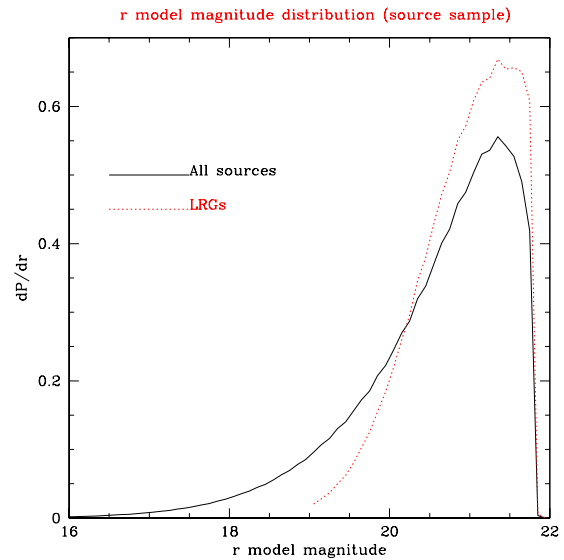}
\caption{\label{F:rhist} The $r$-model magnitude distribution for the
  source catalog, shown for the full catalog (solid) and high-redshift
  LRGs (dotted line).
  The turnover at faint magnitudes occurs because we lose 
  many sources at the faint end due to our
  selection criteria for the shape measurement (e.g., the cut on
  $R_2$).}
\end{figure}

\begin{figure}
\includegraphics[width=3in,angle=0]{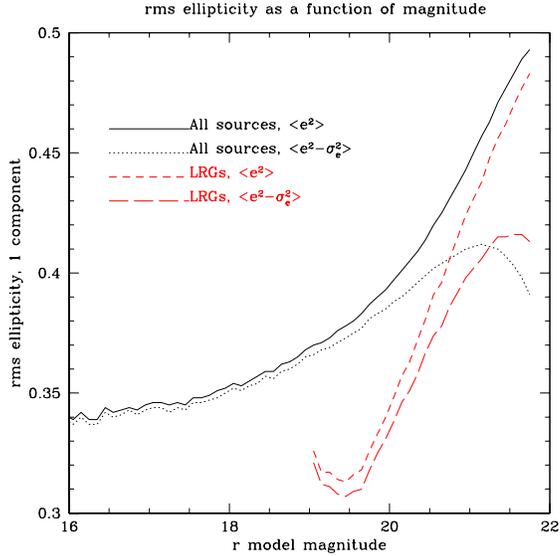}
\caption{\label{F:ehist} The rms ellipticity (for a single ellipticity
  component, averaged over both bands) as a function of magnitude,
  shown for all sources, and 
  for the high-redshift LRG sample.  We show the results both with and
  without measurement noise (including noise from the galaxy itself),
  as labeled.  The increase in the 
  result with noise at faint
  magnitudes shows the
  effects of noise in the shape measurement, but it seems that the
  increase in the result without noise at faint magnitudes may
  indicate a real trend in 
  the rms ellipticities of the galaxies with magnitude.}
\end{figure}

\begin{figure}
\includegraphics[width=3in,angle=0]{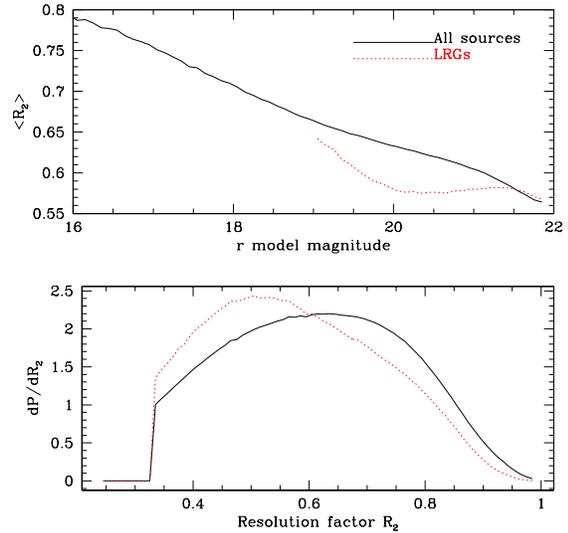}
\caption{\label{F:r2hist} The top plot shows the average $R_2$ value for the
  source catalog as a function of magnitude for all sources (solid)
  and high-redshift LRGs (dashed).  As expected, $\langle
  R_2 \rangle$ is higher for brighter galaxies, indicating that their
  shapes are better-resolved.  The bottom plot shows the
  distribution of $R_2$ values.  The
  plot is shown for the $r$-band $R_2$; results are nearly identical for
  the $i$-band $R_2$.}
\end{figure}

For reference, we include here some plots 
showing information about the source sample (these plots also show
information about the high-redshift Luminous Red Galaxy, or LRG
sample, a subsample of 
the source catalog, that will be described in more detail in
\S\ref{SS:LRG}).  The magnitude 
distribution of sources in the catalog is shown in
figure~\ref{F:rhist}.  The rms ellipticity as a function of magnitude
is shown in figure~\ref{F:ehist}.  The plot of the average $R_2$ as a
function of magnitude, and of the overall distribution of
$R_2$ values, is in Fig.~\ref{F:r2hist}. Some general information
about the catalog is included in table~\ref{tab:sourcecat}.  Note that
for all tests performed in this paper, we divide the sources into
three samples: $r<21$, $r>21$, and high-redshift LRGs.

\subsubsection{Shear calibration bias}\label{SSS:shearcalibration}

Here, we list the sources of shear calibration bias that were
  described in detail in H04, and estimate their magnitudes for the source
  catalog used here; refer to that paper for more detail about
  estimated shear calibration uncertainty.  

\begin{figure*}
\includegraphics[height=6in,angle=-90]{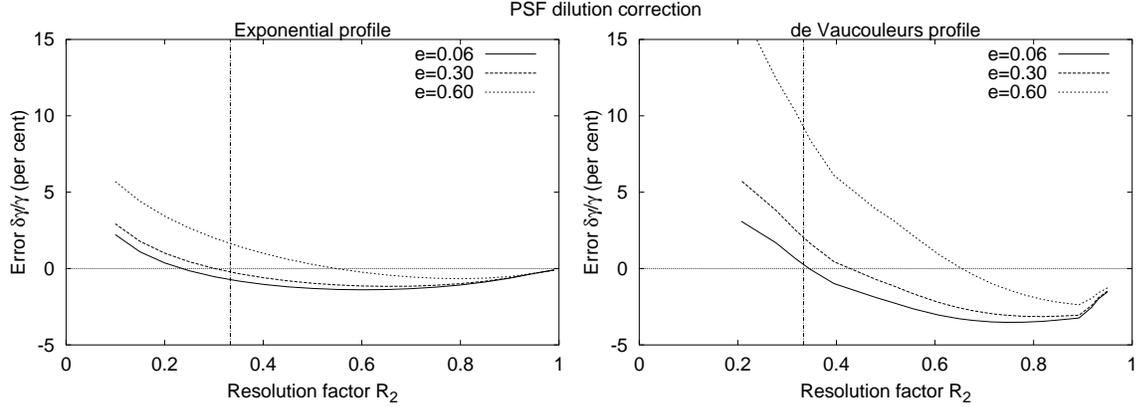}
\caption{\label{F:psfdilution} PSF dilution correction for the two
  types of galaxy profiles, as a function of $R_2$, for different
  values of source ellipticity (with our cut at $R_2=1/3$ shown as a
  vertical dotted line).  For the de Vaucouleurs profile, 
  the usual Gaussian relation $R_2=1/(1+\sigma^{(P)\,2}/\sigma^{(f)\,2})$
  breaks down -- the Gaussian $\sigma^{(f)\,2}$ fits the central cusp of the
  de Vaucouleurs profile and misses much of the light.  This is why the
  values for, e.g., $R_2=1/3$ shown here disagree with the
  $\sigma^{(P)\,2}/\sigma^{(f)\,2}=2.0$ 
  curves in \citet{2003MNRAS.343..459H}.}
\end{figure*}

There are five major sources  of shear calibration bias, as listed in
H04.  First, we consider the PSF dilution correction, the
correction to the measured galaxy image to account for the blurring
due to convolution with the PSF.  Unlike the linear PSF-correction
method using in H04, which has significant shear
calibration uncertainty due to this effect for the less well-resolved
galaxies, the re-Gaussianization method only has a few per cent shear
calibration uncertainty \citep{2003MNRAS.343..459H} even for the lower
limit $R_2=1/3$ studied in that paper.  For this paper, a plot of the
PSF dilution correction as a function of $R_2$, for both exponential and
de Vaucouleurs profiles, for various values of source ellipticity, is
shown in Fig.~\ref{F:psfdilution}.

We model the calibration error due to the PSF dilution correction as
due to a sum of contributions from exponential profile and de
Vaucouluers profile galaxies, 
\begin{equation}
{\delta\gamma\over\gamma}=f_{exp}\left({\delta\gamma\over\gamma}\right)_{exp}
+f_{deV}\left({\delta\gamma\over\gamma}\right)_{deV},
\end{equation}
where $f_{exp}$ and $f_{deV}$ are the weighted fractions of the two
types of galaxies with $f_{exp}+f_{deV}=1$.  A lower bound on
$\delta\gamma/\gamma$ can be obtained from Fig.~\ref{F:psfdilution} by
noting that $\delta\gamma/\gamma\ge -0.014$ for the exponential
profile and $\delta\gamma/\gamma\ge -0.035$ for the de Vaucouleurs profile,
hence
\begin{equation}
{\delta\gamma\over\gamma} \ge -0.014f_{exp}-0.035f_{deV}.
\end{equation}
For the upper bound, we repeat this calculation, except that we use
the $\delta\gamma/\gamma$ corresponding to $R_2=1/3$ and ellipticity
equal to $\sqrt{2}e_{rms}$ (since this is the rms total ellipticity).
This calculation is conservative since most galaxies are at $R_2>1/3$
where $\delta\gamma/\gamma$ is less (at fixed $e$).  (We find that
$R_2$ and $e^2$ are almost completely uncorrelated after the noise
$\sigma_e^2$ is subtracted from $e^2$ for each of our three samples,
and for both the de Vaucouluers and exponential sources within each of
these samples.)  The results are shown in Table~\ref{tab:dgg}; these
should be interpreted as $2\sigma$ bounds, since it is likely that
some cancellation between positive and negative dilution galaxies occurs. 

\begin{table*}
\caption{\label{tab:dgg}Shear calibration biases and other parameters
  for the various source samples, at the $2\sigma$ level.} 
\begin{tabular}{lccc}
\hline\hline
Sample & $r<21$ & $r>21$ & LRG \\
$f_{exp}$ & 0.59 & 0.58 & 0.33 \\
$e_{rms}$(exp) & 0.39 & 0.41 & 0.41 \\
$e_{rms}$(deV) & 0.38 & 0.42 & 0.37 \\
\hline
Calibration bias (per cent)$\!\!\!\!\!\!\!\!$ & & & \\
~~~~PSF dilution & $[-2.2,+2.9]$ & $[-2.2,+4.0]$ & $[-2.8,+3.9]$ \\
~~~~PSF reconstruction & $\pm 2.1$ & $\pm 2.4$ & $\pm 2.5$ \\
~~~~Selection bias & $[0,5.7]$ & $[0,10.3]$ & $[0,11.1]$ \\
~~~~Shear responsivity error & $[0.0, 1.7]$ & $[0.0, 1.7]$ & $[0.0, 1.7]$ \\
~~~~Noise rectification & $[-1.0,0]$ & $[-3.8,0]$ & $[-1.2,0]$ \\
\hline
Total $2\sigma$ $\delta\gamma/\gamma$ (per cent) & $[-5,+12]$ &
$[-8,+18]$ & $[-6,+19]$ \\ 
\hline\hline
\end{tabular}
\end{table*}

Next, we consider errors in PSF reconstruction, which can arise if the
PSF ellipticity or trace are misestimated by the {\sc Photo} PSF pipeline.
This error was considered by H04, which showed that the PSF trace is
accurately reconstructed to within $|\delta T^{(P)}/T^{(P)}|<0.03$. 
We can use Eq.~(20) of H04 to estimate the resulting calibration error;
we find weighted averages $\langle R_2^{-1}\rangle=1.71$, $1.81$, and
$1.90$ for the $r<21$, 
$r>21$, and LRG samples\footnote{The $\langle R_2^{-1}\rangle$ values
  are the same for the $r$ and $i$ bands to within $\pm 0.01$ for the
  $r<21$ and $r>21$   
samples.  For the LRG sample, we find $\langle R_2^{-1}\rangle=1.83$
in $r$ and $1.90$ in $i$; we quote the $i$ value here because the  
signal-to-noise ratio for the LRGs is typically greater in $i$ and
hence this measurement is weighted more heavily.  This is  
certainly conservative as the error increases with increasing $\langle
R_2^{-1}\rangle$.} 
 respectively, resulting in calibration errors of $\pm 2.1$ ($r<21$), $\pm
 2.4$ ($r>21$), and $\pm 2.5$ per cent (LRG). 

We also must be concerned about shear selection bias, the preferential
selection of galaxies at low or high ellipticity.  Considering that
figure~\ref{F:ehist} shows clear evidence for evolution of $e_{rms}$
with magnitude, we do not estimate selection bias using the method from \S3.2.3 of H04, which
assumes no evolution of rms ellipticity with magnitude.
\begin{comment} We applied the method of \S3.2.3 of H04, which estimates the selection
probability as a function of ellipticity by comparing the ellipticity
distributions at different apparent magnitudes, to the $21.5<r<21.8$
versus $r<19$ samples from the new catalog.  (We boosted the noise to
$\sigma_e=0.6$ instead of $0.25$ since the fainter sample has higher
ellipticity noise.)  This gives, using Eq.~(24) of H04, $\delta
\gamma/\gamma = 0.0366$ ($r$ band) or $0.0361$ ($i$ band).  The strength of
this method is that it can be implemented internally to the data without
recourse to theoretical modeling of the selection criteria.  The weakness
of this method is that the underlying assumption -- that the distribution
of ellipticities does not vary with apparent magnitude -- is undoubtedly
violated at some level in the real universe.  Therefore the finding of
$\delta\gamma/\gamma\neq 0$ by this method should not be taken as evidence
that shear selection bias is present.
\end{comment}
An alternative method of estimating the shear selection bias utilizes
a simple model of the selection criteria.    
This method is not dependent on assumptions about the evolution of the
ellipticity distributions.  The main selection criterion that 
can be influenced by shear is the resolution factor cut, which favors
highly elongated galaxies, since these have a larger trace 
$T^{(I)}$ after PSF convolution than a circular galaxy with the same
area.  We model this by noting that for a Gaussian galaxy and  
PSF, the resolution factor obeys
\begin{equation}
R_2 = 1 - {T^{(P)}\over T^{(P)}+2\sigma^{(f)\,2}\sqrt{1-e^{(f)\,2}}},
\end{equation}
where $\sigma_f^2=\sqrt{\det{\mathbfss M}^{(f)}}$ and $e^{(f)}$ is the
ellipticity of the pre-seeing galaxy image.  A gravitational shear along
the $x$-axis leaves $\sigma^{(f)\,2}$ fixed but changes $e^{(f)}$
according to
\begin{equation}
\Delta e^{(f)} = 2(1-e^{(f)\,2})\gamma\frac{e^{(f)}_+}{e^{(f)}},
\end{equation}
where $\gamma$ is the amount of the shear.  Therefore the change in
resolution factor is
\begin{equation}
\Delta R_2 =\left.
{\partial R_2\over\partial e^{(f)}}\right|_{\sigma^{(f)}}
\Delta e^{(f)} = 2e^{(f)}R_2(1-R_2)\gamma\frac{e^{(f)}_+}{e^{(f)}}.
\end{equation}
Now the effect of the $R_2>1/3$ cutoff on the mean ellipticity can be
estimated by averaging the ellipticities of the galaxies that are
accepted into the catalog because of the shear (the integrand is  
negative when galaxies are removed or their $e^{(f)}_+<0$):
\begin{eqnarray}
\Delta\langle e^{(f)}_+\rangle \!\! &=& \int_0^1\rmd e^{(f)}
\left.{\rmd n\over \rmd e^{(f)} \rmd R_2}\right|_{R_2=0.33}
\int_0^\pi {\rmd\phi^{(f)}\over \pi}
e^{(f)}_+\Delta R_2
\nonumber \\
&\approx & 2 R_{2,\rm min}(1-R_{2,\rm min})\gamma e_{rms}^2 n(R_{2,\rm min}),
\end{eqnarray}
where $\phi^{(f)}$ is the position angle\footnote{Defined by
$e^{(f)}_+\pm\rmi e^{(f)}_\times = e^{(f)}\rme^{\pm 2\rmi\phi^{(f)}}$.}
and $\rmd n/\rmd e^{(f)} \rmd R_2$ is the joint ellipticity-resolution
factor distribution. 
In the last line we have approximated $e^{(f)}$ and $R_2$ as
independent, which we have found to be very nearly true, and noted that the
mean  value of $e^{(f)\,2}$ is $2e_{rms}^2$ because the ellipticity
has 2 components. The shear calibration error due to selection at the
$R_2$ cut, assuming that all galaxies are weighted equally, is
\begin{equation}
{\delta\gamma\over\gamma} = \frac{R_{2,\rm min}(1-R_{2,\rm min})}{\cal R} e_{rms}^2 n(R_{2,\rm min}),
\label{eq:rescut1}
\end{equation}
where ${\cal R}$ is the shear responsivity which is discussed further
in subsection~\ref{SS:shearestimator}. Note that $n(R_{2,\rm min})$ cannot be estimated from
Fig.~\ref{F:r2hist} because that plot shows the distribution of $R_2$ values
in the $r$-band.  For this calculation, the relevant quantity is the
distribution of $R_2$ values formed by choosing (for each source) the
lower of the two $R_2$ values, since that number is what determines whether the
object is included in our catalog.  Furthermore, we must use the
distribution of $R_2$ values weighted by the weights used in our
lensing analysis.  
For a typical value of ${\cal R}$ for each sample from
\S\ref{SS:shearestimator},  our cut value $R_{2,\rm
  min}=1/3$, and weighted values $n(R_{2,\rm min})= 1.6$, 2.4,
  and 2.8 for 
  the $r<21$, $r>21$, and LRG samples, respectively, we obtain a
  selection bias estimate $d\gamma/\gamma = 0.057$, 0.103, and 0.111 for
  $r<21$, $r>21$, and LRGs.  

However, the $R_2$ cut is not the only
  one that will cause shear selection bias; the detection requirement that
  $S/N=\nu > 5$ will lead to selection bias in the opposite
  direction, though the magnitude of the effect is not as great (see
  Appendix~\ref{app:sn5}, which shows the calculation of the estimate
  as $d\gamma/\gamma = -0.036$, $-0.066$, and $-0.037$ for the three
  samples respectively). 
  Consequently, we consider the $2\sigma$ estimate of shear selection
  bias to be as low as zero and as high as the values estimated only
  taking into account the $R_2$ selection.

Shear responsivity error is an error in the shear responsivity via a
systematic uncertainty in $e_{rms}$.  We use a value of $e_{rms}({\rm
mag})$ estimated from figure~\ref{F:ehist}, with different results
used for the full source catalog and for the high-redshift LRG
sample.  (Note that in light of our shear selection bias results, we
may consider that the increase of $e_{rms}$ with magnitude is
in part due to shear selection bias, since the average $R_2$ is lower at
fainter magnitudes, and therefore the selection bias should be more
severe there.  Even if this is true, it is still correct to use the
value of $e_{rms}({\rm mag})$ when computing the shear responsivity.)
We estimate the systematic error in $e_{rms}$ assuming that the
uncertainty in $\sigma_e$ is its primary source of uncertainty.  To
estimate uncertainty in $\sigma_e$, we looked at the southern galactic
survey area, for which there are many repeat observations of the same
area (as many as 27 for some areas) that can be used to get empirical
values of $\sigma_e$ that can be compared against the theoretical
value derived from Eqs.~\ref{eq:sigma-sky} and
Eqs.~\ref{eq:sigma-gal}.  We find that $e_{rms}$ is overestimated by
about 0.010 ($2\sigma$ confidence interval $[0.0, 0.02]$), yielding an
estimate of shear calibration bias of $[0.0, 0.017]$ according to
Eq.~25 in H04.

The final major source of shear calibration bias is
noise-rectification bias, whereby the noise in the image leads to a
bias in the ellipticity due to the non-linearity of the PSF correction
process. As shown in H04, equation 26--27 and Appendix
C, the noise-rectification bias can be estimated as
\begin{equation}
\frac{\delta\gamma}{\gamma} \approx K_N \nu^{-2} = 4(1-3R_2^{-1}+R_2^{-2}+2e_{rms}^2)\nu^{-2}
\end{equation}
where $\nu$ is the signal to noise of the detection averaged over
bands:
\begin{equation}
\nu^{-2} = \frac{2}{\nu_r^2 + \nu_i^2}.
\end{equation}
For high-$R_2$ galaxies, $K_N\approx -2.7$, decreasing to $-3.7$ for
$R_2=2/3$ and then increasing to $5.3$ at our $R_{2,\rm min}=1/3$ (and
rising rapidly at lower $R_2$, as high as $21$ for $R_2=1/4$).  For
each source sample, we compute the weighted average value of 
$K_N\nu^{-2}$, to find noise-rectification bias of $-0.005$ ($r<21$),
$-0.019$ ($r>21$), and $-0.006$ (LRG).  To estimate the $2\sigma$ error,
we consider the allowed range of the noise-rectification bias to be
equal to the magnitude of the error estimated above; results are shown
in table~\ref{tab:dgg}.

Those five effects are the major sources of shear calibration bias; 
there are also several minor sources, at the 0.1 per cent level.  These
include camera shear (for which we correct using the
astrometric solutions, as described in \S\ref{SSS:construct}),
errors due to pixelization, and atmospheric 
refraction effects.    We do not attempt to estimate values for these
subdominant sources of error.  The total shear calibration bias (at
the $2\sigma$ level) with
the five main sources of error taken into account is shown at the
bottom of table~\ref{tab:dgg} for the three source samples
individually.  These estimates are conservative in that they do not
assume any distribution for these errors, allowing the actual values
to add, rather than adding them in  quadrature (which assumes some
possible cancellation).

\subsection{Shear estimator}\label{SS:shearestimator}

The weighting
used for this work differs from that of H04 in two ways.
First, rather than the uniform
weighting used in that paper, we weight by measurement error:
\begin{equation}
w_s = \frac{1}{\sigma_{SN}^2 + \sigma_{e}^2}
\end{equation}
where $\sigma_{SN}$, the rms shape noise in one ellipticity
component, was determined as a function of $r$ model magnitude from
Fig.~\ref{F:ehist} for the full source catalog and LRGs separately, and
$\sigma_{e}$ is the error per component on the ellipticity from equation~\ref{eq:sigma-sky}.
This weight is then multiplied by $\scinvtwo$,
downweighting lower redshift lenses
and lens-source pairs with small
redshift separation relative to those at large separation.
Consequently, the weight used for a given 
pair is $w_{l,s}=w_s\scinvtwo(z_l,z_s)$.
The shear responsivity ${\cal R}$ appropriate for this weighting scheme is then
computed using equations (5-33) and (5-35)
from~\cite{2002AJ....123..583B}, with the average value for our 
source samples being 0.86 for the $r<21$ sources, 0.83 for the $r>21$
sources, and 0.85 for LRGs.  

Using these weights, the shear estimator is then
\begin{equation}
\ds = \frac{\sum_{l,s} w_{l,s}e_t/\scinv}{2{\cal R} \sum_{l,s} w_{l,s}}=
  \frac{\sum_{l,s} w_s e_t \Sigma_c^{-1}(z_l,z_s)}{2{\cal R}
  \sum_{l,s} w_{l,s}}
\end{equation}

While we also tried an ellipticity-dependent weight as suggested
in~\cite{2002AJ....123..583B} to increase signal-to-noise, we found it
had minimal effect on the errorbars, so all the work in this paper was
done with the weighting scheme described above.

\begin{comment}
Thus, signal computation involved the following steps:
\begin{enumerate}
\item Around each lens, identify sources within the desired range of
  transverse separations, from 20 \hinvk{} to 2 \hinvm, working in
  logarithmically spaced bins so as to acquire information in smaller
  increments where the signal is varying the fastest, at small
  radius.  (Note that all plots show averaged radial bins for easier
  viewing.) 
\item Accumulate the sums required to compute \ds.
\item Boost this signal to account for contamination by physically
  associated lens-source pairs (see H04 for more in-depth
  description).  In practice, this boost is only significant for
  bright lens samples, and for the small physical separations.
\item Check the values of the errors as will be described in
  \S\ref{SS:errors} 
  against our original computation, which uses equation (5-27)
  in~\cite{2002AJ....123..583B}.   
\end{enumerate}
\end{comment}

\subsection{Error determination}\label{SS:errors}

Several methods of determining the errors on \ds{} were used, each
with its own advantages and shortcomings.  
We describe them here, and in \S\ref{SS:errcomp} we compare the results in
order to determine on which to rely.

\subsubsection{Analytic computation}

Analytic expressions for \ds{} for a given weight function may be
derived from equation (5-27)  in~\cite{2002AJ....123..583B}.  This method
is  the least computationally expensive method of deriving errors, but
suffers from several shortcomings.  First, it gives
incorrect results in the presence of spurious shear power in the source
catalog.  Second, it does not allow an easy way to include errors on
the boost factors, which may be significant.  Finally, it does not
account for correlation of radial bins, which can be significant at
large radius, where the average lens-source separation is larger than
the average lens-lens separation, so a given source contributes to the
measurement in several radial bins.

\subsubsection{Random catalogs}
A more computationally expensive way of determining the errors is
using random lens catalogs.  In the absence of systematic shear, the
average signal around random points 
should be zero, and the rms deviation around the mean gives some measure of the
noisiness of the signal, and therefore of the errors when using the
real lens catalog.  However, as
described in H04, this method is only valid on small distance scales,
less than about 500 \hinvk{} for the faint lens samples and around 1
\hinvm{} for the brighter samples.  Furthermore, this method also
cannot take into account the errors on the boost factors, though it
can account for the correlation of radial bins.  To get a reasonably smooth measure
of the errors, a large number of random catalogs must be used; for
this work, we used 24.

\subsubsection{Bootstrap resampling}\label{SSS:bootstrap}
As described in H04, bootstrap resampling is a useful method
  of determining the errors.  For this purpose, the lens catalog is
  divided up into 200 subregions, and the signal is computed for each
  subregion. The bootstrap-resampled datasets are generated
  by combining the signal from 200 subregions with replacement.  Then,
  a large number of resamplings (for this work, 2500) can be used
  to determine the average signal and its error.  This
  method has several advantages over the other two; for example, it
  naturally incorporates errors in the boost factor, and the
  correlation of radial bins.  However, because of the finite size of
  the subregions, the errors are once again not trustworthy at large
  lens-source separation.  Also, as described in H04, the noise in the
  covariance matrix means that the $\chi^2$ values for fits performed
  using the covariance matrix do not follow a $\chi^2$ distribution.

\section{Redshift distributions}\label{S:zdist}

In this section, we first describe the many commonly-used methods of
source redshift determination, including potential errors.  Then, we
describe the ``reference'' redshift
distributions that we use for calibration.
%
%To highlight the need for a reasonably accurate source redshift
%distribution, particularly for sources that are not very far from the
%lenses in redshift space, we refer the reader to Fig.[??] showing \scinv{} for
%various values of lens redshift, as a function of source redshift.  As
%shown, because of its rapid increase for source redshift near the lens
%redshift, even fairly small errors in source redshift can have a large effect
%on the mass determination.

\subsection{Photometric redshifts}
Surveys that collect photometric information in several passbands
allow for the determination of photometric redshifts, which use the
galaxy colors to extract an approximate redshift, typically based on
the construction of galaxy spectral energy distribution (SED) templates that are evolved with
redshift.  In principle, these photometric redshifts may be used
directly in
the computation of \scinv.  To be accurate, this computation should take into account the
photometric redshift error distribution, which may be highly
non-Gaussian, particularly in certain redshift ranges
and areas of color space.

%%The majority of this section will discuss difficulties with using the 
%%photometric redshifts, with a brief discussion at the end
%%regarding the applicability to photometric redshifts from other
%%sources.  
%%
We studied photometric redshifts computed by two independent groups
for the SDSS, and found that they performed statistically nearly
identically (that is, photometric redshifts for particular galaxies
were not necessarily the 
same, but the mean bias and scatter were nearly the same as a
function of magnitude and of photometric redshift).  One set of
photometric redshifts was computed using the program {\sc kphotoz} as
mentioned in \S\ref{S:sourcecatalog}.  The other set was computed by the SDSS
photometric redshift working group (as discussed in
\citealt{2003AJ....125..580C}, based on work in
\citealt{2000AJ....119...69C} and~\citealt{2000AJ....120.1588B}), for DR1
data only.  Because of the lack 
of photometric redshifts for the full DR2 sample, instead of using the
nearest-neighbor search method from H04 to get photometric redshifts
for the full sample, we 
demonstrate results in this paper using redshifts from {\sc kphotoz}
for the full sample.

We only use photometric redshifts for sources at $r<21$, because of
the large scatter at fainter magnitudes that will be demonstrated in
\S\ref{SSS:pzerr}.  The mean redshift of the source sample at $r<21$
is $\sim$0.35, with a fairly large width.  This fact raises several concerns
for use with the lens sample, which extends out towards the
peak of this 
redshift distribution at the bright end.  Lenses
at higher redshift are weighted more highly because of their higher
$\scinv$, so that if many of the lens-source pairs are at small redshift
separation, and there is a bias on the photometric redshifts, then
even with the $\scinvtwo$ weighting which downweights nearby pairs, we
still end up with a large bias in \ds.   Consequently, the photometric
redshift error distribution, which can be difficult to determine, is
very important.  Previous
studies used the spectroscopic sample, which is brighter, with much
lower photometric errors, supplemented by the CNOC2 spectroscopic
redshifts down 
to $r=21$ \citep{2003AJ....125..580C}, or used stacked images from
the southern SDSS survey due to their lower photometric errors (for
{\sc kphotoz}\footnote{Michael Blanton, private communication.}).
Consequently, the 
applicability of the photometric redshift errors from these studies to a full
catalog based on fainter single images (that will have additional
photometric redshift error due to noise) is questionable.

Fortunately, as will be shown in \S\ref{SS:DEEP2}, we now
have the capability of studying the errors in photometric
redshifts directly for a representative subsection of our source
sample rather than for atypical brighter or noiseless 
samples, and the application of these error distributions in the
computation of the lensing signal can help eliminate the bias in \ds{}
due to
 photometric redshift error.

\subsection{COMBO-17 distribution}

Another commonly-used option for the source redshifts is the 
use of a probability
distribution rather than individual redshifts.  This method has the disadvantage of leading to
a larger boost factor due to the inability to weed out physically
associated pairs.  However, for the $r>21$ galaxies, distributions are
a better option than photometric redshifts due to photometric noise;
consequently, we do not consider photometric redshifts for $r>21$.

The redshift distribution used for this paper is derived from
data from the Classifying Objects by Medium-Band Observations, or
COMBO-17 survey.  COMBO-17
includes photometry in 17 passbands spanning the wavelength range from 350--930~nm, yielding far more information for the
determination of photometric redshifts than the SDSS.
The area of the survey used in the study that derived the
redshift distribution covers 0.78 square degrees, spread over three
disjoint regions (smaller than the full survey), so
while the area is small, the
concern about the derived redshift distribution $p(z | r)$ being unduly
influenced by large-scale 
structure is somewhat lessened.
\cite{2003A&A...401...73W} includes
the luminosity 
functions upon which these distributions were based.  Since the
distributions are for all photometric galaxies rather than those
passing our lensing cuts, we may expect that they lie at slightly
higher redshift than our catalog on average; we show results of tests
of this hypothesis in \S\ref{SSS:zfaint}.  A plot of this distribution
(averaged over $r$) is shown in comparison with other distributions in
Fig.~\ref{F:zdist.DEEP}.  

Note that the COMBO-17 photometric redshifts have been used both
directly for cosmic shear studies in the COMBO-17 survey itself 
(\citealt{2003MNRAS.341..100B}; \citealt{2003astro.ph.10174H}), and to
estimate redshift distributions for other cosmic shear 
investigations \citep{2004astro.ph.11324H}.

\subsection{DEEP2}\label{SS:DEEP2}

Another way to determine the true redshift distribution for
our sources is to find another survey that is flux-limited
and complete to a desired flux.  As shown in
\cite{2005PhRvD..71b3002I} in the context of cosmic shear surveys,
even just 100 spectroscopic redshifts 
may be sufficient to make this determination.  Fortunately, the DEEP2
survey (\citealt{2003SPIE.4834..161D}, \citealt{2003ApJ...599..997M},
\citealt{2004astro.ph..8344D}, \citealt{2004ApJ...609..525C}) provides
results that are useful for this purpose,
with spectroscopic completeness well beyond $r=21.8$, the limits of our
source catalog.  The DEEP2 survey  will eventually include spectroscopy
of $\sim$60~000 galaxies in 4 fields totaling 3.5 square degrees.  While
the targeting in three fields involves the use of
photometric information to select galaxies with $z>0.7$, the targeting
in the extended Groth strip (EGS) does not attempt to place such
restrictions, and 
because it overlaps with the SDSS, it may be used to study redshift
distributions and photometric errors in SDSS data.  Observations are
complete in pointing one of field one (EGS), centred at
Dec. $+52^\circ 12'$ and RA $14^h15^m\!.7$, which has area
approximately 0.15 deg$^2$  (roughly $1/4$
the area in the full EGS).    
The Groth strip is situated $>50$ degrees from the nearest of the
three COMBO-17 fields, so the redshift distribution obtained from it can  
be considered statistically independent from the COMBO-17 results.
The detectors on the Canada-France-Hawaii Telescope (used for imaging
for DEEP2 target selection) saturate at $R_{AB} \approx 
17.6$ at the bright end, so no galaxies brighter than that limit have
spectra, but fortunately those galaxies constitute a very small
fraction ($\sim 2$ per cent) of the source 
sample.  Also, about 1/3 of the galaxies in the Groth strip were
not targeted at all, which reduces the number of potential
matches against SDSS.

The target selection in the Groth strip did involve some color and
magnitude information~\citep{faber2005}.  At $R<21.5$, all galaxies are selected
uniformly; fortunately, the majority of the galaxies in our lensing
catalog fall into this category.  At fainter $R$, galaxies are
classified as low or high redshift via color cuts; low redshift
galaxies are downweighted significantly, which makes the selection
fairly complicated.  To verify that this selection has a
negligible effect on the results in this paper, all redshift
distributions computed using DEEP2 data were recomputed taking into
account selection probability (i.e., weighting each redshift by
$1/p$).  The value of \avgscinv{} for various values of lens redshift
and with the weighted redshift distributions were computed and compared to the
values from the unweighted distributions.  Even for lens redshift
$\sim 0.2$, which is at the high end of the distribution, and should be more
sensitive to the source redshift distributions than most other lenses,
the fractional change in the value of \avgscinv{} was on the order of
a few tenths of a per cent.  Consequently, we consider the selection
function to be of negligible importance for the distributions
presented in this paper, and for the remainder of this work we use the
unweighted results.

To use the DEEP2 redshifts to compute redshift distributions and
photometric redshift error distributions for our catalog, we first
matched between the DEEP2 spectroscopic catalog 
in the Groth strip and our lensing catalog.  This step ensures that
all distributions 
that we derive will apply to lensing-selected
galaxies, which are expected to be at lower average redshift than
all galaxies at the same magnitude, due to our 
requirements on the shape measurement. In principle we could have used
redshift distributions as a function of magnitude presented
in~\cite{2004ApJ...617..765C}, but since those were derived for all
galaxies, they are expected to be at slightly higher average redshift
than the distributions for lensing-selected galaxies.  Once matching
was complete, there were 278 matches, 162 at
$r<21$ and 116 at $r>21$.   Our requirement that there be a
high-quality redshift had eliminated 33 potential matches,
giving a redshift determination success rate of 89 per cent for
lensing-selected galaxies overall, or 91 per cent for $r<21$ and 86
per cent for $r>21$.

We must be concerned about the effects of redshift failures on
our results.  The lack of knowledge
whether or not the failures lie in a particular region of redshift
space (e.g. higher redshift on average) introduces an unknown
systematic into our results.  First, we note that as discussed
in~\cite{2004ApJ...609..525C}, redshifts $z>1.45$ cannot be measured
by the DEEP2 survey.  However, for the magnitude ranges of interest in
this paper, this limit is effectively of no importance.  Second, we find that
the fraction of matches 
with failed redshift determination varies somewhat with magnitude (6 per cent
failed at $18\le r<19$, 8 per cent at $19\le r<20$, 11 per cent at
$20\le r<21$, and 14 per cent at $21\le r<22$), implying an increase
in failures at higher redshift; we  may also have
a problem if the majority of the failures lie at a particular part of
the redshift distribution in a given magnitude range.  We can place
bounds on the effect of such a systematic as follows: we compute
the change in \avgscinv{} that results from assuming that all the
failed redshifts were at 0 and at $\infty$ (which yields the $z\gg
z_l$ asymptotic value of \scinv).  We can compute the fractional
error
\begin{equation}\label{E:redshiftfailure}
\frac{\delta\ds}{\ds} = -\frac{\delta\avgscinv}{\avgscinv} = -f_{failed} \left( 1 - \frac{\avgscinv_{failed}}{\avgscinv_{measured}}\right)
\end{equation} 
where $f_{failed}$ is the fraction of redshift failures (0.11),
$\avgscinv_{failed}$ is the average value of \scinv{} for those galaxies
that had failed redshift determination, and $\avgscinv_{measured}$ is
the average value of \scinv{} for those galaxies that were used to
compute the signal using the observed redshift distribution.  If we
assume all failed redshifts were 0, then $\avgscinv_{failed}=0$, and
therefore we expect a bias of -11 per cent (or $-14$ per cent for
$r>21$ and $-9$ per cent for $r<21$), where our use of the
redshift distribution from the redshift determination successes
overestimated \scinv{} (and therefore underestimated the signal) by that
amount.  If we assume all failed 
redshifts are $\gg z_l$, then the effect depends on
the lens luminosity (where it is less important for lower luminosity
and redshift lenses, for which all sources were essentially at
$\infty$ anyway) and the source sample.  The degree to which the
signal was overestimated in this case is given, for each luminosity
bin and source sample, in table~\ref{T:redshiftfailure}.  As shown,
the effect is at most 2.7 per cent overestimation for L6 with $r<21$
sources, much less for the 
faint bins.  The reality is somewhere in between these two extremes,
most likely towards the higher end of the range quoted since we expect
more redshift failures at higher redshift, but
cannot easily be estimated, so we use these values to define the 95
per cent confidence interval. (This estimate did not take
into account the effect of changes in \scinv{} on the weighting, since
the use of non-optimal weighting should increase the errors without
inducing a bias.)
\begin{table}
\caption{\label{T:redshiftfailure}
The estimated overestimation of the lensing signal if all
redshift failures are at very high redshift, as a function of lens bin
and source sample.}
\begin{tabular}{ccc}
\hline\hline
Lens sample  & \multicolumn{2}{c}{$\delta(\ds)/\ds$, per cent} \\
  & $r<21$ & $r>21$ \\
\hline
L1  & 0.1 & 0.4 \\
L2  & 0.2 & 0.6 \\
L3  & 0.5 & 0.9 \\
L4  & 0.9 & 1.4 \\
L5  & 1.6 & 1.9 \\
L6  & 2.7 & 2.5 \\
\hline\hline
\end{tabular}
\end{table}

We compared the magnitude distribution among
the matches to that in our source catalog.  If they are not comparable,
redshift distributions determined using the DEEP2
sample may not be applicable to our source sample.  
Table~\ref{T:completeness} shows a
comparison of the samples.  The first column shows the fractions of
galaxies in magnitude bins for our source catalog overall, and the
second column shows the fractions in the
Groth strip (with 95 per cent confidence intervals from the binomial
distribution for the latter due to its small size).  Since they agree
within the errorbars, we need not 
worry about the Groth strip being very different from the full
source catalog.  The third
column shows the fractions in magnitude bins for the full SDSS photometric
sample (i.e. no lensing-related cuts), Groth strip only,
restricted to the magnitude range for which the CFHT detector does not
saturate.  The fourth column shows the fractions of galaxies actually
targeted as a function of magnitude (for $r<21.8$); this column is
statistically consistent with the third 
column, implying that the targeting is indeed independent of the $r$
magnitude.  Finally, the fifth column shows the 
fractions of galaxies as a function of magnitude for the matches
between the DEEP2 redshift catalog and our lensing catalog with
successful redshift determination, which is
consistent with column two, the fraction of lensing-selected
galaxies as a function of magnitude in the Groth strip before matching
against DEEP2, with the exception that the DEEP2 galaxies are by
necessity missing the $r<18$ sample, which is roughly 2 per cent of
our catalog.  
Furthermore, the fraction of galaxies categorized as
high-redshift LRGs was similar in the two samples (9 per cent in our
background catalog, versus $11.5_{-3.5}^{+4.3}$ per cent in the
DEEP2 matches at the 95 per cent confidence level), implying that
there is no color-dependence of the targeting and redshift failures with
regards to this particular subsample. 
\begin{table*}
\caption{\label{T:completeness} The fraction of source galaxies in each
  magnitude bin for our source catalog overall and in the Groth strip,
  all SDSS photometric galaxies 
  in the Groth strip with $18<r<21.8$, all DEEP2 galaxies targeted
  with $r<21.8$,  and the matches between DEEP2 and our source catalog.
  (95 per cent confidence intervals on all columns are from the
  binomial distribution; due to the constraint relating the sum of the
  numbers within each column, the errorbars are anti-correlated.)}
\begin{tabular}{cccccc}
\hline\hline
Magnitude & Lensing  & Lensing catalog, & Photometric survey, & DEEP2 & DEEP2 \\
range & catalog & Groth strip & Groth strip & targets & matches \\
\hline
$r\le 18$    &   0.022  & $0.020_{-0.010}^{+0.016}$ & 0.000 & 0.000 & 0.000 \\
$18<r\le 19$ &   0.055  & $0.066_{-0.020}^{+0.025}$ & $0.035_{-0.010}^{+0.012}$ & $0.027_{-0.012}^{+0.017}$ & $0.054_{-0.023}^{+0.033}$ \\
$19<r\le 20$ &   0.153  & $0.127_{-0.028}^{+0.033}$ & $0.096_{-0.016}^{+0.018}$ & $0.096_{-0.023}^{+0.028}$ & $0.133_{-0.038}^{+0.046}$ \\
$20<r\le 21$ &   0.358  & $0.378_{-0.043}^{+0.044}$ & $0.304_{-0.026}^{+0.027}$ & $0.316_{-0.039}^{+0.041}$ & $0.396_{-0.058}^{+0.060}$ \\
$21<r\le 21.8$ & 0.411  & $0.410_{-0.044}^{+0.045}$ & $0.565_{-0.028}^{+0.028}$ & $0.558_{-0.043}^{+0.042}$ & $0.417_{-0.059}^{+0.060}$ \\
\hline\hline
\end{tabular}
\end{table*}

While using the spectroscopic redshift distributions may seem to be
the ideal way of determining \sigmacrit, there are two
caveats that make this solution less promising.  First, the use of
average redshift distributions without any use of photometric redshift
errors
means that many more physically-associated lens-source pairs are
included in the calculation.  While the resulting dilution of signal may be corrected for using
boost factors, boosting is another potential source of
systematic error since we cannot correct for intrinsic shear, and it is desirable that the boost factors be as low
as possible.  Second, because these distributions were determined
using a small portion of the sky, they may be unduly influenced by
large-scale structure, and not representative of the redshift
distribution in the survey as a whole (in \S\ref{SS:zresults},
we show how the effects of large-scale structure increase the
errorbars determined purely using statistics).  The ideal solution to the
second problem would be
to have spectroscopic redshifts determined for a random sample of
galaxies selected from the entire SDSS survey area, which could be used to
determine source redshift distributions or photometric redshift error
distributions. 

Consequently, we also used these DEEP2 redshifts to determine
photometric redshift error
distributions, enabling us to use photometric redshift information to
eliminate physically-associated pairs, yet to also avoid any bias due
to errors in the photometric redshifts.  The determination of these
error distributions was done by dividing the $r<21$ matches into
bins based on photometric redshift (in order to determine error
distributions as a function of photometric redshift).  
Our computation of \scinv{} then takes into account the
photometric redshift error distribution:
\begin{equation}\label{E:deconvolve}
\Sigma_c^{-1}(z_l,z_p) = \int p(z_s | z_p) \Sigma_c^{-1}(z_l,z_s) dz_s
\end{equation}
The signal is then computed as usual, but instead of assuming that
$z_s=z_p$ (i.e., $p(z_s | z_p) = \delta(z_s-z_p)$), we use
the value computed by evaluating the integral in
equation~\ref{E:deconvolve}.

Results from our work
with the DEEP2 data are shown in \S\ref{SS:zresults}.

\subsection{Reference distribution: LRGs}\label{SS:LRG}

There is one subset of SDSS source galaxies for which excellent redshift
information is known: Luminous Red Galaxies (LRGs)~\citep{2001AJ....122.2267E}.  These galaxies have a
well-known color-redshift relation that allows photometric redshifts to
be determined with excellent precision.  

For the LRGs selected according to
criteria shown below, the photometric redshifts perform
significantly better than for the general source sample.   Furthermore, as
shown in  
\citealt{2004astro.ph..7594P}, redshift distributions for LRGs from
the 2dF-Sloan LRG and Quasar Survey (2SLAQ) can be used to
construct very reliable redshift distributions for 
these galaxies.  Note that for LRGs, the redshifts used were from 
a simple template code discussed in~\cite{2004astro.ph..7594P} rather
than from {\sc kphotoz}; however, for most
regions of color space occupied by LRGs, the results are highly correlated.

The selection criteria used for the LRGs are as follows:
\begin{enumerate}
\item $d_{\perp} \equiv (r-i) - 0.125(g-r) > 0.45$
\item $c_{\parallel} \equiv 0.7(g-r) + 1.2(r-i-0.18) > 1.6$
\item $g-r < 2.5$
\item $r-i < 1.5$
\item $0.4 < z_p < 0.65$
\item $r>19$
\end{enumerate}
These criteria were chosen as the combination of those from
\cite{2001AJ....122.2267E} and from
\cite{2004astro.ph..7594P} that best
suited the requirements of this work, namely very low levels of
contamination from non-LRGs, from low-redshift LRGs, and from stars,
yet sufficient numbers that the LRG sample can be used as sources for
weak lensing with reasonable signal to noise.
\begin{comment}
These criteria were designed to avoid
contamination from non-LRGs and from low-redshift LRGs, since the LRG
selection criteria break down at low redshift,
$z<0.2$~\citep{2001AJ....122.2267E}. The first two criteria help
select out the high-redshift LRG sample.  The third and fourth
criteria help avoid color outliers for which selection may not be
reliable, and the fifth criterion helps avoid abnormally low
photometric redshifts (which may not actually be LRGs) and the errors
that can occur at high photometric redshift due to insufficient
knowledge of how LRGs evolve at these redshifts.  The final criterion,
which eliminates a very small fraction of galaxies that pass the other cuts,
is imposed to reduce low-redshift contamination, because it is highly
unlikely that a galaxy that is truly in the desired redshift range
is brighter than $r=19$.
\end{comment}

\begin{figure}
\includegraphics[width=3in,angle=0]{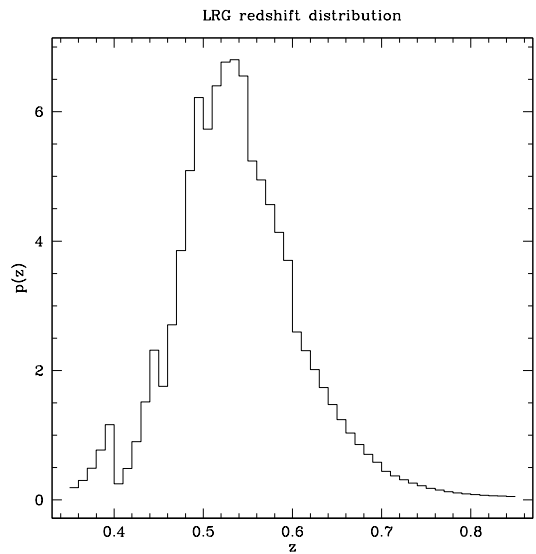}
\caption{\label{F:LRGzdist} The redshift distribution for the
  high-redshift LRG sample derived via the inversion method
  from~\protect\cite{2004astro.ph..7594P}.} 
\end{figure}

Imposing these criteria on our source catalog yields 2~884~242 LRGs
total, though  
only about 65 per cent are in regions with lenses.   A plot of the LRG redshift
distribution derived using 2SLAQ work (\citealt{2004astro.ph..7594P}) is in
Fig.~\ref{F:LRGzdist}; for this distribution, only the 72 per cent of the LRG
sample with
photometric redshift greater than 0.45 were used because the inversion
method gave unreliable results for $0.4<z_p<0.45$.  As 
shown, the redshift distribution peaks around $z \sim
0.5-0.55$, which illustrates another advantage of using LRGs as sources: when
we use SDSS main spectroscopic sample galaxies as lenses, with redshifts mostly
below $z\sim 0.25$, \scinv{} does not vary strongly with source
redshift for sources at such high redshift, so 
an error
in these distributions will make a very small difference in
the results.

Note that approximately 57 per cent of the LRGs in the catalog are fainter
than $i$ model magnitude of 20.  \cite{2004astro.ph..7594P} only includes
redshift distributions down to this limit, so our use of that
inversion method
for this sample is suspect.  However, tests of photometric redshifts using the DEEP2 matches
indicate that that the mean bias is the same for the $i>20$ LRGs
as for the $i<20$ LRGs (0.02), and the rms scatter is larger but not
excessively so
(0.07 vs. 0.05), which suggests that the use of the inversion
method in~\cite{2004astro.ph..7594P} with the same error distributions
for $i<20$ and $i>20$ is justified here.  

In addition to computing the
signal with the full sample using the photometric redshifts directly,
and with the restricted sample using the redshift distribution, we
also computed it with photometric redshifts on two subsamples, those
with $d_{\perp}>0.5$ and $d_{\perp}>0.55$.  Because these stricter cuts
would help avoid contamination by lower-redshift galaxies being
scattered up to higher redshift, we try imposing them on our sample;
if the signal does not differ when we do this, then we know that using
the full sample is relatively safe.

Fig.~\ref{F:rhist} shows the $r$ model magnitude distribution of
sources classified as high-redshift LRGs, and Fig.~\ref{F:r2hist}
shows their $R_2$ distribution, in comparison to the values for the
overall source sample.  As shown, the LRG sample 
is, on average, at lower $R_2$ and fainter magnitudes than the other
source samples, and thus 
we may be concerned that it will have more significant shear
calibration bias, as evidenced by the confidence intervals in Table~\ref{tab:dgg}.
%\begin{figure}
%\includegraphics[width=3in,angle=0]{rhist.LRG.ps}
%\caption{\label{F:rhistLRG} The $r$-model magnitude distribution for the
%  high-redshift LRG sample.  The solid line shows
%  all galaxies with shape measurement in at least one band; the dashed
%  line shows all those with shape measurement in one particular band.}
%\end{figure}
%\begin{figure}
%\includegraphics[width=3in,angle=0]{r2hist.LRG.ps}
%\caption{\label{F:r2histLRG} The top plot shows the average $r$-band $R_2$ value for the
%  LRG catalog as a function of magnitude.  The bottom plot shows the
%  distribution of $R_2$ values in the sample overall.}
%\end{figure}

In principle, given the excellent redshift information for the
high-redshift LRGs, one solution to our 
lack of reliable redshifts for all sources would be to do g-g
lensing entirely with LRGs as sources.  However, we then are faced
with a problem of statistics, since LRGs are such a small fraction of
the sources available (9 per cent), yielding large
statistical errors on \ds.  This problem is the reason why our
test for calibration bias is so useful; it allows
us to use the excellent redshift information from LRGs to check the
calibration of 
the lower-redshift source samples, which contain many more
galaxies, enough to obtain excellent statistical errorbars.

\section{Other Systematics Issues}\label{S:othersys}

Here we consider a number of remaining systematics issues that arise
when computing the lensing signal.  Some are calibration
uncertainties that scale with the signal, similar to the shear and redshift distribution
calibration uncertainties discussed previously; others are systematics
that do not scale with the signal, and have amplitudes that depend on
the angular or physical lens-source separation.

\subsection{Random points test}\label{SS:randpoints}

The random points test
requires computing signal $\ds_{rand}(r)$ using random lens catalogs
(i.e., sets of random positions generated with the angular mask of the 
spectroscopic survey area, 
using the same lens redshift  and magnitude distributions as the
real lens sample).  In practice, these distributions are preserved by
drawing the redshifts and magnitudes from the real sample,
without replacement.  The random points test is useful because a
nonzero signal reveals the presence of spurious shear power
(systematic shear)
in the source catalog which would lead to an additive bias in the
lensing signal.  The random
catalogs used for this work were generated using {\sc mangle}
\citep{2004MNRAS.349..115H}. 

In the absence of systematic shear, we expect the random points test
to show zero signal around random points.  However, 
there is a slight smearing of
images in the scan direction because the charge transfer is not
continuous, instead occurring in very quick transfers from pixel to
pixel, so that the PSF is convolved with a rectangle one pixel (0.4'')
wide along the scan direction only. The result is that the PSF is
not circularly symmetric.  This effect may be exacerbated if the
scan rate and charge transfer rate are not perfectly matched, 
leading to convolution on a scale larger than one pixel.\footnote{We
  thank James Gunn for pointing out this effect.} 
While in  
principle PSF-correction algorithms should correct for 
PSF asymmetry,
in practice this is 
difficult to do perfectly.  In addition, since the charge-transfer efficiency may
also depend on the amount of charge, the
PSF asymmetry may be different for faint objects than for bright ones, but since {\sc
  Photo} fits the PSF using stars around $r\approx 19$, code that uses the
{\sc Photo} PSF may not be able to fully correct for this
effect.  We found that the re-Gaussianization scheme
over-compensates for the effect.

We address the presence of this spurious signal by determining it to
as high precision as possible using $N$ random catalogs (where the number
of random catalogs is limited by processor time), then subtracting
it from the observed signal.  This procedure
results in the errorbars on the signal rising by a factor of
$\sqrt{(N+1)/N}\approx 1 + 1/(2N)$.  Our choice of $N=24$ means that the
errorbars only increase by 2.1 per cent.  Results for the random catalog test
are shown in \S\ref{SS:sigresults}.

However, random catalog signal subtraction, while
widely accepted in the literature, ignores the possibility that
fluctuations in the number density and systematic shear may be
correlated.  We describe this problem as follows: A g-g
weak lensing measurement entails computing the correlation $\langle n
\hat{\gamma}_{+}\rangle$.  The measured number density of
lenses $n$ can be decomposed into $n = \bar{n}+\delta
n$, an average density of sources on the sky plus 
fluctuations.  The measured total shear $\hat{\gamma}_{+}$ can be decomposed
into $\hat{\gamma}_{+} = \gamma_{+} + \gamma_{sys}$, the true shear
field plus systematic shear.  When we compute the signal around random
points, we obtain the quantity $\langle \bar{n}\hat{\gamma}_{+}\rangle$.
Random catalog subtraction thus corresponds to measuring
\begin{equation}
\langle n \hat{\gamma}_{+}\rangle - \langle
  \bar{n}\hat{\gamma}_{+}\rangle = \langle \delta
  n\,\hat{\gamma}_{+}\rangle = \langle \delta n\,\gamma_{+} \rangle +
  \langle \delta n \, \gamma_{\rm sys}\rangle.
\end{equation}
The first term on the right
side, $\langle \delta n \,\gamma_{+}\rangle$, is the correlation that
we hope to measure, but the second term is an
additive systematic error that has not been discussed in previous
works on g-g weak lensing.   We  cannot
assume that $\delta n$ and $\gamma_{\rm sys}$ are completely uncorrelated,
because both quantities are slightly correlated in some
way with the PSF.  There are several ways in which such
an effect could become significant.  For example, since there is a gap between
camcols on the SDSS camera, the same region must be rescanned with
some offset to fill in 
that gap on a different night, which may have very different seeing
and other conditions, such that $\gamma_{\rm sys}$ fluctuates on
small scales.  If the number density of lenses also
fluctuates on the same scale, then we could have some nonzero
contribution from the $\langle \delta n\,\gamma_{\rm
  sys}\rangle$ term.  For this work, we assume that 
this term is negligible, since the correlation should be
small and has 
never been detected before, but this issue should be addressed more
fully in
future work to ensure that this assumption is reasonable.

\subsection{45-degree test}\label{SS:45degree}

Another useful test of systematics in the lensing signal is the 45-degree
test, which requires computing the lensing signal with the coordinate
system rotated by 45 degrees.  By inversion symmetry, the 45-degree
rotated signal $\gamma_{45}$
should vanish, with weak lensing only contributing to the tangential
shear, $\gamma_+$. The presence of such a
signal could indicate a variety of shear systematic errors, since they generally
contribute both to $\gamma_+$ and $\gamma_{45}$.  We computed
the 45-degree rotated signal for all three source samples,
so that together these tests were done for all
sources in the catalog (we assume that the choice of redshift
distribution does not matter for this test, as $\gamma_{45}$ would be
nonzero due to shear systematics, so we only use one choice of
redshift determination method for each of those samples).
Results for this test are in \S\ref{SS:sigresults}.

\subsection{Star-galaxy separation}\label{SS:sgsep}

Star-galaxy separation is an important issue for weak lensing, since
the inclusion of stars in the source sample would dilute the signal.
Thus, a balance must be struck, ensuring the purity of the
galaxy sample used as sources, while avoiding being overly conservative
and eliminating too many 
galaxies in the course of doing this separation (which would lead to
poor statistics).  Star-galaxy separation for this
catalog was accomplished via two cuts: first, the requirement that
the {\sc Photo} flag OBJC\_TYPE be equal to 3, or galaxy; and second,
the requirement that $R_2 > 1/3$ (i.e., the object must be 50 per cent
larger than 
the PSF). The type determination for DR1
and DR2 is described in~\cite{2001adass..10..269L}; in brief, it
utilizes the linear combination of fits of galaxy profiles to two models (exponential and
de Vaucouleurs), then allows the ratio of the flux in a fit to a PSF shape to
that in the linear combination of galaxy models to determine the type,
via the requirement that $m_{PSF}-m_{cmodel}\ge 0.24$ for galaxies.  As shown in
figure 3 in that paper, this procedure ensures a relatively pure
galaxy sample even close to
$r=22$, with stellar contamination fraction (determined using
Hubble Space Telescope, or HST data in the Groth strip) being negligible at $r$ brighter than 21,
and 7 per cent for $21<r<22$.
It is also clear from that
figure that OBJC\_TYPE is much more likely to default to calling
galaxies to stars rather than vice versa, with galaxy contamination of
$\sim 33$ per cent in a sample of ``stars'' for $21<r<22$.
While there are probabilistic methods
that are more accurate at the faint end ($r>21$), such as that used in
\cite{2002ApJ...579...48S} and \cite{2004AJ....127.2544S}, OBJC\_TYPE's
conservative tendency should not cause significant stellar
contamination, though it does reduce the density of sources 
in poor seeing at faint magnitudes. 
%%Also, note that of the 18\%
%%stellar contamination fraction expected from our use of OBJC\_TYPE at
%%$21.5<r<22$, we expect that approximately ??\% should be eliminated by
%%our requirement that $R_2 > 1/3$. [Do a test to determine this fraction?]

In order to check that our shape measurement cuts reduce
stellar contamination due to failures in OBJC\_TYPE, we used
publicly-available catalogs from the Great Observatories Origins Deep
Survey (GOODS), carried out via the 
HST (\citealt{G2004}, \citealt{2004ApJ...600L..93G}).
Characterizations of stars versus 
galaxies are much more accurate to fainter magnitudes in space-based
surveys such as this because the PSF is much smaller.
We found 577 objects at $r<21.8$ (of which 60\% are at $r>21$) that were
matches between the full SDSS photometric catalog and the GOODS north
field, centred at Dec. $+62^\circ 15'$ and RA $12^h37^m\!.7$.   
We found that 
the ``galaxy'' classification is incorrect about 1.5 per cent of the time for
$r<21$ and 7 per cent of the time for
$21<r<21.8$ objects, and the ``star''
classification is incorrect a larger  
fraction of the time (3 per cent at $r<21$ and 40 per cent at
$21<r<21.8$), confirming 
the results from~\cite{2001adass..10..269L} that   
OBJC\_TYPE tends to default to calling small, faint objects stars.  
%These
%statistics are from the full set of matches, both stars and galaxies,
%in the photometric catalog at $r<21.8$
%(577, of which 346, or 60 per cent, are at $21<r<21.8$).  

However, when we
restrict to the subset of 146 objects that passed all shape-measurement
and other criteria to be included in our source catalog, of which 73
are at $r<21$ and 73 at $r>21$, we
find that only 2 (1.4 per cent) of those included are stars, or 0 per
cent contamination 
in the $r<21$ sample and 2.7 per cent contamination in the $r>21$
sample.  
We see that 
the resolution-factor and other cuts  reduce stellar contamination by
a factor of three from the result using OBJC\_TYPE alone. 
Using the binomial distribution to get 95 per cent confidence
intervals on out stellar contamination estimates yields $[0,0.040]$
($r<21$) and $[0.003, 0.096]$ ($r>21$).  Technically, we should take
into account that the GOODS north field is at $1/\sin{b}\sim 1.2$, but
the average value for the full lens sample is $\langle
1/\sin{b}\rangle \sim 1.40$, so we might expect slightly higher
stellar contamination in the full catalog than that computed for the
GOODS field.  However, because the dependence of stellar contamination
on $1/\sin{b}$ is difficult to model, we do not attempt any correction.

To check the signal for contamination of our source catalog by stars, we
computed the signal at low versus at high galactic latitude (cutting
at $\sin{b}<0.7$) to compare the results.  Note that while finding a
lower signal at low galactic latitude may indicate a problem
with star/galaxy separation, it may also indicate the presence of
other systematics.  In particular, since the extinction is greater at
low galactic latitudes, galaxies near the faint end at a given
magnitude were actually lower signal-to-noise measurements at low
galactic latitude than they were for high galactic latitude, so in
principle there could be a shear systematic causing a difference
between these two samples as well.  The
results of this test will be shown in
\S\ref{SS:sigresults}.

\subsection{Seeing dependence of calibration}\label{SS:seeing}

Because our ability to correct the galaxy image for effects due to the
PSF depends on the relative size of the galaxy and the PSF, we must
consider the possibility of a seeing-dependent shear systematic.
Consequently, for each galaxy, we consider the size of the PSF used,
and split our sample into ``good 
seeing'' (PSF size less than the median value, 1.25 pixels
in the $r$ band) and
``bad seeing'' (PSF size greater than the median value).  The signal
was then computed using these two source samples, and compared.  Results for
this test will be shown in \S\ref{SS:sigresults}. 

\subsection{$R_2$ dependence of calibration}\label{SS:r2}

Because some calibration biases may be more prominent at lower $R_2$,
it is important to check for $R_2$-dependence of the calibration.
There are several effects that could lead to apparent $R_2$ dependence
of the calibration: noise
rectification bias, selection biases, and biases due to
PSF-correction, which would be particularly important for
less-well resolved galaxies.  We computed signal using sources with
$R_2$ greater than and less than 0.55 in each band to
check for bias; results of this test are presented in
\S\ref{SS:sigresults}. 

\subsection{Systematic differences between bands}\label{SS:bands}

Like many other studies, we used shape measurements averaged over two
bands, the $r$ and $i$ bands.  While the shape measurements between
the two bands may legitimately differ for individual objects, due to (for
example) spectral differences in emissions from the disk versus from the bulge of
spiral galaxies, we also checked to ensure that the signal computed
with the shape measurements from each band individually gives the same \ds.  The results of this
test will be shown in \S\ref{SS:sigresults}.

\subsection{Boosts}\label{SS:boosts}

As discussed in H04, the lensing
signal at small transverse separations is diluted by the inclusion of
sources that are physically associated with the lens (i.e., are in the
same group or cluster), and therefore are not really lensed.  To
correct for this effect, the signal for a given luminosity bin is
boosted according to the 
weighted number
of galaxies per unit area relative to the number from
random catalogs.  The signal is multiplied by a factor
\begin{equation}
B(r) = \frac{n(r)}{n_{rand}(r)}
\end{equation}
where $n(r)$ is the weighted number of galaxies per unit area when the
signal is computed, and $n_{rand}(r)$ is the same computed with random lens
catalogs.  The number from random catalogs takes into account the
decrease in the number per area with radius due to survey edge effects, so
the boost can accurately account for the dilution of signal by
physically associated pairs.  Consequently, $B(r)-1 = \xi_{ls}$ is 
the lens-source correlation function.

The boosts add two sources of statistical error and two sources of
systematic error.  In the rest of this section we consider each of these.

\subsubsection{Statistical errors}

The statistical error arises because the boost
factor is the ratio of two noisy quantities, $n(r)$ and $n_{rand}(r)$.
The noise in
$n_{rand}(r)$ can be minimized by computing signal from a large number
of random catalogs, where 24 are used for this work.  However, for
subsamples with a small number of lenses, $n_{rand}(r)$ is
still slightly noisy at small separations due to the small size of the
radial bins; this noise is taken into account in the bootstrap by
multiplying the boost from each dataset and radial bin by a Gaussian random number of
mean 1 and standard deviation equal to the fractional error in
$n_{rand}(r)$ as computed from the random catalogs.  The noise in
$n(r)$ is taken into account naturally by the bootstrap, since
each bootstrap resampled dataset will have a slightly different $n(r)$
used for the boost.  Hence, the statistical error due to the boosts is
simple to take into account.

\subsubsection{Systematic error: non-uniformity of boost factor}

One potential systematic error arises because both \ds{} and
the boost factor vary strongly with luminosity at the bright end of
the lens sample, so the luminosity bins that are 1 magnitude
wide may be too wide to properly compute the signal in the innermost
radial bins, $r < 50$ \hinvk, where the boost is most important.  By
averaging over a large range in luminosity, with \dsr{} and $B(r)$
varying with luminosity, we may run into a situation where the product
of two averages (\dsr{} and $B(r)$ separately averaged over
luminosity, then multiplied)
differs significantly from what we really want, the average of
products ($B(r)\dsr$ averaged over luminosity).

One rudimentary method of detecting
the effects of using wide luminosity bins on the boost factor is
to split the brightest luminosity bin in half, compute the signal
separately for each half (boosting each one individually), then
average the signal from each half.  The resulting signal can be
compared against the signal computed using
the full luminosity bin.  Results for this test will be presented in
\S\ref{SS:sigresults}.

\subsubsection{Systematic error: magnification bias}\label{SSS:magbias}

Another boost-related source of error is magnification bias, since the
number of galaxies per 
area around real lenses may not be expected to be the same as that
around the random points.  There are three competing effects: first,
that due to the magnification $\mu=1+2\kappa$ (in the weak lensing
limit), where $\kappa = \Sigma/\Sigma_{c}$, the number 
of lens-source pairs per unit area on the sky will decrease;
second, that the magnification means that fainter sources
will be visible than would have been otherwise, and therefore the number of
lens-source pairs per unit area will increase; and third, the
magnification changes the resolution factors of the source  
galaxies.  The competition
between the first two effects can be quantified by
$s=\rmd\log_{10}N(m)/\rmd m$, 
where $N(m)$ is the total number density of source galaxies given a faint
magnitude limit of $m$.  For the $r>21$ sample, we must taken into
account the loss of sources at the bright magnitude limit of 21, and
compute 
\begin{equation}
s=\frac{\rmd\log_{10}{N(m_-, m_+)}}{\rmd m_+} -
  \frac{\rmd\log_{10}{N(m_-,m_+)}}{\rmd m_-}
\end{equation}
 where $m_+$ is the faint magnitude limit and $m_-$ is the bright magnitude
limit.  We compute $s$ separately for each source sample: for
the $r<21$ source sample, we find $s=0.36$; for $21<r<21.8$, we find
$s=0.47-0.57=-0.10$ (i.e. when the magnitudes are shifted brighter, we
lose more galaxies at the bright end than we gain at the faint end
because of our cuts on the shape measurement);
and for the LRG sample, we find $s=0.27$. 

The resolution factor dependence of this effect has not been previously
evaluated. 
If we take the Gaussian approximation for the galaxy, $T^{(I)}\approx
T^{(f)}+T^{(P)}$, and note that $T^{(f)}\propto \mu$ in the weak  
lensing regime, we find that the effect of magnification is to adjust
the resolution factor by 
\begin{equation} 
\delta R_2 = - \delta\left( \frac{T^{(P)}}{T^{(I)}} \right)
= 2(1-R_2)\kappa.
\end{equation}
The number of galaxies that are gained due to the resolution factor cut is then
\begin{equation}
{\delta N\over N}({\rm res.}) = 2(1-R_{2\,\rm min})n(R_{2\,\rm min})\kappa,
\end{equation}
where $n(R_{2\,\rm min})$ is the resolution factor distribution
normalized to $\int_{R_{2\,\rm min}}^1 n(R_2)\,\rmd R_2=1$.  The total  
change in the number density of galaxies is then
\begin{equation}
{\delta N\over N} = \left[ 5s-2 + 2(1-R_{2\,\rm min})n(R_{2\,\rm min})
  \right]\kappa. 
\end{equation}
For the three source samples, using the values of $s$ given above and
$n(R_{2,\rm min})$ from \S\ref{SSS:shearcalibration}, we estimate
$\delta N/N = 1.9\kappa$, $0.7\kappa$, and $3.1\kappa$ for $r<21$,
$r>21$, and LRG samples, respectively.  (Without taking into account
the effect of the change in $R_2$, we would have had $\delta N/N =
-0.2\kappa$, $-2.5\kappa$, and $-0.6\kappa$, so this effect significantly
changes our sensitivity to magnification bias.)

The convergence can be simply estimated for roughly power-law profiles
$\gamma_t\propto r^{-\alpha}$ as 
\begin{equation}
\kappa(r) = \left({2\over\alpha}-1\right)\gamma_t(r)+\kappa(\infty),
\end{equation}
where $\kappa(\infty)$ represents the mass-sheet degeneracy.  We will
ignore this since our boost factors approach unity at large  
separations.  Since for our galaxies we find $\alpha\approx 0.85$, it
then follows that $\kappa\approx 1.4\gamma_t$.  We can then use
computed values of \ds{} and \avgscinv{} to estimate $\kappa(r)$.
Table~\ref{T:magbias} shows the best-fit power-law $\kappa(r)$ and our resulting
predictions for $\delta N/N$ for L3--L6; L1 and L2 are not used
because the shear is statistically consistent with zero in these bins,
and therefore so is $\kappa$.  
\begin{table}
\caption{\label{T:magbias} Parameters of power-law fits
  $\kappa(r)=\kappa_0 (r/ 1h^{-1}{\rm Mpc})^{-\alpha}$
  to the lensing signal.  The
  resulting value of $\delta N/N$ as a function of lens and source
  sample is shown as well.}
\begin{tabular}{ccccc}
\hline\hline
Lens & $10^4 \kappa_0$ & $\alpha$ & $\delta N/N$ & $\delta N/N$ \\
sample & & & (30 \hinvk) & (100 \hinvk) \\
\hline
\multicolumn{5}{c}{$r<21$ without LRGs} \\
\hline
L3  & $5.5$ & 0.68 & 0.011 & 0.005 \\ 
L4  & $4.7$ & 0.88 & 0.020 & 0.007 \\ 
L5  & $7.9$ & 1.00 & 0.050 & 0.015 \\ 
L6  & $29$ & 0.95 & 0.15 & 0.049 \\ 
\hline
\multicolumn{5}{c}{$r>21$} \\
\hline
L3  & $9.7$ & 0.47 & 0.004 & 0.002 \\ 
L4  & $5.3$ & 0.81 & 0.006 & 0.002 \\ 
L5  & $9.3$ & 0.92 & 0.016 & 0.005 \\ 
L6  & $40$ & 0.85 & 0.055 & 0.020 \\ 
\hline
\multicolumn{5}{c}{LRGs} \\
\hline
L3  & $5.4$ & 0.70 & 0.019 & 0.008 \\ 
L4  & $4.4$ & 0.93 & 0.036 & 0.012 \\ 
L5  & $11$ & 0.95 & 0.095 & 0.030 \\ 
L6  & $44$ & 0.89 & 0.31 & 0.11 \\ 
\hline\hline
\end{tabular}
\end{table}

This effect is not taken into account
explicitly, and therefore may lead to a systematic bias  in the
signal, since we assume that the increase in sources around lenses is
due to physically associated pairs.  For all samples,
we may be overestimating the boosts and therefore
\ds{} (particularly for 
LRGs).  We do not, however, attempt to correct for this attempt
explicitly, since until a concrete detection is made, these estimates
should not be treated as certain.  For the $2\sigma$ confidence
intervals, we use the estimates in table~\ref{T:magbias} as the
expected value, with an expected uncertainty equal to half the
magnitude of the estimate.

\subsection{Intrinsic Alignments}\label{SS:intrinsic}

Satellite galaxies that are physically associated with a lens will not
be lensed; however, it is often difficult to remove them from the
source sample in the absence of good redshift information, which is
why we must use the boost factors described in the previous section.  If the
satellites are actually aligned with the lens in some way (radially or
tangentially elongated), this alignment will cause a false lensing
signal.  This effect is well-motivated theoretically
(\citealt{2001ApJ...555..106L}; \citealt{2002astro.ph..5512H}), and
there are  
several claimed detections in the literature of 
correlations between the galaxy density (or specific features thereof,
such as the Supergalactic Plane) and intrinsic ellipticities  
(e.g. \citealt{1989acfp.proc..418F}; \citealt{2002ApJ...567L.111L};
\citealt{2004ApJ...613L..41N}). 
We use $\ds_{int}$ to represent the signal due to
intrinsic shear, which is the shear signal that would be computed
using our shear estimator with a population of physically associated
galaxies.  Note that the galaxy density-shear correlation that
concerns us is distinct from the intrinsic shear-shear correlations  
that add spurious power to cosmic shear surveys
(\citealt{2000ApJ...545..561C}; \citealt{2000MNRAS.319..649H}; 
\citealt{2000ApJ...532L...5L}; \citealt{2001MNRAS.320L...7C};
\citealt{2001ApJ...559..552C}; \citealt{2002MNRAS.335L..89J}). 

In order to place upper limits on contamination due to
intrinsic shear, we must determine what fraction $f_c$ of the 
lens-source pairs are physically associated, and the typical intrinsic
tangential shear $\Delta\gamma_{int}$ of these sources.  
%For $f_c$, 
%we will use the weighted number of lens-source pairs per unit area 
%computed using both the real lens catalog ($n(r)$) and random lens 
%catalogs ($n_{rand}(r)$), where a comparison of the two  will show the
%effects of physically-associated galaxies by an increase in 
%$n(r)$ relative to $n_{rand}(r)$.  
To compute $f_c$, we use our results for $B(r)$.  Limits on $\Delta\gamma_{int}$ (or
on the intrinsic ellipticity or position angle) have been 
obtained by \citet{2001ApJ...555..106L}, \citet{2002AJ....124..733B}
and H04; we use the results from H04 as these are the tightest 
constraints on relevant scales.  Results for this estimation will be
shown in \S\ref{SSS:intrinsicresults}. 

\begin{comment}
\subsection{Weighting Scheme}

As shown in~\cite{2002AJ....123..583B}, weighting by ellipticity
should cause minimal ($<5$ per cent) bias in the results, we tested
this explicitly, by recomputing the signal using the following
weighting scheme:
\begin{equation}
w = 
\end{equation}
[Finish writing this up: use weighting by meas error only, and
  \scinvtwo weighting.]
[Do this test with new catalog; actually, decide if we want to use ellip weighting at all.]

\subsection{Redshift cutoffs}

Another concern is the effect of constraints 
placed on our photometric redshift sample.  As shown in H04, in order
to avoid contamination by physically associated pairs due to
photometric redshift errors, we imposed a
constraint $z_s > z_l+0.1$ on the photometric redshifts of sources
used to compute the signal.  However, it is necessary to check that
this cut does not cause a clear bias in the results relative to those
with no cut placed on the source redshifts.  (Recall that if the
photometric error distribution is included, then even sources with
$z_s < z_l$ may contribute due to the small probability that they are
actually at $z_l > z_s$).  Results of this comparison are shown in
section ??. [Do this.]
\end{comment}

\subsection{Correction for non-volume limited lens sample}\label{SS:notvol}

In this section, we describe the corrections that 
account for the fact that the lens samples are not volume limited, when
comparing between different source subsamples.  These corrections
required us to compute the mean weighted luminosity  for each
lens-source sample combination.  This computation was done as a
function of transverse separation in order to check for any systematic
effects.  

We found that there is a slight variation in the mean weighted
luminosity with transverse separation, because brighter lenses are  more
clustered, so in the regions where physically-associated pairs are
abundant (i.e., where $\xi_{ls}(r) \gg 0$), the mean weighted luminosity of
the sample was higher than at larger radii where $\xi_{ls}(r)\sim 0$.  We
see the same trend in the mean weighted redshift of the sample, as well.
However, this effect does not really mean that the mean weighted
redshift or 
luminosity corresponding to the computed signal is larger at small radii, since the excess
pairs around the brighter lenses that cause this effect do not
contribute to the signal.

We also expect that $L_{\rm eff}(r)$
and $z_{\rm eff}(r)$ may
increase at large $r$, because for more nearby (less luminous)
lenses, the same transverse separation corresponds to a larger angular
scale, and therefore survey edge effects may cause nearby lenses
to lose pairs faster than distant lenses.  The observation of such an
effect would indicate a problematic selection effect, but fortunately
we do not see it at 
the maximum transverse separations studied in this paper (2 \hinvm).  

The correction when comparing \ds{} at slightly different $L_{\rm
  eff}$ is as follows: we use the results of the
signal averaged over radius in each luminosity bin to derive a
relation $\ds(L)$.  Then, when we compare signals at two slightly different
luminosities (typical differences are less than 1 per cent) $L_{\rm
  eff}$ and $L_{\rm eff}+\delta
L$, we assume that rather than being 1, the ratio of the two signals
should be
\begin{equation}\label{E:deltalum}
\frac{\ds(L_{\rm eff}+\delta L)}{\ds(L_{\rm eff})} = 1+\frac{(\partial
  \ds/\partial L)_{L_{\rm eff}} \delta L}{\ds(L_{\rm eff})}.
\end{equation}
If this correction is non-trivial compared to the statistical error,
then we must apply it; otherwise, we do not.

\section{Application of the systematic test}\label{S:systest}

In this section, we describe our test to
determine the effects of the redshift distributions and systematic errors
described in \S\ref{S:zdist} and \S\ref{S:othersys}, and show
calculations justifying our assertion that cosmology plays a
negligible role.  

\subsection{Methodology}
First, in order to compute bootstrap errors on the results,
we divide our lens catalog into 200 bootstrap subsamples as described
in \S\ref{SSS:bootstrap}.  
For each lens luminosity bin and  source sample, we compute the signal
in each of the bootstrap subsamples, using 46 logarithmically spaced
radial bins from 20 to 2000~\hinvk{} in order to measure \dsr{} more
sensitively where it varies 
the most, at small transverse separation.  (Plots of signal shown in
this paper will have many radial bins averaged at small $r$ so they
are easier to read.)

Next, to account for physically associated pairs
and systematic shear, we used random lens catalogs to compute
$\ds_{rand}(r)$ and $n_{rand}(r)$.  Twenty-four random catalogs were
used, so
that these functions were determined reasonably smoothly.

After this procedure, we performed bootstrap resampling, generating 2500
datasets for each lens luminosity bin $i$ and source sample $\alpha$ as follows
(to avoid confusion, we use Roman letters to denote lens luminosity bins and
Greek letters to denote source samples):
\begin{enumerate}
\item For each lens-source sample combination, we use the same set (in
  order to take into account correlations) of 
  random numbers drawn from a uniform distribution to
  choose 200 bootstrap subsamples randomly, with replacement.  The
  signal from these subsamples are then averaged to get signal for
  each of the lens-source sample combinations, $\ds^{(meas)}_{i,\alpha}(r)$.
\item We compute boost factors $B_{i,\alpha}(r)$ using the weighted
  number of pairs per unit 
  area $n(r)$ relative to the average weighted number from random
  catalogs $n_{rand}(r)$.  The error in $n(r)$ is taken into account
  by the bootstrap procedure itself, and the error in $n_{rand}(r)$ is
  taken into account via multiplication by a Gaussian random number of mean 1
  and standard deviation equal to the fractional uncertainty in
  $n_{rand}(r)$, where a different random number is used for each $i$,
  $\alpha$, and radial bin.
\item We use the measured random catalog signal, taking into account
  its uncertainty via multiplication by a Gaussian random number of mean 1 and
  standard deviation equal to the fractional uncertainty in the random
  catalog signal, where a different number is used for each $i$, $\alpha$,
  and radial bin.  This procedure yields $\ds_{i,\alpha}^{(rand)}(r)$.  
\item Finally, we compute the signal for each dataset,
  $\ds_{i,\alpha}(r)=B_{i,\alpha}(r) (\ds^{(meas)}_{i,\alpha}(r) -
  \ds^{(rand)}_{i,\alpha}(r))$. The random catalog signal (not just
  the signal measured with the real lens catalog) must be
  multiplied by the boost factor because all galaxies, whether
  physically associated with the lenses or not, have systematic shear
  associated with them.
\end{enumerate}
Once the datasets have been generated using this procedure, we can
then do the following:
\begin{enumerate}
\item By averaging over the 2500 datasets, we can compute the mean
  value of $\ds_{i,\alpha}(r)$ (and other quantities such as
  $B_{i,\alpha}(r)$) and its standard deviation.
\item Using the bootstrap errors from the previous step for weighting
  purposes, we  
  can compute a value of $\overline{\ds}_{i,\alpha} \equiv \langle
  \ds_{i,\alpha}(r) 
  (r/ 1 h^{-1}{\rm Mpc})^{0.85}\rangle$  for each bootstrap resampled dataset.  The
  $(r/ 1 h^{-1}{\rm Mpc})^{0.85}$ is included in order to increase signal to noise on the
  averaged value. Only bins with $r>30$ kpc/$h$ are included in this
  calculation.
\item By averaging over the 2500 values of
  $\overline{\ds}_{i,\alpha}$, we can get the 
  mean value $\langle \overline{\ds}_{i,\alpha}\rangle$ and its
  covariance matrix, where its value for overlapping 
source samples will have very high correlations (the extreme
  cases of identical source samples but different methods of computing
  \scinv{} typically have correlation coefficient $r=0.99$).   This
  $\langle\overline{\ds}_{i,\alpha}\rangle$ will 
  be used to compare the signal for different sets of sources, the
  same lens sample.
\item For each luminosity bin $i$, we can compute ratios of the
  averaged signal with different sets of sources, 
\begin{equation}
R_{i,\alpha,\beta} = \frac{\langle\overline{\ds}_{i,\alpha}\rangle}{\langle\overline{\ds}_{i,\beta}\rangle}
\end{equation}
and its error.  The errors on the numerator and denominator, and their
covariance, were used to compute confidence intervals on $R_{i,\alpha,\beta}$
using the full non-Gaussian error calculation, which is important for
those  cases in which either number may be consistent with zero.  
 (For details of the
calculation of confidence intervals on the ratio of two correlated
variables, see Appendix~\ref{app:confint}.)  Due to the high
correlations between some of the source samples, the ratio could often
be computed to extremely high accuracy. 
\item For many of the tests described below, the ratio test was also
  done for the signal averaged over luminosity bins as well.  This
  comparison must be done with care; since the relative weights of the
  signal in the different luminosity bins varies from sample to sample
  (e.g., all LRGs contribute to each luminosity bin, but there
  are many more $r<21$ galaxies in the faint bins than in the bright
  ones due to the $z_p>z_l+0.1$ cut, so for $r<21$ the fainter bins
  get more weight than the bright ones relative to the LRG sample), we must be careful that when
  comparing samples $\alpha$ and $\beta$ averaged over luminosity, we
  use the same weights for averaging each sample over luminosity.  If the weights
  differ, then we could mistakenly be led to believe that there is a
  calibration error.  So, for each bootstrap dataset, we compute the
  average value of \ds{} via
\begin{equation}
\overline{\ds}_{{\rm all}, \alpha} = \sum_{i} w_i \overline{\ds}_{i,\alpha}
\end{equation}
where
\begin{equation}
w_i = \frac{\overline{\ds}_i v_i}{\sum_i \overline{\ds}_i v_i}.
\end{equation}
Here, $\ds_i$ is some approximation to the actual value of
 $\overline{\ds}_i$ (in practice, the value averaged over subsamples
 $\alpha$ and $\beta$), and 
\begin{equation}
v_i = \frac{1}{\sigma^2_{i,\alpha} + \sigma^2_{i,\beta} -2{\rm Cov}(\overline{\ds}_{i,\alpha},\overline{\ds}_{i,\beta})}.
\end{equation}
The values of $\sigma$ used in this expression are from the covariance
 matrix already obtained for $\langle
 \overline{\ds}_{i,\alpha}\rangle$.  By averaging over bootstrap
 datasets, we can then get values $\langle
 \overline{\ds}_{{\rm all},\alpha}\rangle$ that allow us to compute ratios
\begin{equation}
R_{{\rm all},\alpha,\beta} = \frac{\langle\overline{\ds}_{{\rm
      all},\alpha}\rangle}{\langle\overline{\ds}_{{\rm all},\beta}\rangle}
\end{equation}
and obtain non-Gaussian confidence intervals as usual.
\end{enumerate}

\subsection{Cosmology dependence}

In this section, we show why the systematics test is essentially 
cosmology-independent.  We compute \ds{} by
computing $\gamma_t$ (which is independent of cosmology, except via the
weighting scheme since $w_{ls} \propto \scinvtwo$, which should not change
the results, only the errorbars) and $\Sigma_{c,\rm guess}^{-1}$ with our assumed
cosmology.  Now, we propose that our cosmological model is wrong, so
the true value of \ds{} can be computed from the same $\gamma_t$ that
we computed, but with a different \scinv{} that we will call
$\Sigma_{c,\rm true}^{-1}$.  Our results for a given lens and source sample are
then related to the true \ds{} by the relation
\begin{equation}
\ds_{\rm measured} = \ds_{\rm true}\left(\frac{\Sigma_{c,\rm
    true}^{-1}}{\Sigma_{c,\rm guess}^{-1}}\right)
\end{equation}
so we can define a ``cosmology factor'' $f_{cos} \equiv \Sigma_{c,\rm
  true}^{-1}/\Sigma_{c,\rm guess}^{-1}$
relating the true and measured signal.  Our concern, then, is that
$f_{cos}$ for a given lens redshift will be different for each source
sample, so that our comparison of \ds{} computed with different
source samples is invalid.

We do a simple test for three lens redshifts ($z_l=0.02$, the minimum
value; 0.1, the typical value; and
0.25, on the high-redshift tail of the lens redshift distribution),
and two 
source redshifts ($z_s=z_l+0.1$ and 0.7) that span the range of source
redshifts used.  Furthermore, we test two cosmologies that are
drastically different from the one assumed: a flat
cosmology with $\Omega_m=1$, and an open cosmology with
$\Omega_m=0.3$.  For each cosmology, lens redshift, and source
redshift, we can compute $f_{cos}$ relating our measured \ds{} and the
true value, and compare the values of $f_{cos}$ between the different
source samples for a given lens redshift and cosmology.  The results
of this test are shown in table~\ref{T:fcos}.
\begin{table}
\caption{\label{T:fcos}
The ratio $f_{cos}$ relating the true value of \ds{} for two different
cosmologies to the one measured using
the assumed cosmology, shown for several lens redshifts and source redshifts.}
\begin{tabular}{cccc}
\hline\hline
$z_l$ & $z_s$ & $f_{cos}$ & $f_{cos}$ \\
 & & (flat $\Omega_m=1$) & (open $\Omega_m=0.3$) \\
\hline
0.02 & 0.12 & 0.980 & 0.987 \\
0.02 & 0.70 & 0.981 & 0.988 \\
0.10 & 0.20 & 0.910 & 0.941 \\
0.10 & 0.70 & 0.915 & 0.946 \\
0.25 & 0.35 & 0.814 & 0.878 \\
0.25 & 0.70 & 0.821 & 0.885 \\
\hline\hline
\end{tabular}
\end{table}

We can see from Table~\ref{T:fcos} that while the measured value of
\ds{} may be off from the true one by a significant fraction, the
differences between $f_{cos}$ for the same lens redshift but different
source redshift are less than 1 per cent even for these extreme cosmologies
for all cases.  In reality, most of the lenses are near $z\sim
0.1$, most of the sources are at more intermediate values of redshift,
and the allowed ranges of cosmology is much smaller than the extreme
cases considered here.  Consequently, we can state with confidence
that cosmology plays a negligible role in the comparisons of source
samples for the allowable range of cosmology and lens redshift range
probed by the SDSS.

There were several over-simplications involved in this simple
calculation.  First, we did not take into account the lens or source
redshift distributions; however, since we found that cosmology is not
important for the range of lens redshifts used, and for the most
extreme values of source redshift, this over-simplification is
unimportant.  Second, we did not account for the change in weighting
when \scinvtwo{} changes from the assumed to the true cosmology, but
this change simply means that the weighting scheme was not optimal, which
would lead to larger errorbars but no change in the result.  Finally,
we did not account for another change in \ds{} due to the change in
cosmology: since it is computed as a function of transverse comoving
separation, $r=\theta_{ls}D_A(z_l)(1+z_l)$, cosmology also comes into
the computation of $r$ for a given lens-source pair.  The ratio
$r_{guess}/r_{true}$ is nontrivially different from 1 for the two cosmologies considered,
and since $\ds \propto r^{-\alpha}$, rescaling $r$ leads to \ds{} being
off by some factor.  However, this effect is only a function of lens
redshift, so the rescaling will be the same for all source samples,
and therefore does not affect our use of the ratio test for systematics.

Finally, it is worth noting that for futuristic surveys with lens
redshifts around $z_l \sim 0.5-0.7$ and two source samples at $z_s\sim 1$
and $z_s\sim 1.5$, the differences between $f_{cos}$ for the two
source samples for the 
cosmologies considered here is more than 1 per cent, so if the
systematics are better under control, then this sort of difference may
be detectable with future datasets.

\section{Results}\label{S:results}

In this section, we present results for DEEP2 redshift distributions
and photometric redshift error distributions; compare the different
methods of error determination; and demonstrate the results of the
systematics test.

\subsection{Redshift distributions and photometric redshift
  performance}\label{SS:zresults}

Here we describe the redshift distributions and photometric redshift error
distributions used for the rest of the paper.  As mentioned in
\S\ref{SS:DEEP2}, several tests were performed using the 
DEEP2 redshifts.  First, they were used to determine the
redshift distribution of the source galaxies, and then they were used
to study the photometric redshift error distributions.

\subsubsection{Redshift distributions}
The redshift distribution
determination using DEEP2 galaxies was done by choosing a 
common functional form for the redshift distribution, the $\Gamma$
distribution: 
\begin{equation}
p(z)\equiv \frac{\rmd P}{\rmd z} = \frac{z^{\alpha-1} e^{-z/z_s}}{z_s^{\alpha}\Gamma(\alpha)}
\label{eq:gammadist}
\end{equation}
This probability distribution has mean $\meanz = \alpha z_s$, variance $\langle
(z-\meanz)^2\rangle = \alpha z_s^2$, and mode $(\alpha-1) z_s$. 
A maximum-likelihood fit was done for parameters $\alpha$ and  
$z_s$ for $18<r<21$ and $r>21$ galaxies separately, with fit results
shown in table~\ref{T:fitzdist}.  The fit was performed by minimizing 
\begin{equation}
\chi^2 = \sum_i [-2\ln p({z_i})],
\end{equation}
where the summation is over all DEEP2 matches in the appropriate
magnitude range.  This ``$\chi^2$'' function can be minimized
to give the best-fit redshift distribution, but it does not  
have the same distribution as the usual $\chi^2$ function; its
statistical properties are summarized in Appendix~\ref{app:chi2}. 
For example, goodness-of-fit cannot be measured in an absolute sense
directly from the $\chi^2$, since the ``zero-point'' is undetermined;
only differences between the fits for different
models are significant. 
Note that since the majority of
the galaxies are located in the peak of the probability distribution,
where $p(z) > 1$, the best-fit $\chi^2<0$.  Table~\ref{T:fitzdist} 
summarizes the fit results, and 
includes errorbars both with and without large-scale structure (LSS)
taken into account; LSS is taken into account as described in
Appendix~\ref{app:chi2}, 
assuming angular correlation function $w(\theta)$
from~\cite{2002ApJ...579...42C}. 
Because that $w(\theta)$ may not be correct for lensing selected
galaxies, this model may not perfectly capture the increase in
errorbars due to LSS, but should compute them to within reasonable
accuracy.  For the $r>21$ sample, we used amplitude of the angular
correlation function for the $21<r<22$ sample
from~\cite{2002ApJ...579...42C}.  For the $r<21$ sample, our errorbars
including LSS are a worst-case estimate, where we combined the
amplitudes $A_1$, $A_2$, and $A_3$ for $18<r<19$, $19<r<20$, and $20<r<21$
sources (with fractions of galaxies $f_1$, $f_2$, and $f_3$) from that work to
get an overall amplitude $A=( f_1\sqrt{A_1} + f_2\sqrt{A_2}
+ f_3\sqrt{A_3} )^2$, which assumes perfect correlation between the
samples.  While a perfect correlation is unrealistic, we use this
result to place 
conservative errorbars with LSS taken into account.
\begin{table}
\caption{\label{T:fitzdist} Parameters of best-fit redshift
  distributions for several source samples. For both samples,
  $r(z_s,\alpha)=-0.94$, so the fit 
  parameters are highly correlated.  Results for errorbars and the
  variance of the $\chi^2$ are shown both with and without large-scale
  structure taken into account.}
%\begin{tabular}{ccccccccccccccc}
\begin{tabular}{rcc}
\hline\hline
%Sample  & $N_s$ & $z_s$  & $\sigma(z_s)$ & $\sigma_{\rm LSS}(z_s)$ &
%$\alpha$  & $\sigma(\alpha)$ & $\sigma_{\rm LSS}(\alpha)$ &
%\meanz & 
%$\sigma(z)$ & mode($z$) & $\langle\chi^2\rangle$ & $\langle\chi^2\rangle_{theory}$  & Var($\chi^2$) &
%Var$_{\rm LSS}(\chi^2$) \\
Sample & $18<r\le21$ & $r>21$ \\
\hline
$N_s$ & 162 & 116\\
$z_s$ & 0.100 & 0.105\\
$\sigma(z_s)$ & 0.012 & 0.014\\
$\sigma_{\rm LSS}(z_s)$ & 0.047 & 0.017\\
$\alpha$ & 3.52 & 4.36 \\
$\sigma(\alpha)$ & 0.38 & 0.55 \\
$\sigma_{\rm LSS}(\alpha)$ & 1.77 & 0.70 \\
\meanz & 0.35 & 0.46 \\
$\sigma(z)$ & 0.19 & 0.22 \\
mode($z$) & 0.25 & 0.35 \\
$\langle\chi^2\rangle$ & -117 & -43\\
$\langle\chi^2\rangle_{theory}$ & -115 & -42\\
Var($\chi^2$) & 283 & 194 \\
Var$_{\rm LSS}(\chi^2)$ & 4917 & 361 \\
%$18<r\le21$ & 162 & $0.100$ & 0.012 & 0.047 & $3.52$ & 0.38 &
%1.77 & 0.35 & 0.19 & 0.25 & -117 & -115 & 283 & 4917 \\
%$r>21$ & 116 & $0.105$ & 0.014 & 0.017 & $4.36$ & 0.55 & 0.70 & 
%0.46 & 0.22 & 0.35 & -43 & -42 & 194 & 361 \\
\hline\hline
\end{tabular}
\end{table}

\begin{figure}
\includegraphics[width=3in,angle=0]{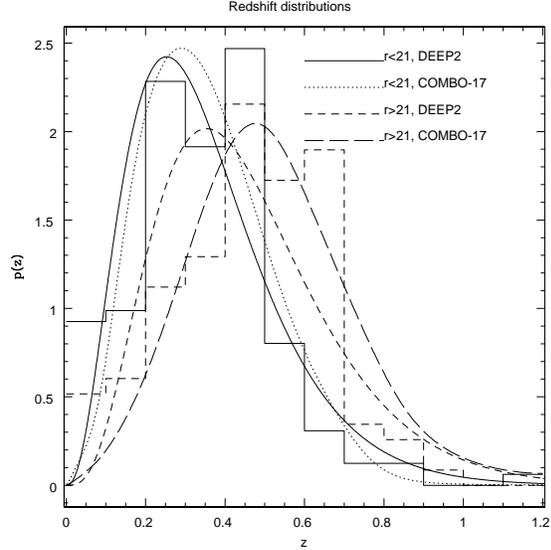}
\caption{\label{F:zdist.DEEP}The redshift distribution from COMBO-17
  and DEEP2, shown separately for $18<r<21$ and for $r>21$ galaxies.
  The lines are, as labeled, the COMBO-17 $p(z|r)$ averaged over
  $p(r)$ for all galaxies, and the best fit $\Gamma$ distribution from
  DEEP2 for lensing-selected galaxies.  The
  histograms are derived from the actual data.}
\end{figure}

The resulting 
distributions for lensing-selected galaxies are shown in
figure~\ref{F:zdist.DEEP}, along with the 
histogram of redshifts from DEEP2, and the COMBO-17 distribution for
all galaxies. While the fits
are reasonably good, it is apparent that
the $\Gamma$ distribution may not be the best choice of
distribution because its shape is not well-suited to matching the shape of the
histogram of redshifts from DEEP2 for $r>21$. However, we have not found
another functional form that would be more appropriate. Note that this
plot shows the 
actual redshift distributions; when the signal is computed, the
effective source redshift distribution, which involves the weights
used for each lens-source pair, is more important.  These weights tend
to emphasize the higher-redshift portion of the curve via the
inclusion of \scinvtwo, but that portion is also downweighted due to
the fact that more of the higher-redshift sources are detected with
lower significance, and therefore have higher $\sigma_e$.  All
plots of redshift distributions shown in this paper are the unweighted
versions.  As shown, the COMBO-17 distribution is at slightly higher
redshift than the best-fit DEEP2 distributions; the discrepancy may be
attributed to an offset in the magnitudes used 
to determine the distribution, and the fact that the galaxies involved
in its determination were not required to pass our selection criteria.

As compared to DEEP2 matches for the full photometric sample,
including those that had failed shape determination and therefore were
not in the source catalog, these redshift distributions are at
slightly lower redshift on average, as would be expected.

We can estimate the possible variation of the signal due to
statistical error in the redshift distribution determination.  This
calculation can be done using the statistical errorbars from the fits,
and the larger ones that include LSS, for both $r<21$ and $r>21$
sources.  We determine errorbars on \scinv{} via propagation of the
fit covariance matrix ${\mathbfss C}$ for the vector of parameters
$\vec{a}=(z_s,\alpha)$ via 
\begin{equation}
\sigma_{\scinv}^2 = \left(\frac{\rmd\scinv}{\rmd\vec{a}}\right)^T
\cdot {\mathbfss C} \cdot \left(\frac{\rmd\scinv}{\rmd\vec{a}}\right).
\end{equation}
%Since the correlation coefficient for $\alpha$ and $z_s$ is
%$\sim -0.94$, the final term in the second
%line is negative and nearly as large as the sum of the first two
%terms.  
For this computation, we use two-sided 
derivatives with respect to $z_s$ and $\alpha$ evaluated at
separations $\pm 0.1\sigma$ around the
best-fit parameters $\vec{a}_{fit}$.  We can then compute the
calibration uncertainty 
on the signal, $\delta (\ds)/\ds = -\delta(\scinv)/\scinv$ for a given
lens redshift.  Results
for this computation are shown in table~\ref{T:zcalibration}.
\begin{table}
\caption{\label{T:zcalibration} 
  Estimated calibration uncertainty on \ds{} for two
   values 
  of $z_l$ and $p(z)$ computed for $r<21$ and $r>21$
  sources.  The uncertainties are shown both with and without LSS
  taken into account.}
\begin{tabular}{cccc}
\hline\hline
$z_l$ & & $\frac{\delta(\ds)}{\ds}$ & $\frac{\delta(\ds)}{\ds}$ \\
 & & ($18<r<21$) & ($r>21$) \\
\hline
0.03 & without LSS & $\pm 0.007$ & $\pm 0.0055$ \\
0.03 & with LSS    & $\pm 0.040$ & $\pm 0.0073$ \\
0.20 & without LSS & $\pm 0.053$ & $\pm 0.045$ \\
0.20 & with LSS    & $\pm 0.25$ & $\pm 0.057$ \\
\hline\hline
\end{tabular}
\end{table}
As shown, the uncertainty in the signal
due to the redshift distribution statistical uncertainty is larger for
higher $z_l$ since those lenses are closer to the source redshift
distribution; for lenses at $z_l=0.03$, the majority of the
distribution is essentially at $z=\infty$, so $\rmd\scinv/\rmd\vec{a}$
is small.  Furthermore, the errors on the distribution at $r<21$ have
a greater effect than those on the distribution at $r>21$ because the
brighter sources are, on average, closer to the lenses.  As shown, the
errors for the $18<r<21$ sources with LSS are quite large (large
enough for $z_l=0.2$ that we may worry about the accuracy of this
linear calculation), but that is due to the fact that we used a very
conservative estimate of $w(\theta)$ for this sample as described
above.  Consequently, those errorbars are very conservative, and in
reality are most likely significantly smaller and closer to the
errorbars without LSS.  Nonetheless, we can see that using the
redshift distribution for $r<21$ may be a bad idea because 
errors in its determination may have a large effect on \ds.

\subsubsection{Photometric redshift errors}\label{SSS:pzerr}

In order to be able to use photometric redshifts to
eliminate physically associated pairs, we also computed photometric redshift
error distributions using the matches from DEEP2.  Of the 278
matches, 162 (58 per cent) were at $r<21$ (with 9 of those, or 6 per cent, having 
failed photometric redshift determination).  Because high-redshift LRGs
have different properties from the overall photometric sample, we
explicitly exclude them from this comparison, leaving 135 galaxies
with which to study photometric redshifts. 

First, we define the terminology used in this section.  For a set of
true redshifts $z$ and photometric redshifts $z_p$, we define the
photometric redshift errors as $\delta=z-z_p$, and can 
construct average distributions $p(\delta|z)$ and $p(\delta|z_p)$ in bins.
The statement that the 
photometric redshifts are biased means that
\begin{equation}
\int p(\delta | z)\,\delta\, \rmd\delta \ne 0.
\end{equation}
For our sample of 135 redshifts at $r<21$, we find that overall there is a small
photometric redshift bias, where the overall bias is 0.04, with
$\langle z\rangle=0.34$ and $\langle z_p\rangle=0.30$.  When we look at the
subsample of 64 galaxies below the mean redshift, the bias is zero to
two significant figures
and the scatter is 0.11.  When we restrict to the subsample of 71
galaxies above the mean redshift, the mean bias is 0.08, and the
scatter is 0.17.

However, the calculation that is actually more relevant for the
purpose of this paper is the ``conditional bias,''
\begin{equation}
\langle\delta\rangle \equiv \int p(\delta | z_p)\,\delta\, \rmd\delta
\end{equation}
In particular, even if the bias is zero, then for a redshift distribution
that peaks at redshift $\langle z \rangle=0.34$, we will have
$\langle\delta\rangle>0$ for $z_p<\langle z \rangle$ and
$\langle\delta\rangle<0$ for $z_p>\langle z \rangle$ simply due the
scatter in the photometric redshifts.  We also define a ``conditional
scatter,''
\begin{equation}
\sigmad \equiv \sqrt{\int p(\delta|z_p) (\delta-\meand)^2 \,\rmd\delta}.
\end{equation}

Table~\ref{T:errmag} shows the mean conditional bias and scatter as a
function of 
magnitude; table~\ref{T:errph} shows the same information as a
function of 
photometric redshift.  Information is included for $r>21$ in
Table~\ref{T:errmag} for informational purposes, but since we did not use
photometric redshift information for those sources due to the large
bias and scatter, they are not included in the calculations
for Table~\ref{T:errph}.  A likely explanation for the increase in \sigmad{}
at fainter magnitudes is the larger photometric errors which make
photometric redshift determination more difficult.  The fact that our
test of this program on the SDSS spectroscopic sample at $r<18$, on
galaxies with negligible photometric errors, shows
$\meand=-0.003$ and $\sigmad=0.048$, with $\sigmad$ increasing with
$r$ magnitude if the sample is split into subsets, supports this
hypothesis.  Also, because   
we require $z_l>0.02$ and $z_s>z_l+0.1$, we only use
photometric redshifts greater than 0.12 in our analysis, though error
information has been included in the tables for those less than 0.12.  
Our finding that photometric redshifts are, on average,
biased low is in accordance with the same finding in H04 (based on the
fractions of physically associated pairs found using $z_p<z_l-0.1$).
%Furthermore,
%because photometric redshift errors for LRGs have been well-studied
%in~\cite{2004astro.ph..7594P} with much better statistics as described in \S\ref{SS:LRG}, high-redshift LRGs
%have been excluded explicitly from the study of photometric redshift
%errors in this section.
\begin{table}
\caption{\label{T:errmag} 
Conditional bias \meand{} and scatter \sigmad{} as a function of $r$ model
magnitude for non-LRGs with successful photometric redshift
determination}
\begin{tabular}{ccccc}
\hline\hline
Magnitude range & $\langle r\rangle$ & $N_{gal}$ & $\meand$
& $\sigmad$ \\
\hline
\multicolumn{5}{c}{All $z_p$}\\
\hline
$18\le r<20$ & 19.3 & 46 & 0.035 & 0.14 \\
$20\le r<20.5$ & 20.2 & 39 & 0.008 & 0.13 \\
$20.5\le r<21$ & 20.8 & 50 & 0.084 & 0.17 \\
$21\le r<22$ & 21.5 & 100 & 0.11 & 0.25 \\
\hline
\multicolumn{5}{c}{$z_p>0.12$ only}\\
\hline
$18\le r<20$ & 19.2 & 39 & 0.032 & 0.15 \\
$20\le r<20.5$ & 20.2 & 32 & -0.022 & 0.12 \\
$20.5\le r < 21$ & 20.8 & 44 & 0.058 & 0.16 \\
\hline\hline
\end{tabular}
\end{table}
\begin{table}
\caption{\label{T:errph} 
Photometric redshift conditional bias $\meand$ as a function of $z_p$
for non-LRGs with successful photometric redshift determination.
Only $r<21$ sources are included; conditional biases and scatter become
worse for $r>21$ sources, so their photometric redshifts were not used.}
\begin{tabular}{ccccc}
\hline\hline
$z_p$ range & $\langle z_p \rangle$ & $N_{gal}$ & $\meand$
& $\sigmad$ \\
\hline
All & 0.30 & 135 & 0.045 & 0.15 \\
$z_p < 0.12$ & 0.04 & 20 & 0.15 & 0.13 \\
$z_p\ge 0.12$ & 0.34 & 115 & 0.027 & 0.15 \\
$0.12\le z_p<0.25$ & 0.17 & 28 & 0.074 & 0.13 \\
$0.25\le z_p<0.35$ & 0.31 & 28 & 0.076 & 0.21 \\
$0.35\le z_p<0.42$ & 0.38 & 28 & 0.017 & 0.07 \\
$z_p\ge 0.42$ & 0.49 & 31 & -0.051 & 0.11 \\
\hline\hline
\end{tabular}
\end{table}

\begin{figure}
\includegraphics[width=3in,angle=0]{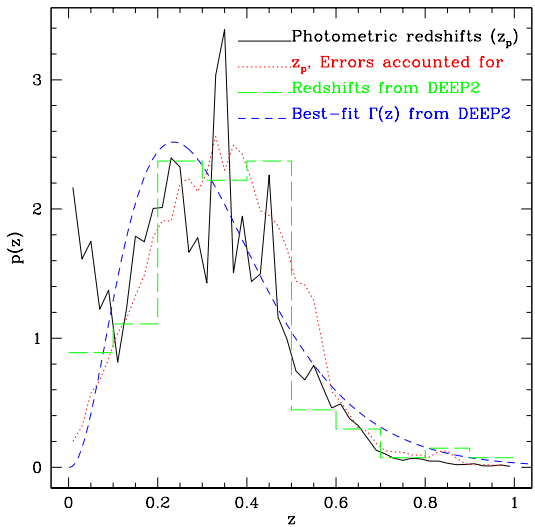}
\caption{\label{F:rLT21dists}The redshift distribution for $r<21$ non-LRG
  sources with successful photometric redshift
  determination, shown computed with spectroscopic redshifts and with
  photometric redshifts, with and without error distributions taken
  into account.}
\end{figure}

To illustrate the effects of photometric redshift errors,
Fig.~\ref{F:rLT21dists} shows the redshift distribution
for the $r<21$ sample computed using photometric redshifts, with and
without the errors taken into account according to Eq.~\ref{E:deconvolve}, and using the
spectroscopic redshift distributions from DEEP2 directly.  As shown, the
photometric redshift distribution has some strange features, including
the peak near $z=0$ and at $z=0.38$; the distribution with the errors
taken into account looks significantly more reasonable, and is
an excellent match for the histogram of redshifts from DEEP2.  In
fact, the distribution with photometric redshift errors taken into account 
appears to be a better match than the best-fit $\Gamma$-distribution.
(Any discrepancies between the distributions in
Figs.~\ref{F:zdist.DEEP} and~\ref{F:rLT21dists} arise from the fact
that figure~\ref{F:zdist.DEEP} includes LRGs in the $r<21$ sample but
Fig.~\ref{F:rLT21dists} does not.)  While we may have been concerned
about using so few galaxies to determine the error distribution as a
function of photometric redshift (by dividing 135 galaxies into five
bins), it is apparent that even with so few galaxies, enough of the
error distribution can be determined that this procedure is
reasonably successful.
%\begin{figure}
%\includegraphics[width=3in,angle=0]{pzerrdist.mag.zp.ps}
%\caption{\label{F:pzerrdist}
%Plots of the photometric redshift error distribution for non-LRGs as a
%function of photometric redshift (top) and $r$ model magnitude (bottom).
%}
%\end{figure}

While we have determined the photometric redshift error distributions, we are
still far from an understanding of how they affect the lensing
signal.  There are two effects that must be included:
\begin{itemize}
\item The conditional bias, \meand, must be included in our computation of
  \scinv.  Because $\meand>0$ for low $z_p$, the
  corrected redshift is larger than $z_p$, raising
   \scinv{} (lowering \ds) once we make this correction.  This effect is
  particularly important at low values of $z_p$, because \scinv{} is
  varying more rapidly with source redshift at low redshift.
  (However, since \scinv{} is lower at low source redshifts,
  these sources get less weight, so the error they cause is not as
  high as one might naively expect.)  At high
  redshift, $\meand<0$, so the correction will lower the
  redshift and \scinv{} (raise \ds), but because \scinv{} varies so slowly with
  redshift in that regime, we expect that this will only be a minor
  correction.  The majority of the photometric redshifts used are 
  around 0.35-0.4,  where the conditional
  bias
  goes to zero, so we may expect that its effects will not be
  too large.
\item The conditional scatter \sigmad{} must also be included.  At photometric
  redshift more than \sigmad{} above the lens redshift (i.e., for
  most sources in this work, due to our requirement that
  $z_p>z_l+0.1$), because \scinv{} is an
  increasing function of source redshift with negative second
  derivative in this regime, the inclusion of the width of the error
  distribution will 
  lower the estimate of \scinv{} (raise \ds).  This effect is counter
  to the effect 
  of the conditional bias \meand{} at low $z_p$, and in the same
  direction as it at 
  high $z_p$.   However,
  at low $z_p$, close to $z_l$, the effect of the $\delta$-function
  in the second derivative at $z_s=z_l$ is that \scinv{} is actually
  higher than it would have been without the scatter, i.e. this effect
  goes in the same direction as the bias at very low photometric redshift.
\end{itemize}

\begin{figure}
\includegraphics[width=3in,angle=0]{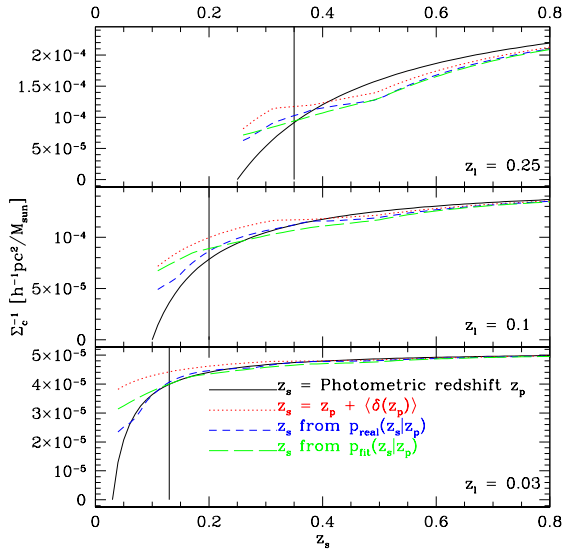}
\caption{\label{F:comparescinv}
$\scinv(z_l,z_s)$ for several values of lens redshift.  For each case,
  a line showing our cutoff at $z_p=z_l+0.1$ is shown.  As labeled on
  the plot, \scinv{} was computed four different ways: using photometric
  redshifts directly; using them corrected for the conditional bias
  \meand{} only; and with
  corrections for   the actual and the best-fit Gaussian photometric redshift error
  distribution.}
\end{figure}

Figure~\ref{F:comparescinv} shows a plot of \scinv{} for lenses at several redshifts, as a
function of source redshift, for several different possibilities:
straight acceptance of source photometric redshift, use of the
photometric redshift corrected for the conditional bias, and corrected
for the full
photometric redshift error distributions (actual, and best-fit).  The
fit for the distribution was done by assuming it is a Gaussian with a
fixed width but with a mean that is a linear function of $z_p$.  This
form was assumed because it is a simple parametrization that can still
account for the fact that the error distribution changes sign as we go
from low to high $z_p$.  The
effects of the mean correction (which raises \scinv{} for $z_p<0.4$
and lowers it for $z_p>0.4$)
and the width of the error distribution can clearly be seen in this
plot.  

Vertical lines show our cutoff of $z_p=z_l+0.1$; fortunately, as shown,
this limitation on the photometric redshifts means that even just
using photometric redshifts directly, without accounting for the
errors, is less likely to cause errors.  Furthermore, as shown, the
errors in \scinv{} for $z_p \sim z_l+0.1$ are downweighted in importance due
to our weighting scheme, with weights $\propto \scinvtwo$.
Consequently, we can expect to see some cancellation, whereby the
large increase in \scinv{} (decrease in signal) for small $z_p-z_l$,
which has small weight due to the weighting scheme, is cancelled by the
modest decrease in \scinv{} (increase in signal) for large $z_p$ and
$z_l$, which get considerable weight. The extent of the cancellation
will be shown in \S\ref{SS:sigresults}.  

The difference in results between the best-fit Gaussian error
distribution versus the actual error distribution is quite small, except
for the range $z_p \sim z_l$ that is not used in this work; probably, the
best-fit distribution gives higher values of \scinv{}
in this range because the fit distribution underestimates the tails
that are present in the actual distribution, and the tails help to
lower \scinv{} with the actual error distribution for $z_p \sim z_l$.
This result implies that when restricting to the range $z_p>z_l+0.1$
as in this work, it is unimportant whether or not the method of
parametrizing the photometric redshift error distribution is able to
account for the tails, as long as the basic shape (mean and width) is
correct.  However, for analyses that include lower values of $z_p$, a
more careful treatment of the error distribution is necessary.

It is clear from these results that
for different limits on source photometric redshift (e.g., without
requiring $z_p > z_l+0.1$), the
results of photometric redshift errors will vary, and must be
re-evaluated if we choose to relax this restriction.

As illustrated by this discussion, the effect of photometric redshift
errors is complicated, with several possible effects that will tend to
push the lensing signal in opposite directions, and it is difficult to
anticipate which effects will turn out to be more important than
others for any given lens and source sample.  Consequently, we must
turn to the systematics test described in 
\S\ref{S:systest} to determine the effects of
photometric redshift errors on \ds. 

We also must consider the effects of errors in the photometric
redshift error distribution.  Statistical (Poisson) error can be
determined via a bootstrap resampling of the DEEP2 matches.
The effects of LSS and systematic error are more complex, because correlations
between photometric redshift errors are more difficult to model than
the correlations between the redshifts used to derive $p(z)$.  There
are several concerns: statistical errors on photometry; systematic
errors on colors; and deviations of galaxy SEDs from templates.  For
example, if the colors of galaxies depend on environments, we may
expect different photometric redshift error distributions for field
galaxies versus those in groups or clusters. 

\subsection{Error comparison}\label{SS:errcomp}

Because we used three methods to compute error estimates on the
signal (the analytic expression for the error, the random catalog
errors, and the bootstrap errors), we compare the results of the
three methods to ensure that they gave comparable results.  In
general, the results of the three methods were quite comparable,
with the exception of two situations.  First, for the
lower luminosity bins, the bootstrap and random catalog errors were
larger than those computed analytically by 15--20 per cent.
Second, for the higher luminosity bins, the radial bins that required
a sizable boost had larger errors computed using the bootstrap than
via the other methods.  Because there is some uncertainty on the boost
factors (which was also computed using the bootstrap), the uncertainty
on the product $B(r)\ds(r)$ is approximately
$B(r)\ds(r)\sqrt{((\sigma_B/B)^2+(\sigma_{\ds}/\ds)^2)}$ when computed
using the bootstrap, raising the fractional error by roughly 5 per
cent at small pair separations.  The larger errors from the bootstrap are more
likely to be correct since they take into account errors on the boost
factor itself, so these are the errors that were used for the
computation of ratios $R_{i,\alpha,\beta}$.  

The analytic errors are essentially noiseless since they are computed
over millions of lens-source pairs.  The random catalog errors had the
most noise, since they were computed with only 24 random catalogs due
to the computational expense involved.  If the error amplitude $E(r)$
is fit to a power-law $Ar^{-1}$, and the noise
amplitude is defined as the fractional difference between the actual errors
and the best-fit value, then the random catalog errors
typically had a noise amplitude of 15 per cent.  The bootstrap errors have
little to no noise, with a maximum noise amplitude of 5 per cent even for
L1 and L6 (which had the fewest lenses) and 1 per cent or lower for the others.

We also checked the covariance between the signal in different radial
bins using the bootstrap for all luminosity bins and source samples.
We can anticipate that these correlations may become important on
angular scales for which the lens-source separation is comparable to
the lens-lens separation, since on those scales a given source will
contribute to the calculation for more than one lens.  We find a
typical lens-lens separation to be 7', which corresponds to comoving
transverse separation of 200 \hinvk{} in L1, and as high as 1100 \hinvk{}
in L6.  Using the bootstrap resampling for 2500 datasets, we find that
for L1, there are correlations as high as 0.2--0.3 starting at radii of 800
\hinvk{} between nearby radial bins (generally the 3 nearest radial bins
in either direction) but little covariance between the bins for lower
radii.  The bootstrap procedure naturally introduces noise with
standard deviation
$1/\sqrt{M}$ ($M$ is the number of bootstrap regions) in the
correlation coefficients when they are $\ll 1$ (the noise is less for
higher correlations), so the statement that there is little
covariance means that the correlation coefficients were consistent
with being drawn from a Gaussian distribution $N(0,1/M)$.  For L6, no
covariance is observed within the noise, even for the outermost bins
that should be most correlated, since these correspond to a smaller
angular scale  than for the fainter (more nearby) lenses.  Note that
our procedure 
described in \S\ref{S:systest} automatically takes any
covariance between the radial bins into account because it uses the
bootstrap.

We also found that our measure of the signal in a given
bootstrap-resampled dataset, $\overline{\ds}$ as  defined in
\S\ref{S:systest}, has a significant covariance between
luminosity bins.  This result is unsurprising considering that many of
the same sources contribute to the result in different luminosity
bins, though the 
weighting may differ significantly.  We found correlations between the
results in different luminosity bins as high as 0.30 for adjacent
luminosity bins, 0.10 for luminosity bins that were separated by one
(e.g. L4 and L6), and consistent with zero for further separations.
This kind of correlation must be taken into account 
(as it is with the bootstrap) for any analysis that utilizes the
results in several luminosity bins simultaneously.

As an additional test, we also examined the distribution of values of
$\overline{\ds}$ for each luminosity bin and source sample, computed
from the 2500 bootstrap-resampled datasets.  This distribution was
found to be Gaussian to high accuracy. The Gaussianity is quantified
via calculation of the skewness
\[
s_3 = \frac{\langle (\overline{\ds}-\langle\overline{\ds}\rangle)^3\rangle}{(\mbox{Var}(\overline{\ds}))^{1.5}}
\]
and kurtosis excess
\[
s_4 = \frac{\langle (\overline{\ds}-\langle\overline{\ds}\rangle)^4\rangle}{(\mbox{Var}(\overline{\ds}))^2} - 3
\]
both of which should be zero for a Gaussian distribution within the
noise.  For $M$ bootstrap subregions, and $N$ resampled datasets, the
expected variance of the skewness is $\mbox{Var}(s_3)=6(M^{-1}+N^{-1})\approx 6/M$
($M=250$ and $N=2500$ in our case) and of the kurtosis is
$\mbox{Var}(s_4)=24(M^{-1}+N^{-1})\approx 24/M$ \citep{cramer}.  The values of
$s_3$ and $s_4$ calculated for a number of different sources samples
and luminosity bins were consistent with being drawn from random
distributions with mean zero and approximately those variances, and
therefore consistent with Gaussianity.

\subsection{General systematic tests}\label{SS:sigresults}

In this section, we present results of tests for systematics described
in \S\ref{S:othersys}.

\subsubsection{Random points test}\label{SSS:randpointsresults}

As mentioned in \S\ref{SS:randpoints}, we found a nonzero signal
computed around random points in the survey region,
indicating the presence of a systematic shear $\gamma_{\rm sys}$ in
the source catalog that must be accounted for.  The shape and
amplitude of this signal varies with luminosity bin and source sample,
 because it is related to angular scale (and therefore
 different transverse separations for lens samples at different
 redshifts) and likely varies with source $R_2$ as well.

In principle, if the systematic shear is uniform across the survey,
then it will only be revealed on large scales.  If the distribution of
sources around lenses is circularly symmetric, then the contribution
of the systematic shear to the signal averages out, so only edge
effects on large scales will show the presence of systematic shear.
The effect of this spurious ellipticity is most noticable for the
narrow southern stripes, of approximate width 2.5 degrees; our
decision not to use them for the 
work in this paper reduced the observed random catalog signal by
approximately a factor of four.
However, it is unlikely that the systematic shear is perfectly
uniform, and if it fluctuates on small scales then we may also see a
random catalog signal at small transverse separations.  

\begin{figure}
\includegraphics[width=3in,angle=0]{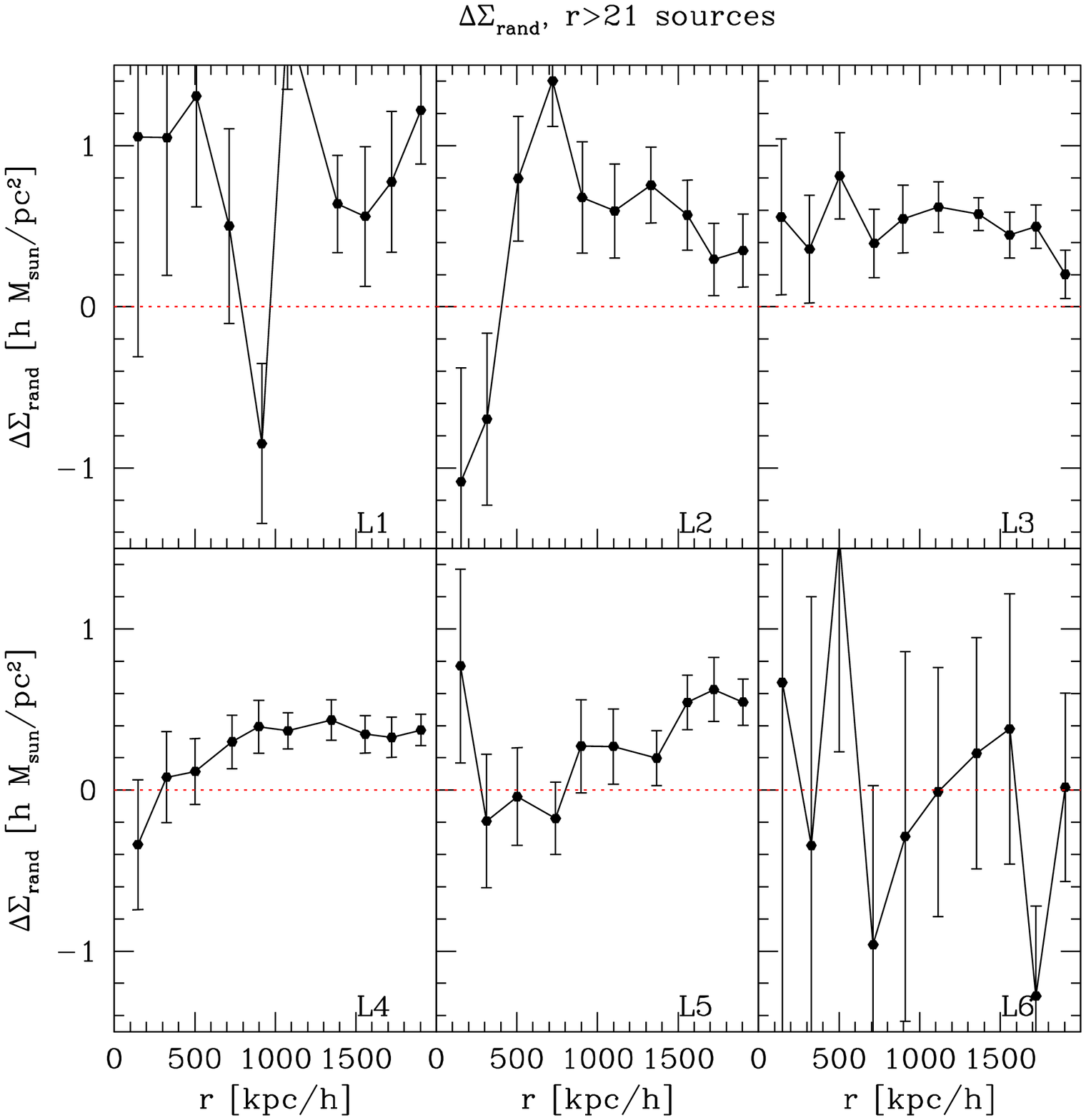}
\caption{\label{F:randsigfaint}
Average signal from 24 random catalogs, with 
$r>21$ sources, and the dashed line showing $\dsr=0$
for reference.}
\end{figure}

\begin{figure}
\includegraphics[width=3in,angle=0]{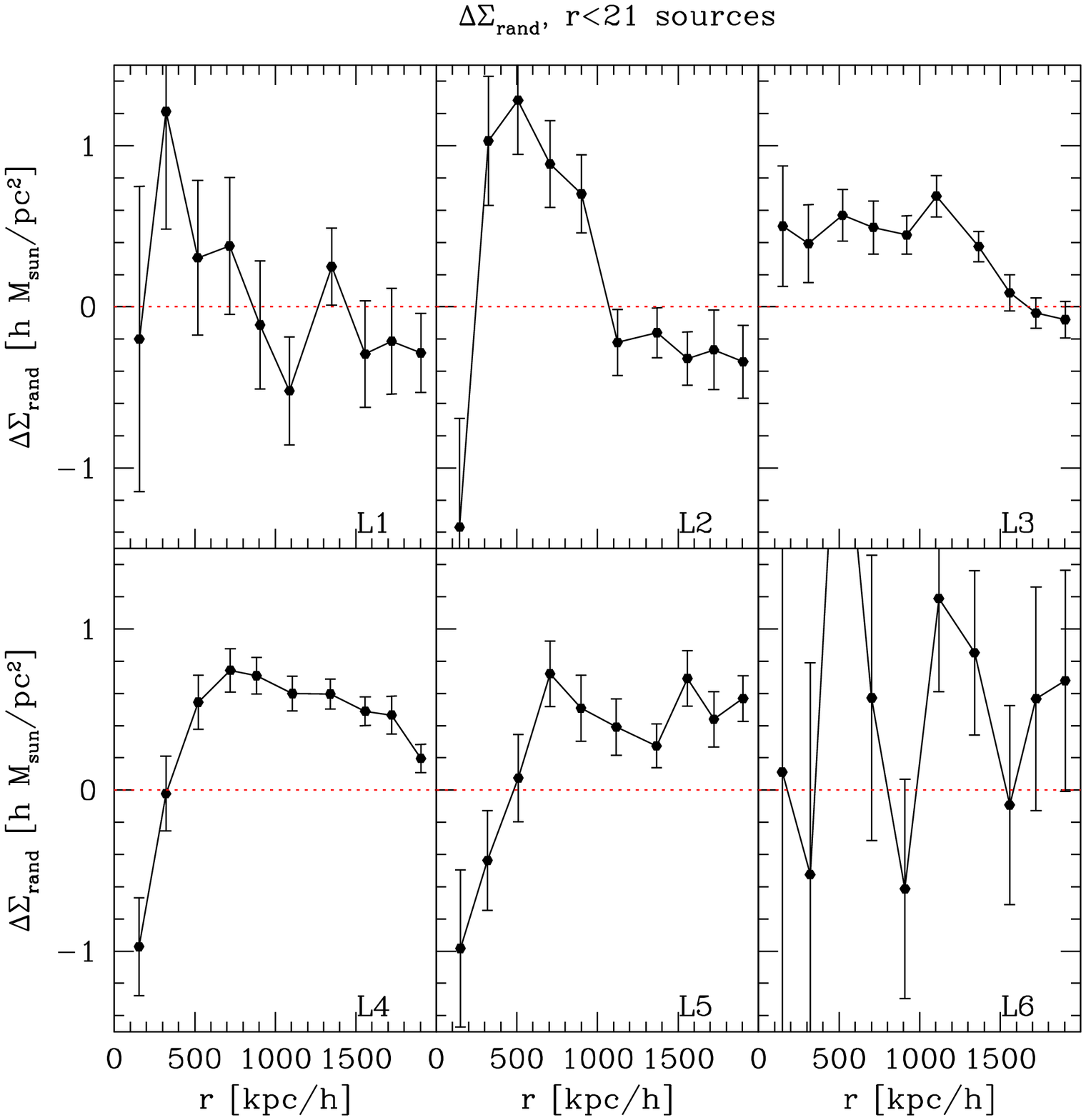}
\caption{\label{F:randsigbright}
Average signal from 24 random catalogs, with $r<21$ sources, and the
dashed line showing $\dsr=0$ for reference.}
\end{figure}

As an illustration,
figure~\ref{F:randsigfaint} shows the mean random catalog signal for 
all 6 luminosity bins, computed with $r>21$ sources. 
As shown in this figure, there is a clear signal from the random
catalogs at large radius; the statistical error due to this signal
can be taken into account by the bootstrap.  For the brighter sources,
$r<21$, the random catalog signal has slightly different shape, and is
shown in figure~\ref{F:randsigbright}.
As shown there, the signal is zero at low transverse separation, then
becomes positive in the fainter bins, 
decreasing to zero at larger radii.  While this occurs at different
radii for L1--L3, the corresponding angular scale is same: roughly
25'.  This number is of significance in the SDSS  as the separation
between adjacent camcols; however, the reason it is showing up here is
not entirely clear.  For the characteristic redshift of L4, 25'
corresponds to 2.2 \hinvm, which is why the signal does not go to zero
in that bin; however, it is clearly declining by the maximum radius of
2 \hinvm.   

If the systematic shear is constant across the survey, then one way to
eliminate or at least reduce it would be to only keep lenses that have
a reasonably circular distribution of sources around them, since the
systematic shear along the scan direction will then cancel out of the
calculation of \ds.   We implemented this cut, requiring the
ellipticity of the distribution of sources around the lens to be less
than 0.2, but found that this cut did not significantly decrease the
random catalog signal; consequently, we conclude that the systematic
shear fluctuates on a scale smaller than the typical distribution of
sources around a lens, and did not use this cut for the work in this paper.

\subsubsection{45-degree test}\label{SSS:45degreeresults}

In this subsection we describe results of the 45-degree test described
in \S\ref{SS:45degree}.    This test was performed using the
bootstrap with 2500 subsamples.  In order to check for a systematic
dependent on transverse separation, we did the test in 3 radial bins:
30--100 \hinvk, 100--600 \hinvk, and 600--2000 \hinvk, as well as for
the full range 30--2000 \hinvk.  The test was done for each lens
luminosity sample separately, and the results were also checked for
the combined result averaged over lens luminosity.  

Note that the
45-degree random catalog signal was nonzero for all lens-source
subsamples, and therefore had to be subtracted just like for the
non-rotated signal.  Unlike for the real signal, rather than averaging
$\overline{\ds} = \langle \ds (r/1h^{-1}{\rm Mpc})^{0.85}\rangle$, we average
$\overline{\ds}_{45} = \langle \ds_{45} \rangle$ since we do not expect
this quantity to vary widely over the radial range used.  We checked
the error distribution for $\overline{\ds}_{45}$ with the bootstrap and, since it was
found to be Gaussian, use a simple $\chi^2$ test to determine the
statistical significance of the 45-degree signal.  Since the test
involves using just a single variable ($\ds_{45}$ averaged over a given
radial range), we use the $\chi^2$ distribution with a single degree
of freedom to calculate $p$-values, the probability for
$\overline{\ds}_{45}$ to exceed the given value by chance assuming
that it really is consistent with zero. Problems that cause a pure
calibration bias alone have no effect on $p$-values since they change
both the amplitude of the signal and the errors.

\begin{figure}
\includegraphics[width=3in,angle=0]{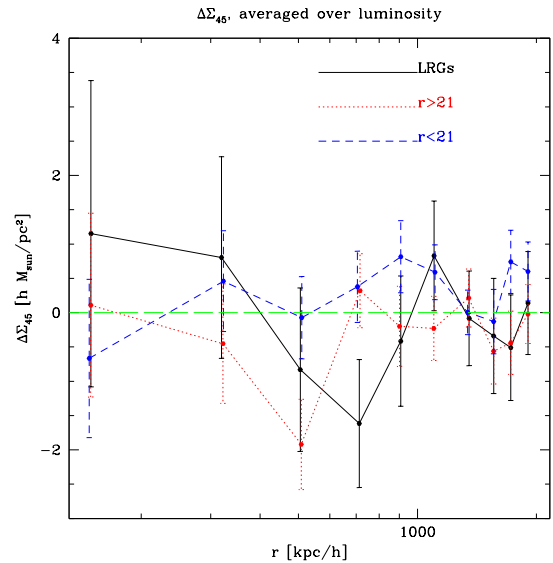}
\caption{\label{F:45degree}
45-degree rotated signal for each of the three source samples,
averaged over all luminosities.  Many radial bins are averaged to
reduce the noise and make the plot easier to understand.}
\end{figure}

Results for the 45-degree test for all source
samples and radial ranges are shown in table~\ref{T:45degree} averaged
over luminosity bins.  As
shown there, we have no evidence for a 45-degree signal, which
indicates that we are not dominated by systematics, since most
contribute to the 45-degree signal on the same level as the
non-rotated signal.  Results for each luminosity bin are not shown, in
order to simplify the presentation of results,
but we also have no evidence of 45-degree signal in individual
luminosity bins.  Fig.~\ref{F:45degree} shows the 45-degree signal
averaged over luminosity for each source sample.

\begin{table}
\caption{\label{T:45degree} 
$\overline{\ds}_{45}$, its standard deviation $\sigma_{45}$, and
  the probability to exceed the measured value by chance if it is truly
  consistent with zero, for all source samples and
  radial ranges.  The results are shown for the signal averaged over
  lens luminosity.  Because the $r>21$ sample includes some LRGs, the
  results shown for those two samples are not independent; the results
for all source samples combined take their covariance into account.}
\begin{tabular}{cccc}
\hline\hline
Radial range (\hinvk) & $\overline{\ds}_{45}$ $(h M_{\odot}/pc^2)$ & $\sigma_{45}$ &
$p$-value \\
\hline
\multicolumn{4}{c}{$r<21$ sources, without LRGs}\\
\hline
$ 30 < r <  100$ &  1.70 &   2.45 &  0.49 \\ 
$ 100 < r <  600$ &  -0.03 &   0.45 &  0.96 \\ 
$ 600 < r <  2000$ &   0.38 &   0.24 &  0.12 \\ 
$ 30 < r <  2000$ &   0.33 &   0.22 &  0.13 \\ 
\hline
\multicolumn{4}{c}{$r>21$ sources}\\
\hline
$ 30 < r <  100$ &  -1.72 &   3.04 &  0.57 \\ 
$ 100 < r <  600$ &  -1.17 &   0.55 &  0.04 \\ 
$ 600 < r <  2000$ &   -0.13 &   0.25 &  0.62 \\ 
$ 30 < r <  2000$ &   -0.25 &   0.23 &  0.26 \\ 
\hline
\multicolumn{4}{c}{LRGs}\\
\hline
$ 30 < r <  100$ &  6.37 &   5.31 &  0.23 \\ 
$ 100 < r <  600$ &  -0.17 &   0.91 &  0.85 \\ 
$ 600 < r <  2000$ &   -0.21 &   0.45 &  0.64 \\ 
$ 30 < r <  2000$ &   -0.19 &   0.42 &  0.66 \\ 
\hline
\multicolumn{4}{c}{All}\\
\hline
$ 30 < r <  100$ &  1.44 &   1.90 &  0.45 \\ 
$ 100 < r <  600$ &  -0.47 &   0.34 &  0.17 \\ 
$ 600 < r <  2000$ &   0.09 &   0.18 &  0.63 \\ 
$ 30 < r <  2000$ &   0.02 &   0.17 &  0.88 \\ 
\hline\hline
\end{tabular}
\end{table}
\subsubsection{Star-galaxy separation}\label{SSS:sgsepresults}

As described in \S\ref{SS:sgsep}, we have good reason to
believe that stellar contamination in the source catalog, even at 
faint magnitudes, is minimal ($\sim 2$ per cent for $r>21$).  However, we tested this hypothesis by
computing signal for lenses at low galactic latitude, defined as
$\sin{b}<0.7$, and compared the results against the signal at high
galactic latitude.  In principle, stellar contamination can cause the
signal at low galactic latitudes to be lower than the signal at high
galactic latitudes.  For this test, in order to maximize
signal-to-noise, the signal was computed using lenses in L3--L6
combined, since most of the signal comes from those bins.  This 
process was done for both the $r<21$ and $r>21$ sources separately, where we
expect most stellar contamination to be in the $r>21$ sample.  

The
results of the test are shown in table~\ref{T:systests}.  As
shown, while the signal in the low galactic latitude sample is
lower
than that in the high galactic latitude sample, the signals are
statistically consistent with each other for $r<21$ and $r>21$ cases.  This result, combined with
the results from GOODS shown in \S\ref{SS:sgsep}, supports
our assertion that stellar contamination is negligible in our
catalog.  For $2\sigma$ bounds on the error in the signal due to
stellar contamination, we use the results from GOODS with the binomial
distribution as in \S\ref{SS:sgsep}.
%we allow the 2 per cent contamination found
%with GOODS for $r>21$ to vary according to the binomial distribution, so the bound is
%$[-0.066, 0]$ for $r>21$, $[-0.033, 0]$ for LRGs (for which about half
%are at $r>21$), and negligible for $r<21$.
\begin{table*}
\caption{\label{T:systests}
Results of the systematics test for various subsamples, including
non-Gaussian 68 per cent and 95 per cent confidence intervals.  In all
cases shown 
here the lenses are from luminosity bins L3--L6. } 
\begin{tabular}{ccccccc}
\hline\hline
Sources $\alpha$ & Sources $\beta$ & $R_{\alpha,\beta}$ & \multicolumn{2}{c}{68\% CL} &
\multicolumn{2}{c}{95\% CL} \\
 & & & lower & upper & lower & upper \\
\hline
\multicolumn{7}{c}{Stellar contamination test, \S\ref{SSS:sgsepresults}} \\
$r<21$ without LRGs, $\sin{b}<0.7$ & $r>21$ without LRGs,
$\sin{b}>0.7$ & 
0.86 & 0.78 & 0.94 & 0.70 & 1.03 \\
$r>21$, $\sin{b}<0.7$ & $r>21$, $\sin{b}>0.7$ & 
0.92 & 0.84 & 1.01 & 0.76 & 1.11 \\
\hline
\multicolumn{7}{c}{Seeing comparison, \S\ref{SSS:seeingresults}} \\
$r<21$ without LRGs, bad seeing & $r<21$ without LRGs, good seeing &
0.99 & 0.92 & 1.08 & 0.85 & 1.17 \\
$r>21$, bad seeing & $r>21$, good seeing & 
1.00 & 0.92 & 1.09 & 0.84 & 1.19 \\
\hline
\multicolumn{7}{c}{$R_2$ comparison, \S\ref{SSS:r2results}} \\
$r<21$ without LRGs, $R_2<0.55$, $r$-band & $r<21$ without LRGs,
$R_2>0.55$, $r$-band & 
1.11 & 1.02 & 1.20 & 0.94 & 1.30 \\ 
$r<21$ without LRGs, $R_2<0.55$, $i$-band & $r<21$ without LRGs,
$R_2>0.55$, $i$-band & 
1.12 & 1.04 & 1.21 & 0.96 & 1.30 \\ 
$r>21$, $R_2<0.55$, $r$-band & $r>21$, $R_2>0.55$, $r$-band & 
0.89 & 0.80 & 0.99 & 0.71 & 1.10 \\
$r>21$, $R_2<0.55$, $i$-band & $r>21$, $R_2>0.55$, $i$-band & 
1.24 & 1.13 & 1.37 & 1.03 & 1.50 \\ 
\hline
\multicolumn{7}{c}{$r$ versus $i$ band comparison,
  \S\ref{SSS:bandsresults}} \\ 
$r<21$ without LRGs, $i$-band & $r<21$ without LRGs, $r$-band &
1.06 & 1.00 & 1.12 & 0.95 & 1.18 \\
$r>21$, $i$-band & $r>21$, $r$-band & 
1.04 & 1.00 & 1.07 & 0.97 & 1.11 \\
\hline
\end{tabular}
\end{table*}

\subsubsection{Seeing dependence of
  calibration}\label{SSS:seeingresults}

As described in \S\ref{SS:seeing}, we also used the systematics
test to compare the signal computed with
sources that had PSF sizes greater than and less than the median.  The
results of the test using L3--L6 lenses are shown in table~\ref{T:systests}.  As
shown there, the signals are statistically consistent with each other,
indicating that we do not have to worry about seeing-dependent
calibration of the shear.

\subsubsection{$R_2$ dependence of calibration}\label{SSS:r2results}

As described in \S\ref{SS:r2}, we used the systematics test
to compare the signal computed with
sources with $R_2<0.55$ versus with
$R_2>0.55$.  This exercise was done in both the $r$ and $i$ bands
separately, and results are shown in table~\ref{T:systests}.  As shown
there, there is no evidence at the $2\sigma$ level for a systematic calibration offset that
depends on resolution factor $R_2$.

This result is important because recent tests have indicated that the
source catalog used in H04 did have an $R_2$-dependent calibration,
with the calibration changing by as much as 50 per cent
from the lowest $R_2=1/3$ to the highest.  When averaging over all
$r<21$
sources, this effect caused the calibration for that paper to be too low by
approximately 15 per cent (only $r<21$ sources were used for that work;
since the LRG sample and the $r>21$ sample is, on average, at lower
$R_2$, the offset is even worse for those samples).  Tests to
determine the source of the problem were not definitive, but seemed to
rule out selection biases.  The basic conclusions of that paper, namely
the 99.9 per cent stat+sys confidence intervals on the intrinsic shear, are
essentially unaffected by this problem, since we assumed up to 20 per cent
shear calibration bias when computing the confidence intervals.
However, this finding highlights the importance of our results here,
which is that this catalog does not have a statistically significant
$R_2$-dependent calibration.

\subsubsection{Systematic differences between bands}\label{SSS:bandsresults}

While for most tests we used the shape measurements averaged over both
$r$ and $i$ bands, we also compared the signal computed
using only one band versus the signal computed using the other.  This
test was done separately for $r<21$ and $r>21$ sources.
As shown in table~\ref{T:systests}, there is no
sign of a systematic discrepancy between the signal computed using the
shape measurements in either band individually, so to
improve statistics we will henceforth use the shape
measurement averaged over bands.

\subsubsection{Boosts}\label{SSS:boostsresults}

The boost factors $B_{i,\alpha}(r)$ generally follow expected
trends, decreasing with radius and increasing with luminosity.
Furthermore, they are largest with the $r<21$ sources, which have the
largest overlap in redshift
range with the lens sample; less with the $r>21$
sources; and the smallest with the LRG sample, since those are chosen
specifically to avoid contamination from low-redshift ($z<0.35$)
galaxies.

A systematic effect at the smallest radii ($r<30$ \hinvk) is revealed
by the boost factors.  While they should, in principle, be
monotonically decreasing, we found that for the brightest lens
samples, they actually increase from 20--30 \hinvk, then follow the
expected trend of decreasing for $r>20$ \hinvk.  In H04, this trend
appeared for even larger radii; 
however, for that work, our source catalog did not include deblended
children, so this result is expected.  This source catalog does
include deblended children, so we do not expect a loss in lens-source
pairs for small transverse separation due to the loss of deblended
children.  However, very bright galaxies, for which this
effect is noticable, are large enough that at those
separations, their light could make it impossible to detect much
smaller, faint galaxies.  This could explain the fact that this effect
is more noticable for the fainter $r>21$ sources than it is for the
$r<21$ sources.  In order to avoid strange selection effects due to
this problem, this work only uses $r>30$ \hinvk{} for the analysis.

\begin{figure}
\includegraphics[width=3in,angle=0]{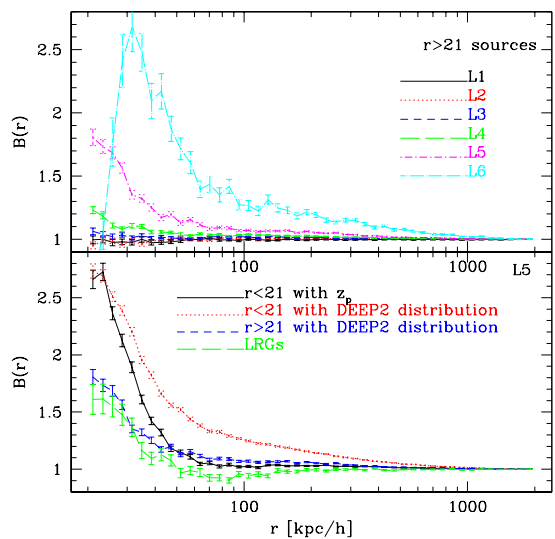}
\caption{\label{F:boosts}
The top plot shows $B(r)$ for $r>21$ sources and all luminosity bins.
The bottom shows $B(r)$ for L5 with all source samples as labeled on
the plot.} 
\end{figure}

Plots of the boost factor are shown in Fig.~\ref{F:boosts}.  The top
plot shows the trends of the boost with luminosity bin for the $r>21$
sources; the bottom shows the variation of boost with source sample
for L5 lenses.  The problem with $B(r)$ at $r<30$ \hinvk{} is clearly
visible in this plot, especially for L6.

\begin{figure}
\includegraphics[width=3in,angle=0]{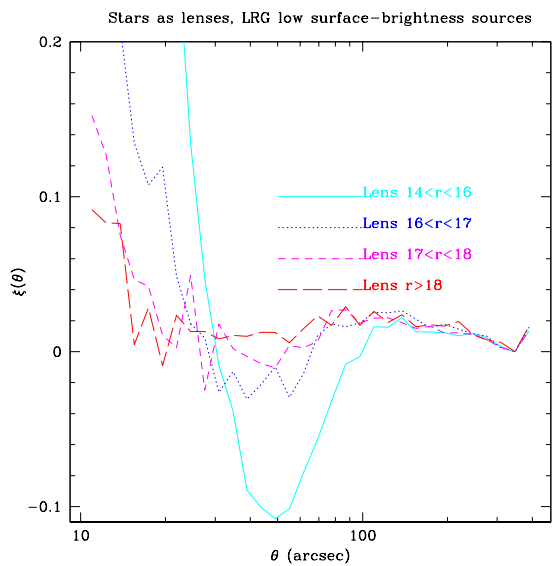}
\caption{\label{F:negcf}
Correlation function between bright stars and the high-redshift LRG
sample, for low surface-brightness LRGs only, as a function of star
apparent magnitude.}
\end{figure}

A close examination of the boost factors reveals a systematic effect
at small transverse separations that affects any analysis of
galaxy-galaxy weak lensing done using 
SDSS data.  From plots of the boost factor at small transverse
separation computed using LRGs as sources, it became apparent that there is a
deficit in lens-source pairs below a certain radius.  This deficit
manifests itself in the boost factor actually dipping below 1 (i.e.,
$\xi_{ls}<0$), which is not physical.  This deficit also is present for the $r<21$ and
$r>21$ samples, though it is harder to notice because of the fact that
there are physically-associated pairs at the affected radii.  We
determined that this effect is due to either an instrumental effect or
something in the {\sc Photo} pipeline (rather than due to some
selection effect in our catalog, error in signal computation, or
actual physics) due to the following factors:
(1) it occurs at a particular angular (not physical) scale, most
noticably around 
50'' though it extends up to about 90'', (2) the effect is noticable
with two different 
catalogs, the one from this work and from H04, which were processed
with completely independent software pipelines from the {\sc Photo}
outputs, (3) it is most noticable for lenses that are bright in
apparent magnitude, (4) it is most noticable for low
surface-brightness sources, and (5) we see the effect even if we
compute the signal around bright stars instead of around galaxies.
While a hint of this effect can be seen for $B(r)$ from LRGs in
Fig.~\ref{F:boosts}, it is more obvious in Fig.~\ref{F:negcf}, which
shows a plot of the correlation function between 
bright ($r<19$) stars and high-redshift LRGs, where there is a
non-zero correlation function at small separations because of
low-redshift quasar contamination in the star sample.  This plot was
computed using only low surface-brightness sources (those with a value
below the median), and is shown as a function of star magnitude.  

The effect turns out to be due to a problem in determination of the
sky level around bright objects (stars or galaxies).  The sky level is
determined in overlapping boxes of size 256 pixels (approximately 100'') and
linearly interpolated on the 128 pixel (50'' scale), using
a $2.3\sigma$ clipped median\footnote{For more information about the
  sky level determination, see {\slshape http://www.sdss.org/dr3/algorithms/sky.html}.}.  We expect the sky
level near bright lenses (within about 1') to be higher than the
global sky because light from the lens galaxies extends out to a large
angular separation, and we do not want that extra light to be included
in the flux from the sources in that region.  So, we do want the sky
level to be influenced by the presence of the bright objects.
However, our results suggest that the sky level 
determination is overly influenced by the presence of bright objects,
 with the sky level being set too high near them.  The problem is
different for each band.  This
change in the sky level has several effects: (1) some sources are not
detected at all by the {\sc Photo} software, (2) some are
detected, but due to the faulty sky level, the magnitudes, colors, and shape
measurements determined by model fits to the light profile are off,
and therefore they are not included in our 
source catalog because of magnitude or $R_2$ cuts or OBJC\_TYPE
failure, or are included but 
with these quantities computed wrong, and (3) our
computed redshift distribution of these 
sources is likely wrong because of the incorrect fluxes.  Furthermore,
because this effect can vary from band to band, the colors are
incorrect as well, affecting both photometric redshift
determination and selection of the high-redshift LRG sample.

\begin{figure}
\includegraphics[width=3in,angle=0]{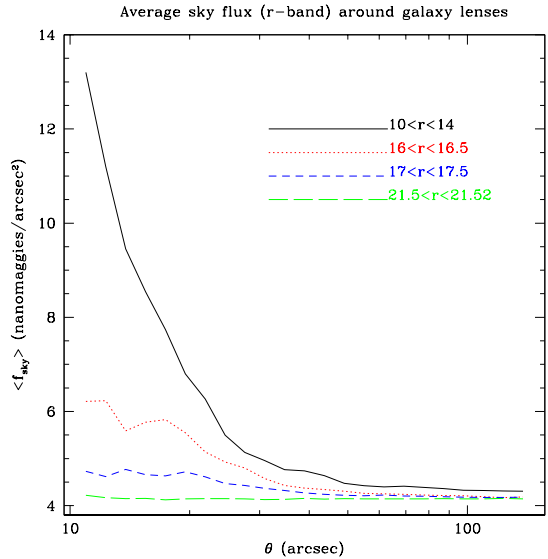}
\caption{\label{F:skyfluxr}
Average sky flux in nanomaggies per square arcsecond as a function of
angular separation from bright galaxies, computed in the $r$ band.
The lowest line, the sky flux around very faint galaxies, is shown as
a reference, and is flat, as expected.} 
\end{figure}

To test that this effect is indeed the cause of the observed effect,
we used the {\sc Photo} reductions to compute the averaged sky level
as a function of angular separation from bright galaxies (we except
the effect to differ around stars versus galaxies, because galaxies
are extended and therefore affect more pixels).  This computation was
done in each band, and we verified that the sky flux level $f_{sky}$
around bright galaxies is indeed 
higher than the baseline value computed around faint galaxies, by a
quantity we call $\Delta_{sky}$ in 
each band. For such small angular separations, we expect that
$f_{sky}$ should be higher than that baseline, so much of the
rise at small angular separations is not spurious.
Fig.~\ref{F:skyfluxr} shows a plot of the average sky flux 
in the $r$ band as a function of angular separation from lens galaxies
with apparent magnitudes as labeled on the plot.  
%%As shown, even if the spurious
%%$\Delta_{sky}$ is only 25 per cent of the measured one, it could still
%%have a major effect on
%%galaxies at faint magnitudes.  
We can make a naive calculation of the
effect of this problem (assuming that all of $\Delta_{sky}$ is
spurious and that the only effect was to change the surface brightness
of the faint source without changing its size) at 40'' for a circular
source galaxy 4'' in 
diameter at magnitude 21.5. 
The flux for such an object is 2.5 nanomaggies, and since its area is
12.5 square arcsec, then if the lens is at $r\sim 16$, with
$\Delta_{sky}=0.4$ nanomaggies per square arcsecond, its true flux
should have been 5.0 nanomaggies, a full factor of two higher, giving it a true magnitude of 20.8,
a highly significant offset from the observed value.  For lenses
around $r\sim 17.5$, a large fraction of the sample, $\Delta_{sky}$ is
smaller, but we still expect a ``true'' magnitude of 21.0.  Of course,
this is just a crude estimate, and a more careful calculation of what
happens to the fits to the light profiles when the sky level is
changed is necessary.  This calculation is actually quite complicated
because the change in sky flux changes the effective radius of the
source from the fits, so not only is the flux underestimated within
the area of the galaxy, light is also lost due to the underestimated
size.  However, when we tried applying this naive correction, we found
that we overcompensated for this effect, so it seems clear that some
of $\Delta_{sky}$ is not spurious (as would be expected for such
large, bright lenses, for which some of the light does indeed extend
to these radii).

There are several implications of this effect for our work.  First,
while figure~\ref{F:negcf} shows what looks like a dip in the boost
factor below 90'' and most prominently at 
50'', with it then rising at lower separations, our findings suggest
that the boost factor is actually underestimated for all radii below
the 50'' level, though it is impossible to actually see this effect
due to the physically-associated pairs.  (Note, however, that if the
problem is due to the lens galaxies light profile falling off faster
than the linear interpolation scale of the sky level, then the effect
should be worst around the 50'' scale and, if anything, somewhat
better at smaller separations.)  It seems that we have lost
pairs either because the sources were not identified by {\sc Photo} at
all, because their fluxes and sizes were miscalculated so they failed
our magnitude, $R_2$, or OBJC\_TYPE cuts, or (for the high-redshift LRG sample)
because their colors were wrong.  Furthermore, for the pairs that were
included, we must be concerned about incorrect shape measurement and
redshift distributions.  In general, this error will tend to lower the
measured signal, since the underestimated boosts lower the signal, and
the effect of underestimating the flux is, on average, to make us
think the galaxies are fainter and more distant, overestimating \avgscinv{} and
underestimating \ds.  Since at 50'' we see a 5 per cent effect, and it
may be worse at smaller $\theta$, though likely not much worse since
the problem is due to the interpolation, we place a $2\sigma$ bound on the
error due to this 
effect of $[-0.15, 0]$, but we emphasize that more study is necessary
to quantify this effect more precisely.

We attempted to correct for this effect by excluding regions near
objects (stars or galaxies) with saturated pixels, or about 1.5 per
cent of the sample, and regions within
60'' of $r<16$ galaxies, or about 8 per cent of the
sample.\footnote{We thank Ryan Scranton for providing 
  the necessary masks and mask utilities to allow us to perform this
  test.}  However, with these restrictions, we still see the effect at
approximately the same level as before.  This fact indicates that most
of the problem is due to the $r>16$ galaxies that comprise the
majority of our lens sample rather than due solely to the very
brightest subsample of lenses.

In future work, we will attempt a more detailed modeling of this
  problem.  For most of our results, we do not attempt to correct the
  boost factors, 
  which are clearly underestimated, though we will show what happens
  to the results if we try a naive correction; since only small
  separations are involved, the change in the signal averaged over radius due to
  this effect is small.  Furthermore, while we have estimated rather
  large values
  for the magnification bias at small separations
for the $r<21$ and LRG samples, this
  sky level problem makes it effectively impossible to make any detection of
  magnification bias with the SDSS using MAIN spectroscopic sample lenses
  and working at angular separations of less than approximately 90''. 

The final test performed in this section was to compare the signal
computed for the bright and faint halves of the brightest bin (L6)
separately and then averaging them (i.e., taking the average value of
$B(r)[\dsr-\ds_{rand}(r)]$) versus the signal computed by averaging
over the whole luminosity bin, as we usually do, to get the average
values of $B(r)$ and $\dsr-\ds_{rand}(r)$.  The concern is that due to the
non-uniformity of the boost factor, with both the boost factor and
signal rapidly increasing in this bin, our averaging process may bias
the signal.  Our tests indicate that for the both $r<21$ and $r>21$
sources, the effect of this averaging is less than a 1 per cent effect
even for L6; consequently, it must be negligible for the other bins,
for which the signal and boosts are varying less rapidly.
 
\subsubsection{Intrinsic Alignments}\label{SSS:intrinsicresults}

When computing the expected contamination due to intrinsic alignments,
we must do the computation separately for each luminosity bin $i$ and
source sample $\alpha$ because the contributions may
vary with lens mass (and therefore luminosity) and 
with source sample (more distant source samples will have fewer
physically associated lens-source pairs).  Furthermore, the procedure
should be done as a function of radius, since limits on the intrinsic shear
$\Delta\gamma_{int}$ vary with radius. We split the
data into 3 radial ranges $j$ with limits $r_{min,j}$ and $r_{max,j}$.
This estimate assumes that the boost factors $B(r)$ are
entirely due to physically associated pairs, and ignores the
magnification bias estimates in table~\ref{T:magbias}.

The signal contamination due to intrinsic alignments can be computed
as follows.  First, we determine the fraction of lens-source pairs that are
physically associated.  For each lens sample $i$, source
sample $\alpha$, and radial bin $j$, we compute
\begin{equation}
f_c^{(i,\alpha)}(r_j) = \frac{\sum_{r_{min,j}}^{r_{max,j}}
  (B_{i,\alpha}(r)-1)A(r)n_{rand}^{(i,\alpha)}(r)}{\sum_{r_{min,j}}^{r_{max,j}}A(r)n_{rand}^{(i,\alpha)}(r)} 
\end{equation}
where $A(r)$ is the area of the annular bin centered at $r$, and the
value of $f_c$ is determined via the bootstrap.  

Then, for a given radial range, we can compute the estimated
contribution of intrinsic alignments to the lensing signal via
\begin{equation}
\ds_{int}^{(i,\alpha)}(r_j) = f_c^{(i,\alpha)}(r_j)\frac{\Delta\gamma_{int}^{(i,\alpha)}(r_j)}{\avgscinv_{i,\alpha}}
\end{equation}

In order to avoid
excessively low (or even non-physical negative) values of $f_c$, we
had to correct for the sky level effect that
suppresses $B(r)$ at small radii.  This correction was done as follows.
For the LRG sample, for which this effect is most noticable,
 we examined $B(r)$ for all lens samples.
We found the minimum value $B_{min}$ at $r=r_{min}$, and assumed that
the effect was actually the same at all $r<r_{min}$ (and could not be
observed because $B(r)>1$ for those radii).  So, for $r \le r_{min}$, we
multiplied the signal and $B(r)$ by $1/B_{min}$; for $r>r_{min}$, we multiplied
the signal and $B(r)$ by $1/B(r)$ if $B(r)<1$, and by $1$ otherwise.  This
correction is very rough, since it assumes that the sky level effect
is the same for all source samples, and that the sky level effect only
changes the boosts without having an effect on the shear or redshift
calibration.  However, our estimate of contamination from intrinsic
alignments in table~\ref{T:intrinsic} is still conservative, because
we ignore the fact that
magnification bias contributes to the boost factors at
small radii.  For the purpose of this calculation, the 99.9 per cent
stat+sys confidence intervals from H04 have been reduced by a factor
of 0.6 so they will be $2\sigma$ rather than $3.4\sigma$ bounds.

\begin{table*}
\caption{\label{T:intrinsic}
Estimated contamination from intrinsic alignments for each lens
luminosity and source sample at the $2\sigma$ level; the average
signal in that radial range, $\langle \ds\rangle$, is
shown for comparison.} 
\begin{tabular}{ccccccccccc}
\hline\hline
Lenses & $r$ & $\langle \ds\rangle$ & \multicolumn{2}{c}{$r<21$, $z_p$} &
\multicolumn{2}{c}{$r<21$, dist} & \multicolumn{2}{c}{$r>21$} &
\multicolumn{2}{c}{LRG} \\
 & \hinvk & $h M_{\odot}/pc^2$ & $f_c$ & $\ds_{int}$ & $f_c$ & $\ds_{int}$ & $f_c$ & $\ds_{int}$ &
$f_c$ & $\ds_{int}$ \\
\hline
L1 & $[30, 446]$ & $3$ & $0.003$ & \cl{-1}{12}{24} & $0.012$ &
\cl{-5}{44}{79} & $0.002$ & \cl{-0.6}{6}{11} & $0.002$ &
\cl{-0.7}{6}{13} \\ 
L1 & $[30, 100]$ & $-1$ & $0.016$ & \cl{-6}{24}{35} & $0.032$ & 
\cl{-12}{50}{71} & $0.015$ & \cl{-5}{22}{32} & $0.007$ &
\cl{-2}{9}{16} \\
L1 & $[100, 300]$ & $5$ & $0.004$ & \cl{-2}{13}{19} & $0.013$ &
\cl{-6}{44}{59} & $0.002$ & \cl{-1}{6}{9} & $0.002$ &
\cl{-1}{5}{8} \\
L1 & $[300, 446]$ & $2$ & $0.002$ & \cl{-0.6}{\infty}{\infty} &
$0.009$ & \cl{-3}{\infty}{\infty} & $0.00007$ &
\cl{-0.02}{\infty}{\infty} & $0.002$ & \cl{-0.5}{\infty}{\infty} \\
L2 & $[30, 493]$ & $4$ & $0.009$ & \cl{-0.1}{1.5}{1.9} & $0.021$ &
\cl{-0.2}{3.8}{4.9} & $0.009$ & \cl{-0.1}{1.5}{2.0} & $0.002$ &
\cl{0.0}{0.4}{0.5} \\
L2 & $[30, 100]$ & $14$ & $0.021$ & \cl{-5}{6}{9} & $0.051$ &
\cl{-13}{17}{23} & $0.026$ & \cl{-6}{8}{11} & $0.007$ & \cl{-1}{2}{2}
\\ 
L2 & $[100,300]$ & $1$ & $0.011$ & \cl{-1}{2}{4} & $0.026$ &
\cl{-3}{6}{9} & $0.010$ & \cl{-1}{2}{3} & $0.001$ &
\cl{-0.1}{0.3}{0.4} \\
L2 & $[300,493]$ & $5$ & $0.006$ & \cl{0.5}{2}{2} & $0.016$ &
\cl{1}{5}{5} & $0.007$ & \cl{0.6}{2}{2} & $0.002$ &
\cl{0.2}{0.7}{0.6} \\
L3 & $[30, 545]$ & $5$ & $0.004$ & \cl{0.0}{0.2}{0.3} & $0.029$ &
\cl{0.3}{1.6}{2.1} & $0.011$ & \cl{0.1}{0.5}{0.7} & $0.001$ &
\cl{0.01}{0.05}{0.05} \\
L3 & $[30, 100]$ & $19$ & $0.026$ & \cl{1.6}{4}{4} & $0.092$ &
\cl{6}{14}{14} & $0.044$ & \cl{2.6}{6}{6} & $0.005$ &
\cl{0.3}{0.6}{0.7} \\
L3 & $[100, 300]$ & $5$ & $0.006$ & \cl{0.0}{0.4}{0.5} & $0.038$ &
\cl{-0.3}{2.8}{3.6} & $0.014$ & \cl{-0.1}{0.9}{1.4} & $0.002$ &
\cl{-0.02}{0.14}{0.21} \\
L3 & $[300, 545]$ & $5$ & $0.003$ & \cl{0.05}{0.24}{0.26} & $0.023$ &
\cl{0.4}{1.5}{2.1} & $0.008$ & \cl{0.1}{0.5}{0.6} & $0.0003$ &
\cl{0.005}{0.021}{0.027} \\
L4 & $[30, 735]$ & $5$ & $0.007$ & \cl{0.1}{0.2}{0.2} & $0.035$ &
\cl{1}{1}{1} & $0.014$ & \cl{0.3}{0.4}{0.5} & $0.0006$ &
\cl{0.01}{0.02}{0.02} \\
L4 & $[30, 100]$ & $31$ & $0.041$ & \cl{2}{3}{4} & $0.181$ &
\cl{11}{15}{18} & $0.077$ & \cl{4}{5}{6} & $0.002$ &
\cl{0.10}{0.15}{0.18} \\
L4 & $[100, 300]$ & $10$ & $0.015$ & \cl{-0.1}{0.5}{0.8} & $0.067$ &
\cl{-0.6}{3.1}{4.6} & $0.028$ & \cl{-0.2}{0.8}{1.6} & $0.002$ &
\cl{-0.01}{0.04}{0.08} \\
L4 & $[300, 735]$ & $4$ & $0.005$ & \cl{0.1}{0.2}{0.2} & $0.026$ &
\cl{0.8}{1.2}{1.2} & $0.010$ & \cl{0.3}{0.4}{0.4} & $0.0005$ &
\cl{0.01}{0.02}{0.02} \\
L5 & $[30, 992]$ & $10$ & $0.017$ & \cl{0.3}{0.4}{0.5} & $0.051$ &
\cl{1.1}{1.5}{2.1} & $0.021$ & \cl{0.4}{0.5}{0.7} & $0.001$ &
\cl{0.02}{0.02}{0.03} \\
L5 & $[30, 100]$ & $66$ & $0.21$ & \cl{5}{12}{14} & $0.55$ &
\cl{16}{40}{46} & $0.23$ & \cl{5}{13}{15} & $0.070$ & \cl{1}{3}{3}\\
L5 & $[100, 300]$ & $27$ & $0.042$ & \cl{0.9}{1.4}{1.9} & $0.16$ &
\cl{4}{7}{9} & $0.067$ & \cl{1}{2}{3} & $0.0$ & $0.0$ \\
L5 & $[300, 992]$ & $8$ & $0.013$ & \cl{0.2}{0.4}{0.5} & $0.037$ &
\cl{0.8}{1.2}{1.7} & $0.015$ & \cl{0.2}{0.4}{0.5} & $0.0004$ &
\cl{0.006}{0.009}{0.013} \\
L6 & $[45, 992]$ & $26$ & $0.071$ & \cl{2}{3}{3} & $0.19$ &
\cl{8}{11}{11} & $0.081$ & \cl{2.5}{4}{4} & $0.007$ &
\cl{0.16}{0.2}{0.2} \\
L6 & $[45, 100]$ & $162$ & $0.87$ & \cl{-3}{86}{122} & $1.8$ &
\cl{-9}{246}{364} & $0.74$ & \cl{-3}{78}{110} & $0.26$ &
\cl{-1}{21}{30} \\
L6 & $[100, 300]$ & $70$ & $0.17$ & \cl{3}{8}{10} & $0.58$ &
\cl{16}{41}{51} & $0.23$ & \cl{5}{12}{15} & $0.021$ &
\cl{0.3}{0.9}{1.1} \\
L6 & $[300, 992]$ & $23$ & $0.056$ & \cl{2}{3}{2} & $0.14$ &
\cl{7}{10}{9} & $0.062$ & \cl{2}{4}{3} & $0.003$ & \cl{0.1}{0.1}{0.1}
\\
\hline
\end{tabular}
\end{table*}

For this calculation, we assumed that $\Delta\gamma_{int}$ for
the $r>21$ and LRG samples is the same as that for the $r<21$ sample.
While this assumption is not fully justifiable, and means that the
actual estimates of the contamination should not be taken too
seriously, the size of the confidence intervals are likely still
trustworthy.

Here we discuss the results in table~\ref{T:intrinsic}.  The entries
with $\pm\infty$ as bounds are those for which H04 could not
place any constraint on $\Delta\gamma_{int}$.
As shown, because of the poor constraints at low luminosities (due to
the fact that the contamination fractions $f_c$ are so low that there
were very few lens-source pairs with which to study the intrinsic
alignments), no strong constraints on the
contamination of signal due to intrinsic alignments could be placed in
H04.  However, the
fact that the average signal $\langle \ds\rangle$ for these ranges is
entirely reasonable suggests that the contamination by intrinsically
associated pairs is not causing a large problem.  For $r<21$, use of
$p(z)$ rather than 
photometric redshifts leads to very large contamination fractions
$f_c$ and therefore very loose constraints on $\ds_{int}$.  The fact
that the contraints on intrinsic alignments for $r<21$, $r>21$,
and LRG samples are quite different, generally tightest on LRGs and
loosest on $r>21$, yet the results agree quite well (as will be shown
in \S\ref{SSS:all}) suggests once again that we do not have a major
problem with contamination by intrinsic alignments.  
%For large radii,
%this contamination is considered to be negligible at the $2\sigma$
%level. 
For small radii, the $r<21$ (with $z_p$) and $r>21$ samples, we
assign $2\sigma$ 
uncertainties due to intrinsic alignments of $\pm 20$ per cent for
L3--L5 and $\pm 60$ per cent for L6; for LRGs, $\pm 5$ per cent for
L3--L5 and $\pm 15$ per cent for L6.  For large radii, the $r<21$
(with $z_p$) sample, we assign $2\sigma$ uncertainties of $\pm 5$ per
cent for L3--L5 and $\pm 15$ per cent for L6; for $r>21$, $\pm 15$ per
cent for L3--L5 and $\pm 25$ per cent for L6; and negligible for LRGs.
These constraints were assigned by comparing the $2\sigma$ constraints
on $\ds_{int}$ to the characteristic \ds{} in the table.

\subsubsection{Corrections for non-volume limited sample}

As described in \S\ref{SS:notvol}, we must consider the effects of differences in the
mean weighted luminosity of the subsample when computed with different
source samples or redshift distributions for the same sample.  Using a
simple model for $\ds(L)$, we can compute the expected change in the
signal from Eq.~\ref{E:deltalum}, and expected ratios 
\begin{equation}
R_{i,\alpha,\beta} = 1+\frac{\delta(\ds)}{\ds}.
\end{equation}  

We have no need for these corrections for L1 and L2, since for those
samples all source samples are essentially at infinite redshift, and
$\delta L$ is exceedingly small.  Using our results for L3--L6, we
find $\ds'(L)/\ds(L)=0.19$, 0.43, 0.48, and 0.33 respectively.

We must consider several types of corrections.  First, we consider
signal 
comparisons for  $r<21$ sources, with different methods of
determining \scinv.  The maximum variation in luminosity $\delta L$
found when comparing among the methods
is 0.0023, 0.0028, 0.0046, and  0.0027 for L3--L6
respectively, which gives expected ratios of 1.0004, 1.0012, 1.0022,
1.0008, far too small to be detected considering our
statistical errorbars; consequently, we do not apply any correction
when comparing between the various methods of computing \scinv{} for
the $r<21$ sample.  Next, we consider comparisons within the $r>21$
source sample, different methods of comparing \scinv.  The maximum
variation in luminosity $\delta L$ is 0.0017, 0.0033, 0.0041, and
0.0017, again giving expected ratios similar to those for $r<21$, far
too small to measure. For comparisons within the LRG sample, there was
no detectable difference in the weighted luminosity since the source
redshift distributions varied very little and are all essentially at
infinity.

The final test  is the comparison of the luminosities
between the different source samples ($r<21$, $r>21$, and LRGs) to
ensure that the ratio test should not have some correction due to
their different weighted luminosities.  In general, the weighted
luminosity of the $r<21$ sample was lowest, and the LRG
sample was highest.  We find maximum $\delta L$ of
0.0092, 0.016, 0.016, and 0.006 (comparing LRG to $r<21$), giving
expected ratios of 1.002, 1.007, 1.006, 1.002.  Given that these
ratios represent a difference in overall calibration of less than 1
per cent, much lower than our $2\sigma$ bounds on the calibration, we
find that we need not apply any correction for the change in mean weighted
luminosity when comparing the signal from different source samples.

\subsection{Redshift systematic tests}

In this section, we describe our application of the systematics test
to assess which methods of redshift distribution
determination are optimal.  First, for each group of sources ($r<21$
without LRGs, $r>21$, and LRGs), we compared the various methods of
determining the redshift distribution.  Next, we compared the signal
determined using each of these three samples, to make sure that our
results are consistent.  In each case, this test was done individually
for each luminosity bin, because we can expect that different lens
redshift samples will sample different parts of the source redshift
distribution, and therefore results may differ systematically across
the luminosity bins. However, we also use the results averaged over
luminosity bins for greater statistical power.  Plots of the signal will
be shown in \S\ref{SSS:all}. 

All errorbars shown in this section are statistical errorbars from the
computation of signal only; they do not include systematic error on
the shear or redshift distributions, or statistical error in the
determination of redshift distributions or photometric redshift error
distributions.

\subsubsection{Bright ($r<21$) sources}

For the $r<21$ sources, several methods of redshift distribution
determination are compared:
\renewcommand{\labelenumi}{\arabic{enumi}.}
\begin{enumerate}
\item The redshift distribution $p(z|r)$ from COMBO-17,
\item The average redshift distribution $p(z)$ from DEEP2,
\item Photometric redshifts from {\sc kphotoz} directly,
\item Photometric redshifts, corrected for the conditional bias
  $\langle\delta(z_p)\rangle$ only,
\item Photometric redshifts, corrected for the best-fit error
  distribution, and
\item Photometric redshifts, corrected for the actual (noisy) error distribution.
\end{enumerate}
The first two methods involve averaging over a redshift distribution
to determine \scinv{} without the use of photometric redshift
information to eliminate physically associated pairs.  The final four
methods use photometric redshifts to select sources with
$z_p>z_l+0.1$ to eliminate physically
associated source-source pairs.  The two methods involving
distributions use all $r<21$ galaxies, including those for which
photometric redshifts were not available and those passing the
high-redshift LRG cuts; the last four methods only involve non-LRGs,
since the photometric redshift error distributions for non-LRGs and
LRGs are quite different.  

In this section, we compare the results using 30--2000 \hinvk, without
subdividing this region, since
changes in the redshift distribution only affect the amplitude of
the signal, and not its slope.  Furthermore, we only use luminosity
bins L3--L6 since the majority of the signal is in those bins; adding
L1 and L2 actually lowers the $S/N$ slightly.  Since changes in
redshift distribution will tend to have a greater effect on the
 higher-redshift lenses, we are also interested in the results
as a function of luminosity bin.  Table~\ref{T:rLT21} shows the
results for the ratios $R_{i,\alpha,\beta}$ for the selected ranges of
radii, including 95 per cent confidence intervals computed via the
non-Gaussian error computation for correlated variables (correlations
between the signal computed using method 1 or 2 and methods 3--6 were
of order 0.75, and 
correlations among methods 3--6 were of order 0.98).  Because of the
high correlations, the ratios were determined to high
accuracy.  Results are shown for L3, L6 (so that comparison can
illustrate trends in lens luminosity), and the average signal in
bins L3--L6.
\begin{table}
\caption{\label{T:rLT21}
A comparison of the overall calibration of the signal when computed
using the 6 methods considered in this section for the $r<21$ sources,
for $30<r<2000$ \hinvk, all luminosity bins with significant signal.  The ratios $R_{i,\alpha,\beta}$
are given along with the 95 per cent confidence interval.} 
\begin{tabular}{lllll}
\hline\hline
Sources $\alpha$ & Sources $\beta$ & \multicolumn{3}{c}{$R_{i,\alpha,\beta}$} \\
 & & L3 & L6 & L3--L6 \\
\hline
DEEP2 & $z_p$                  & \cl{0.89}{0.16}{0.13} & \cl{1.06}{0.17}{0.15} & \cl{0.99}{0.07}{0.06} \\
DEEP2 & $z_p$, fit $p(\delta)\!\!$ & \cl{0.87}{0.15}{0.13} & \cl{0.97}{0.16}{0.13} & \cl{0.94}{0.06}{0.06} \\
DEEP2 & $z_p$, bias            & \cl{0.92}{0.16}{0.13} & \cl{1.07}{0.16}{0.15} & \cl{1.02}{0.07}{0.06} \\
DEEP2 & $z_p$, $p(\delta)$     & \cl{0.89}{0.16}{0.13} & \cl{1.01}{0.16}{0.14} & \cl{0.97}{0.07}{0.06} \\
DEEP2 & COMBO17$\!\!\!\!$               & \cl{1.01}{0.06}{0.05} & \cl{1.01}{0.09}{0.09} & \cl{1.00}{0.03}{0.02} \\
$z_p$ & $z_p$, fit $p(\delta)\!\!$ & \cl{0.97}{0.05}{0.04} & \cl{0.92}{0.05}{0.04} & \cl{0.95}{0.02}{0.02} \\
$z_p$ & $z_p$, bias            & \cl{1.03}{0.05}{0.05} & \cl{1.01}{0.05}{0.05} & \cl{1.03}{0.02}{0.02} \\
$z_p$ & $z_p$, $p(\delta)$     & \cl{1.00}{0.04}{0.05} & \cl{0.95}{0.05}{0.04} & \cl{0.98}{0.02}{0.02} \\
$z_p$ & COMBO17$\!\!\!\!$               & \cl{1.13}{0.20}{0.17} & \cl{0.95}{0.14}{0.12} & \cl{1.01}{0.07}{0.06} \\
$z_p$, fit $p(\delta)\!\!$ & $z_p$, bias & \cl{1.06}{0.05}{0.05} & \cl{1.10}{0.05}{0.05} & \cl{1.08}{0.02}{0.02} \\
$z_p$, fit $p(\delta)\!\!$ & $z_p$, $p(\delta)$ & \cl{1.02}{0.05}{0.04} & \cl{1.04}{0.04}{0.05} & \cl{1.03}{0.02}{0.01} \\
$z_p$, fit $p(\delta)\!\!$ & COMBO17$\!\!\!\!$ & \cl{1.16}{0.21}{0.17} & \cl{1.03}{0.16}{0.13} & \cl{1.06}{0.07}{0.06} \\
$z_p$, bias & $z_p$, $p(\delta)$ & \cl{0.97}{0.03}{0.04}$\!\!\!\!\!\!\!\!$ & \cl{0.95}{0.04}{0.05} & \cl{0.95}{0.02}{0.01} \\
$z_p$, bias & COMBO17$\!\!\!\!$        & \cl{1.10}{0.19}{0.17} & \cl{0.94}{0.14}{0.12} & \cl{0.98}{0.06}{0.06} \\
$z_p$, $p(\delta)$ & COMBO17$\!\!\!\!$ & \cl{1.14}{0.20}{0.18} & \cl{1.00}{0.15}{0.13} & \cl{1.03}{0.06}{0.06} \\
\hline 
\hline
\end{tabular}
\end{table}

We can see from table~\ref{T:rLT21} that at 95 per cent confidence
level (CL), the 
results from these different ways of computing \scinv{} all give
results that have overall calibration within 15 per cent of each other.
First, we compare the two redshift distributions shown in the 
table; this comparison is on line 5 of the table.  As shown, the
distributions from COMBO-17 and DEEP2 give results that have no
difference in overall 
calibration, and this calibration can only differ by 
$-2$ to $+3$  per cent at the 95 per cent CL when averaged over luminosity
bins.  Since Fig.~\ref{F:zdist.DEEP} shows how similar these
distributions are for $r<21$, this result is not too surprising.

Next, we compare the results for the different ways of using
photometric redshifts.  First, we can consider the difference between
using photometric redshifts directly and correcting for their mean
bias (but not the width of the error distribution), shown in line
7.  As demonstrated by the results for L3--L6, correcting for the mean bias
lowers the signal by about 3 per cent, and the results are
statistically inconsistent with the results without any correction for
photometric redshift error at the 95 per cent CL.  We expected the
correction for the mean bias to lower the signal because $\meand<0$
for low photometric 
redshift, so correcting for the mean bias will raise the assumed
source redshift and  \scinv{}, lowering \ds{} (while $\meand>0$ for
high photometric redshift, so correcting for it should raise the
signal, the effect at low photometric redshift is apparently more
important since \scinv{} varies more strongly with $z_p$ at low $z_p$ despite the weights proportional to \scinvtwo{} that make
higher redshift sources more important).  We next consider what
happens when we go from including the mean bias only, to including the
full error distribution (line 13).  As shown, including the width of
the error distribution then raises the signal by 5 per cent; the
results corrected for only the mean bias versus using the full error
distribution (bias and scatter) are statistically inconsistent at
approximately the $4\sigma$ level, 
which means that including the width of the error distribution is very
important.  This increase in signal when we include the scatter is
consistent with our expectation that scatter will lead to decreased
\scinv, at least for the values $z_p>z_l+0.1$ used here (at lower
$z_p$ it may actually raise \scinv).  The next comparison is using
photometric redshifts directly versus including the full error
distribution, line 8; consistent with combining the aforementioned
results from lines 7 and 13, we find that 
including the full error distribution increases the signal compared to
no correction at all by approximately 2 per cent, though the results are 
marginally consistent at the $2\sigma$ level.  This near agreement is
due to the 
effects from the mean bias and width of the distributions cancelling
out; the nearly exact cancellation was impossible to predict in
advance, which is why this systematics test
is so useful.

We can also use these results to place approximate $2\sigma$ bounds on
the calibration uncertainty due to statistical error in the
photometric redshift error distribution.  We place these bounds using
the comparison between no correction at all, correction for \meand,
and correction for the full error distribution, which gives
approximately $\pm 3$ per cent calibration uncertainty (95 per cent
confidence level).  This result
is comparable to the $2\sigma$ Poisson errorbars (determined via
bootstrap resampling of the photometric redshift error distribution)
which are $\pm 2.6$ per cent for lens redshift $\sim z_{\rm eff, L6}$,
lower for lower lens redshift; we use the $\pm 3$ per cent figure to
include extra uncertainty due to the effects of LSS.

Another informative test is the comparison between the results using
the best-fit error distribution versus the actual error distribution.
When plotted against each other, the best-fit distribution did show
some significant differences from the actual one; for example, it did
not account for the tails in the distribution very well.  This
comparison is on line 11 of table~\ref{T:rLT21}; as shown, use of the
best-fit error distribution rather than the actual one increases the
signal by about 3 per cent, and the two results are inconsistent at
the $3\sigma$ level.  The sign of the discrepancy is in accordance
with the results for \scinv{} in Fig.~\ref{F:comparescinv}.  While the
magnitude of this bias is clearly not 
large, since the difference is statistically significant, it seems
that it would be preferable to use the actual error distribution to
avoid this bias altogether.

Finally, we compare the results using redshift distributions versus
using photometric redshift error distributions, lines 4 (DEEP2) and 15
(COMBO-17).  We find that the signal is 3 per cent lower when we use
either redshift distribution than when we use the photometric
redshift error distributions; however, the two results only differ by
about $1\sigma$, so the difference in calibration is not statistically
significant. 
However, as mentioned previously and demonstrated in
Fig.~\ref{F:boosts}, there is significantly more contamination from
physically associated pairs when we use a distribution for $r<21$
rather than using photometric redshift information to remove that
contamination. Since there
is no clear discrepancy due to the use of the redshift distributions
versus photometric redshifts with error distributions, but the use of
redshift distributions can 
lead to higher systematic error due to intrinsic alignments, we will
henceforth use photometric redshifts with the actual error
distributions for the $r<21$ sample when comparing against the signal
from other source subsamples.  

Another systematics comparison that was completed was the computation
of signal with $z_p > \langle z_p\rangle$ versus with $z_p < \langle
z_p\rangle$ without correcting for photometric redshift errors.  If
our understanding of photometric redshift errors is 
correct, then we expect the signal at high $z_p$ to be lower than the
signal at low $z_p$ by a statistically significant amount.  Indeed, the ratio
$R_{{\rm all},\alpha,\beta}=0.79$; the $3\sigma$ confidence interval
is $[0.63, 1.00]$, so the discrepancy between the two samples is
$3\sigma$ and has the correct sign.  When we correct for photometric
redshift error, the discrepancy still exists at the $2\sigma$ level;
it is 0.86, with $2\sigma$ confidence interval $[0.74, 1.00]$.  The
fact that correcting for photometric redshift errors helps reduce the
discrepancy indicates that we are understanding these errors properly.
The fact that there is still a discrepancy at the $2\sigma$ level is
not too alarming, because as mentioned previously, the errors used
here are purely statistical.  Once we add the overall calibration
uncertainty, the discrepancy between the two samples is only $1\sigma$
and therefore not of concern.  

\subsubsection{Faint ($r\ge 21$) sources}\label{SSS:zfaint}

For the $r>21$ sample, two methods of redshift distribution
determination were considered:
\begin{enumerate}
\item The redshift distributions from COMBO-17, $p(z|r)$, and
\item The average redshift distribution $p(z)$  from DEEP2 for $21<r<21.8$.
\end{enumerate}
In both cases, distributions were used, so we had no way of
eliminating physically-associated lens-source pairs; however, since
the $r>21$ sample is, on average, at higher redshift than the lenses,
this procedure is less important than for the $r<21$ sample.  In both
cases, high-redshift LRGs are included in the sample.  As in the
previous subsection, we consider results from 30--2000 \hinvk.
Results for $R_{i,\alpha,\beta}$ are shown in table~\ref{T:rGE21};
since only one comparison is being made, we show results for all
luminosity bins.

\begin{table}
\caption{\label{T:rGE21}
A comparison of the overall calibration of the signal
$R_{i,\alpha,\beta}$, where $\alpha$ refers to the distribution from
DEEP2 and $\beta$ to the distribution from COMBO-17, for the
$r\ge 21$ sources, 
$30<r<2000$ \hinvk, all luminosity bins with significant signal.} 
\begin{tabular}{cc}
\hline\hline
Lens luminosity & $R_{i,\alpha,\beta}$ with 95 per cent CL bounds \\
\hline
L3 & \cl{1.02}{0.07}{0.06} \\
L4 & \cl{1.04}{0.04}{0.05} \\
L5 & \cl{1.05}{0.04}{0.03} \\
L6 & \cl{1.09}{0.05}{0.05} \\
L3--L6 & \cl{1.05}{0.03}{0.02} \\
\hline 
\hline
\end{tabular}
\end{table}

Not surprisingly, the minor discrepancy ($<2\sigma$) that exists for
L3--L4 becomes worse for the higher redshift
lenses, since they are closer to the peak of the distribution and are
therefore more sensitive to details of the distribution.  For L5, the
discrepancy is $3\sigma$ and for L6, $4\sigma$, with the result
averaged over luminosities a 5 per cent discrepancy, and the two
signals being different at the $4\sigma$ level.  Consequently, while
the difference in signals is relatively small, we are able to use the
systematics test to determine it to very high accuracy.  The fact that
the distribution from DEEP2 gives slightly higher results implies that
it gives on average slightly lower $\avgscinv$ and therefore is at
lower redshift on average.  This result is not entirely surprising,
since the DEEP2 distribution is specific to our lensing catalog,
but the COMBO-17 distribution was for photometric galaxies in general,
and our lensing cuts will tend to eliminate higher-redshift galaxies
more than lower redshift ones.  This understanding is in accordance
with Fig.~\ref{F:zdist.DEEP}, which shows both distributions for
$r>21$.  Consequently, for the
remainder of this paper, the DEEP2 redshift distribution will be used
for the $r>21$ sample.

\subsubsection{LRG sources}

For LRGs, the signal was computed two ways:
\begin{enumerate}
\item Using photometric redshifts directly, for $0.4<z_p<0.65$, and\label{zLRG}
\item Using redshift distributions $p(z|z_p)$ determined via the
  inversion procedure from~\cite{2004astro.ph..7594P} for
  $0.45<z_p<0.65$.\label{distLRG}
\end{enumerate}
Besides these two methods of comparing the redshift distribution, we
did several tests to ensure that our cuts on the LRG sample were
optimal.  Using method~\ref{zLRG} to determine redshifts, we recompute
the signal twice, once with the requirement that $d_{\perp}>0.5$ and
then with
$d_{\perp}>0.55$; these stricter cuts should help eliminate more of
the low-redshift contamination (this contamination is a concern
because we would be vastly overestimating its redshift and 
\scinv, and  underestimating \ds).  If the signal computed using these
cuts is similar to that computed with our cut, $d_{\perp}>0.45$, then
we can be assured that the role of low-redshift contamination is
minimal.  In addition, we also compute the signal using photometric
redshifts directly from $0.45<z_p<0.65$ to ensure consistency with the
full sample, in case there is contamination in the $0.4<z_p<0.45$.
Results for the ratio test are shown in table~\ref{T:LRGratio}.
\begin{table}
\caption{\label{T:LRGratio}
A comparison of the overall calibration of the signal when computed
using the methods considered in this section for the LRG sources,
for $30<r<2000$ \hinvk, all luminosity bins with significant signal.
The error bars are 95 per cent confidence limits.} 
\begin{tabular}{llc}
\hline\hline
Sources $\alpha$ & Sources $\beta$ & $R_{i,\alpha,\beta}$ \\
\hline
$z_p$, $d_{\perp}>0.5$  & $z_p$                  & \cl{1.02}{0.06}{0.05} \\
$z_p$, $d_{\perp}>0.55$ & $z_p$                  & \cl{1.00}{0.08}{0.07} \\
$z_p$                   & $0.45<z_p<0.65$        & \cl{0.98}{0.10}{0.07} \\
$z_p$                   & $p(z|z_p)$             & \cl{1.00}{0.10}{0.08} \\
\hline 
\hline
\end{tabular}
\end{table}

We can come to several conclusions from this table.  Lines 1--3 show
the results of the systematics tests: using photometric redshifts
directly for the full sample with $d_{\perp}>0.45$ and $0.4<z_p<0.65$, versus requiring
$d_{\perp}>0.5$ (line 1), $d_{\perp}>0.55$ (line 2), or
$0.45<z_p<0.65$ (line 3).  As shown, we see no sign of any systematic
discrepancy.  This result is useful because it suggests that our
photometric redshift and color cuts are sufficient to eliminate low-redshift
contamination to a negligible level, so there is no need to impose
stricter cuts such as those listed above, which also lower the
statistical power of this subsample.  Line 4 shows a comparison of the
results using photometric redshifts directly for $0.4<z_p<0.65$ versus
using the inversion method from \cite{2004astro.ph..7594P} for $0.45<z_p<0.65$.  As
shown, the results are completely statistically consistent. This
result is not surprising, since the LRG sample is at much
higher redshift than the lenses, and therefore small changes in the
redshift distribution have little  effect on the final results.  Because
of the larger numbers of LRGs in the $0.4<z_p<0.65$ sample, we will
henceforth use the full LRG sample with photometric redshifts directly
when comparing against other subsamples.  When considering the
variation in calibration between the samples in
table~\ref{T:LRGratio}, we conclude that the calibration uncertainty
in the LRG signal due to redshift distribution determination is $\pm
10$ per cent at the $2\sigma$ level.

\subsubsection{All}\label{SSS:all}

In this subsection we compare the results from the three source
samples ($r<21$, $r>21$, and LRGs) to check that they are consistent.
As mentioned in the previous sections, we will use the photometric
redshifts with the full error distributions from DEEP2 for $r<21$
(without LRGs), the redshift distribution from DEEP2 for the $r>21$
sample, and photometric redshifts directly for the LRG sample.

Furthermore, in contrast to the previous sections, our concern here is
not merely overall calibration; since the different source samples
have different boost factors and different random catalog signals, it
is important to check for radius-dependent effects.  In principle, we could
also measure the ratios $R_{i,\alpha,\beta}$ for several radial
ranges.  However, the $S/N$ for these comparisons was not high enough
to give reasonable errorbars.  Consequently, we only consider larger
ranges of 150--2000
\hinvk{} and 30--2000 \hinvk.  The reason for using the 150--2000
\hinvk{} range is that even for L6 it excludes the angular separations
$<90$'' for which the boost factor has the difficult to quantify
systematic due to sky subtraction, and it excludes the radii for which
magnification bias may be important.  Consequently, 150--2000 \hinvk{}
is a more
systematic-free ranges.  Results are shown in table~\ref{T:allratio}.
\begin{table}
\caption{\label{T:allratio}
A comparison of the overall calibration of the signal when computed
using the three main source samples, all luminosity bins with
significant signal.  The results for the two radial ranges are, of
course, highly correlated.  We show $R_{i,\alpha,\beta}$ and the
95 per cent CL.}   
\begin{tabular}{ccc}
\hline\hline
Lenses & 30--2000 \hinvk & 150-2000 \hinvk \\
\hline
\multicolumn{3}{c}{$r<21$ without LRGs versus $r>21$} \\
\hline
L3 & \cl{1.12}{0.48}{0.40} & \cl{1.09}{0.51}{0.32} \\
L4 & \cl{1.08}{0.30}{0.22} & \cl{1.07}{0.34}{0.24} \\
L5 & \cl{1.07}{0.19}{0.16} & \cl{1.03}{0.22}{0.17} \\
L6 & \cl{0.90}{0.22}{0.18} & \cl{0.85}{0.21}{0.18} \\
L3--L6 & \cl{1.03}{0.13}{0.11} & \cl{0.99}{0.14}{0.12} \\
\hline 
\multicolumn{3}{c}{$r<21$ without LRGs versus LRGs} \\
\hline
L3 & \cl{1.11}{0.93}{0.38} & \cl{1.03}{0.92}{0.38} \\
L4 & \cl{1.07}{0.44}{0.27} & \cl{1.10}{0.59}{0.33} \\
L5 & \cl{1.02}{0.25}{0.18} & \cl{1.01}{0.30}{0.21} \\
L6 & \cl{0.93}{0.27}{0.20} & \cl{0.90}{0.33}{0.23} \\
L3--L6 & \cl{1.01}{0.29}{0.20} & \cl{1.00}{0.19}{0.16} \\
\hline
\multicolumn{3}{c}{$r>21$ versus LRGs} \\
\hline
L3 & \cl{0.99}{0.75}{0.34} & \cl{0.94}{0.79}{0.34} \\
L4 & \cl{0.99}{0.36}{0.24} & \cl{1.02}{0.46}{0.26} \\
L5 & \cl{0.96}{0.19}{0.15} & \cl{0.98}{0.25}{0.18} \\
L6 & \cl{1.03}{0.27}{0.21} & \cl{1.06}{0.33}{0.23} \\
L3--L6 & \cl{0.98}{0.15}{0.11} & \cl{1.01}{0.18}{0.14} \\
\hline
\end{tabular}
\end{table}

We can draw several conclusions from this table.  First, the fact that
the ratios using 30--2000 and 150--2000 \hinvk{} are so similar
illustrates that even with our concern about the systematic in the
boosts at low values of transverse separation, use of the full radial
range does not cause obvious problems.  Second, there seems to be some
luminosity-dependence of the overall calibration difference between
the samples (especially $r<21$ versus $r>21$), but the errorbars on individual luminosity bins are large
enough that no definite statement about this can be made.  Finally, when
averaged over all luminosities, the results for the three source
subsamples agree well within errorbars.  We see no suggestion that
there is anything fundamentally wrong with any of these three samples,
either in the shear calibration, redshift distributions, or other
systematics, and consider the consistency of the samples to be
established at the 10\% level ($1\sigma$).

\begin{figure}
\includegraphics[width=3in,angle=0]{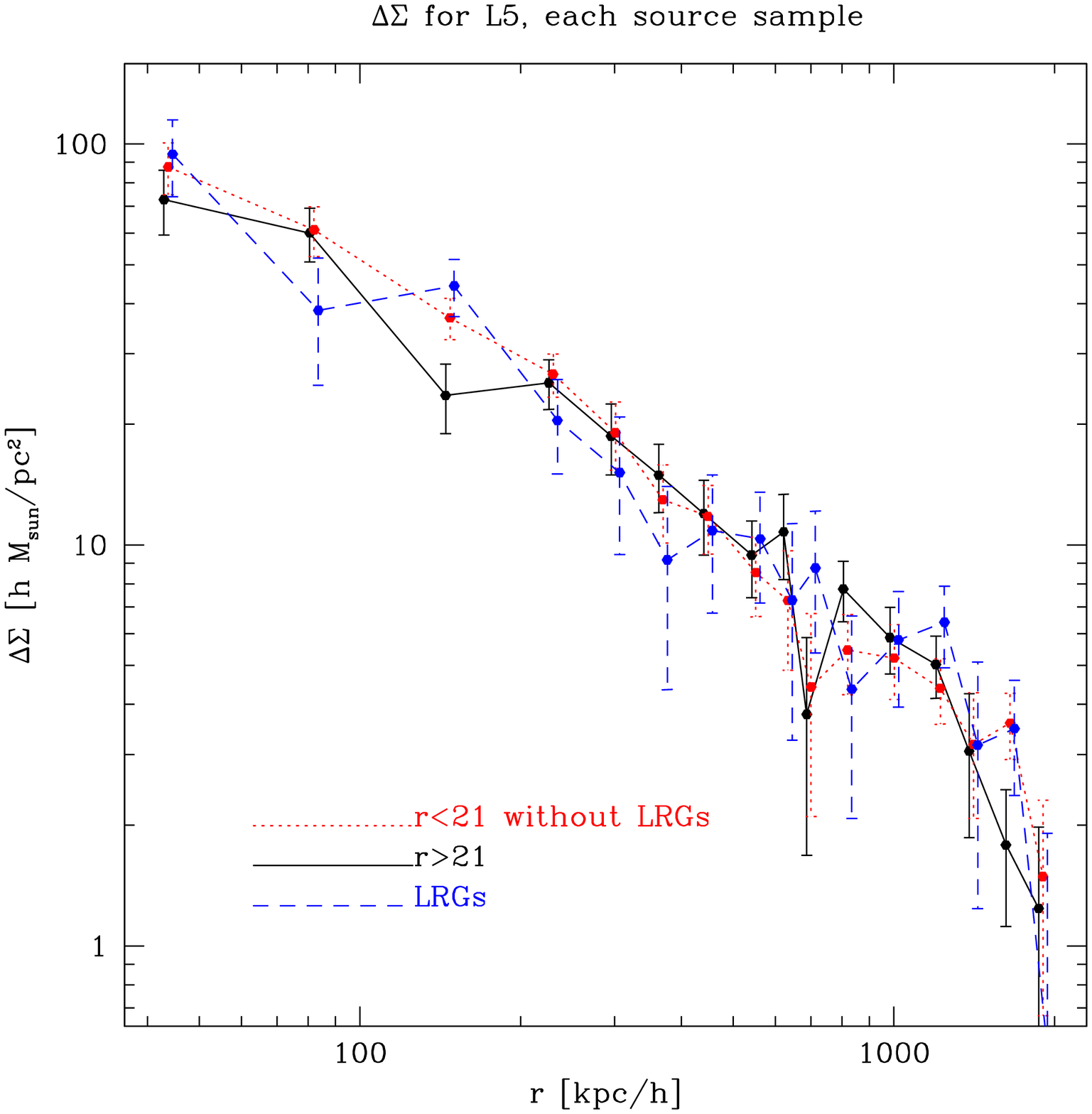}
\caption{\label{F:dsL5}
\dsr{} in L5 for each source sample as labelled on the plot.  Points
have a slight horizontal offset so that errorbars on all three signals
are visible.
}
\end{figure}

\begin{figure*}
\includegraphics[width=6in,angle=0]{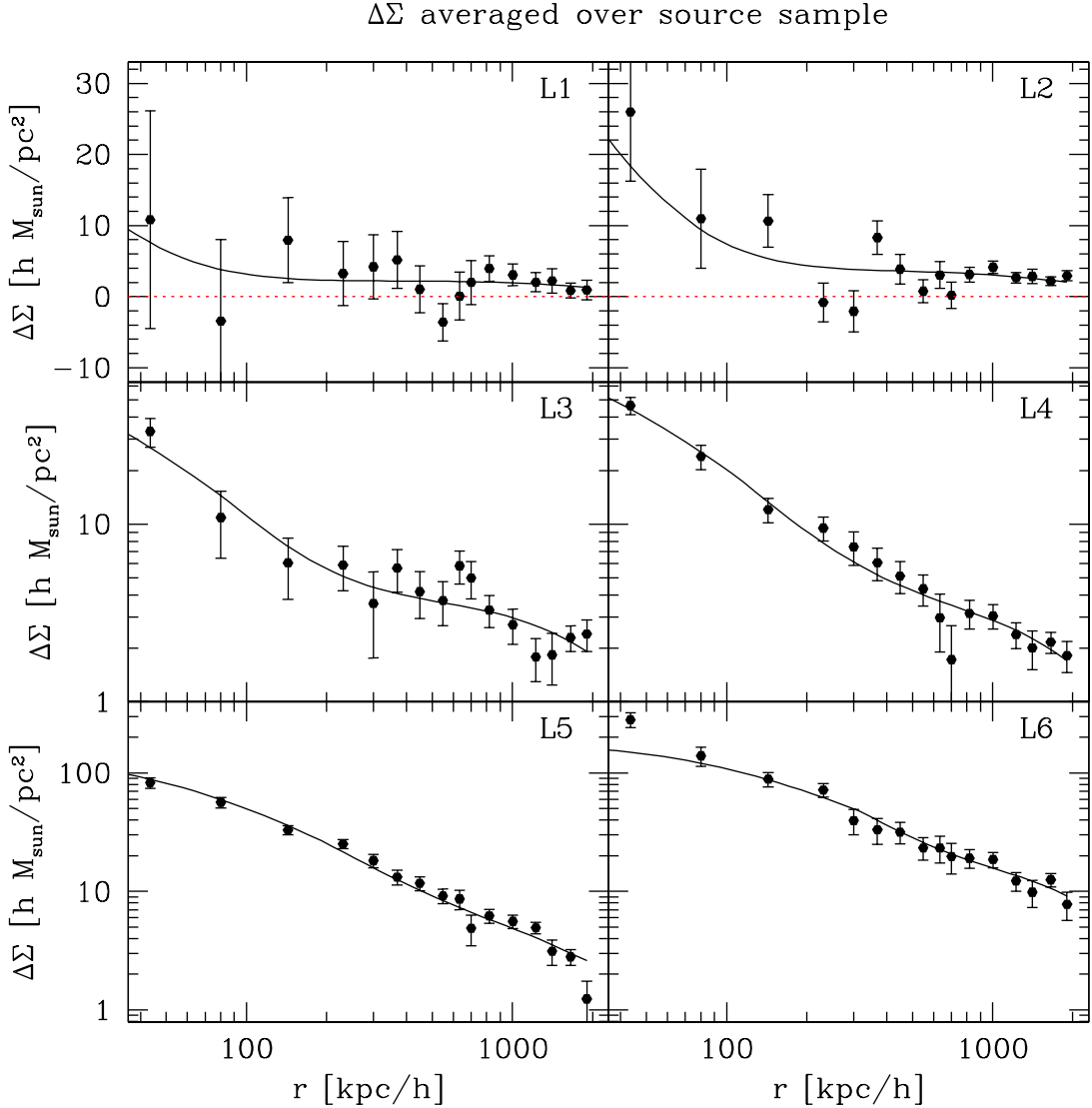}
\caption{\label{F:ds6lum}
\dsr{} averaged over source subsample as a function of lens
luminosity.  For L1 and L2, which have signal statistically consistent
with zero, the vertical scale is not logarithmic, and the zero level
is shown as a dashed line; for the other luminosity bins, a
logarithmic scale is used for \ds.  The errorbars shown are
statistical only; the lines are 
the results of halo model fits as described in the text.}
\end{figure*}

Here we show plots of the signal.  Fig.~\ref{F:dsL5} shows the signal
from L5 for the three samples to illustrate our finding from the
systematics test that there is no systematic calibration offset
between the three samples.
Fig.~\ref{F:ds6lum} shows the signal 
averaged over source samples for all 6 luminosity bins, with
statistical errorbars only, and the results of halo model fits to the
signal as described in \cite{2005PhRvD..71d3511S} and \cite{2004astro.ph.10711M}.  
In order to allow the \ds{} values shown on the
plots to be converted to shear $\gamma_t$, table~\ref{T:avgscinv} includes
\avgscinv{} for each lens and source sample combination. 
\begin{table}
\caption{\label{T:avgscinv}
For reference, \avgscinv\ values for all luminosity bins and source samples.}
\begin{tabular}{cccc}
\hline\hline
Lenses & \multicolumn{3}{c}{$10^5 \avgscinv$, $pc^2/hM_{\odot}$} \\
 & $r<21$ & $r>21$ & LRG \\
\hline
L1 &  5.2 & 5.3 & 5.5 \\
L2 &  7.2 & 7.4 & 7.8 \\
L3 & 10.2 & 10.5 & 11.4 \\
L4 & 13.3 & 13.5 & 15.4 \\
L5 & 16.1 & 15.9 & 19.0 \\
L6 & 18.1 & 17.1 & 22.0 \\
\hline
\end{tabular}
\end{table}

\section{Conclusions}\label{S:concl}

Using the systematics test, we have estimated the effects
of redshift distribution errors and other systematics on the lensing
signal.  As a summary, table~\ref{T:summary} lists those systematics that are
significant at approximately the 1 per cent level or higher.  
Systematics are only estimated in this table for the redshift
determination methods for each sample that we have selected to use in
this and future works (e.g., we do not show results for the use of
redshift distributions for the $r<21$ sample). We must
emphasize that the estimates of the calibration uncertainty due to these effects is
generally only relevant for this work, and their effects for other
works must be assessed using the lens/source catalogs used for those
works.  For those systematics that have some nontrivial dependence on
lens  sample and/or radial range, rather than listing a value in this
table, we refer the reader to the appropriate table/section.  We have
not included the following systematic effects because they are
expected to be at the 0.1 per cent level: atmospheric refraction,
camera shear, the
density-systematic shear correlation $\langle  \delta n \cdot
\gamma_{sys}\rangle$, and non-uniformity of the boost factor.

\begin{table*}
\caption{\label{T:summary} 
Summary of the main systematics investigated (and found to be
significant at the 1 per cent level or greater) in this work, and
estimates of their significance. Bounds on the error are given at the
$2\sigma$ level for each effect and for all systematics combined.}
\begin{tabular}{lcccc}
\hline\hline
Systematic & Lenses & Sources & Radii & $2\sigma$ bounds \\
\hline
\multicolumn{5}{c}{Redshift distribution calibration} \\
\hline
Error in $p(\delta | z_p)$ & All & $r<21$ & All
& $\pm 0.03$ \\
Redshift failures from DEEP2 (table~\ref{T:redshiftfailure}) & All &
$r>21$ & All & $[-0.14$, $+0.025]$ \\  
Statistical error in $p(z)$  & All & $r>21$ & All &
Table~\ref{T:zcalibration} \\
LRG redshift distribution failure & All & LRGs & All & $\pm 0.10$ \\
\hline
Overall redshift calibration & All & $r<21$ & All & $[-0.03, 0.03]$ \\
Overall redshift calibration & All & $r>21$ & All & $[-0.14, 0.07]$ \\
Overall redshift calibration & All & LRGs & All & $\pm 0.10$ \\
\hline
\multicolumn{5}{c}{Other systematic effects} \\
\hline
Shear calibration bias (\S\ref{SSS:shearcalibration}) & All &
$r<21$ & All & $[-0.05, 0.12]$ \\
Shear calibration bias  & All &
$r>21$ & All & $[-0.08, 0.18]$ \\
Shear calibration bias  & All &
LRGs & All & $[-0.06, 0.19]$ \\
Stellar contamination (\S\ref{SSS:sgsepresults}) & All & $r<21$ & All & $[-0.040, 0.0]$ \\
Stellar contamination & All & $r>21$ & All  & $[-0.096, -0.003]$ \\
Stellar contamination & All & LRGs & All & $[-0.068, 0]$ \\
Intrinsic alignments (\S\ref{SSS:intrinsicresults}) & L3--L6 & $r<21$ & All & Table~\ref{T:intrinsic} \\
Intrinsic alignments & L3--L6 & $r>21$ & All &
Table~\ref{T:intrinsic}  \\
Intrinsic alignments & L3--L6 & LRGs & Small &
Table~\ref{T:intrinsic} \\
Magnification bias (\S\ref{SSS:magbias}) & L3--L6 & $r<21$ & Small
& Table~\ref{T:magbias} \\
Magnification bias & L3--L6 & $r>21$ & Small 
 & Table~\ref{T:magbias} \\
Magnification bias & L3--L6 & LRGs & Small &
Table~\ref{T:magbias} \\
{\sc Photo} sky level error (\S\ref{SSS:boostsresults}) & All & All & Small & 
$[-0.15,0]$ \\
\hline
\hline
Overall uncertainty & L1--L4 & $r<21$ 
 & Small & $[-0.15, 0.16]$ \\ 
Overall uncertainty & L1--L4 & $r<21$ 
 & Large & $[-0.09, 0.11]$ \\ 
Overall uncertainty & L5--L6 & $r<21$ 
 & Small & $[-0.41, 0.42]$ \\ 
Overall uncertainty & L5--L6 & $r<21$ 
 & Large & $[-0.13, 0.14]$ \\ 
Overall uncertainty & L1--L4 & $r>21$ & Small &
$[-0.22, 0.21]$ \\ 
Overall uncertainty & L1--L4 & $r>21$ & Large &
$[-0.19, 0.19]$ \\ 
Overall uncertainty & L5--L6 & $r>21$ & Small &
$[-0.44, 0.44]$ \\ 
Overall uncertainty & L5--L6 & $r>21$ & Large &
$[-0.27, 0.26]$ \\ 
Overall uncertainty & L1--L4 & LRGs & Small & $[-0.18,
  0.19]$ \\
Overall uncertainty & L1--L4 & LRGs & Large & $[-0.16,
  0.18]$ \\
Overall uncertainty & L5--L6 & LRGs & Small & $[-0.21,
  0.23]$ \\
Overall uncertainty & L5--L6 & LRGs & Large & $[-0.16,
  0.18]$ \\
\hline\hline
\end{tabular}
\end{table*}
The first section of table~\ref{T:summary} includes a summary of the
findings regarding the redshift distribution calibration, and ends
with a calculation of the overall calibration uncertainty for each
sample due to
redshift distribution-related factors.  The second section includes a
list of other factors that lead to calibration uncertainty that have
been discussed in this paper.  The table ends with an estimate of the
overall calibration uncertainty at the $2\sigma$ level for various
lens, source combinations and ranges of radii.  The estimates are
somewhat rough, in the 
sense that they cannot include all details of the error estimates
completed in the text, and consequently a better idea  of the
magnitudes of some of the errors can be found by reading the
appropriate sections of the text.

In this table, we refer to ``small'' versus ``large'' radii, where the
distinction between the two is that on ``large'' radii the estimates
only include the calibration uncertainties such as stellar
contamination, shear calibration, and redshift distributions, but for
``small'' radii, we must also be
concerned about systematics such as intrinsic alignments, sky
subtraction effects, and 
magnification bias.  The definition of small versus large radii varies by
luminosity bin, and the approximate transition between the two occurs
at 40, 60, 80, 100, 140, and 200 \hinvk{} for L1--L6 respectively.

The overall uncertainty estimates in table~\ref{T:summary}
were computed using one of several possible methods. The difficulty is
that by definition systematic errors are those with unknown
probability distributions, so it is not clear how to combine
confidence intervals for multiple effects.  The first
possible method, the most conservative, would be to linearly add
together the upper bounds for all the independent sources of error to
get an overall upper bound, and similarly for the lower bound.  This
method does not require us to make any assumptions about the
probability distributions for these systematic errors, most of which
are very likely not remotely Gaussian.  The second approach would be to
assume a uniform distribution of the systematic error between the
$2\sigma$ bounds $[a,b]$ given in the table.  With this assumption, we can
find the mean $(a+b)/2$ and the standard deviation $(b-a)/\sqrt{12}$
for each systematic error, average the means and add the standard
deviations in quadrature,
and use those results to find combined $2\sigma$ bounds.  This
approach is also justifiable, since the use of a uniform distribution
leads to rather generous confidence intervals, but the results are
still smaller than the very conservative bounds from the first
approach.  The final 
approach would be to add the lower bounds in quadrature, and add the
upper bounds in quadrature.  For table~\ref{T:summary}, we used the
bounds from the second method (assuming a uniform distribution); the
results using the first method are about 50--100 per cent larger, and using
the third method are about 0--10 per cent larger.

In this paper, we have sought to understand the various sources of
calibration uncertainty and other systematics in the galaxy-galaxy weak lensing signal
computed using our lens and source catalogs.  
%In addition, we have
%tested the effects of several types of calibration uncertainty
%(overall, and as a function of transverse separation) on a simple
%application of our results, halo model fits for the central halo mass
%and satellite fraction as a function of lens luminosity.  
As more data
is collected, we will be able to perform these tests to constrain the
systematics with greater precision.  The
understanding we have gained from the systematics test will allow
us to use this catalog for many
scientific applications (studies of the relation between mass and
luminosity, mass and stellar masses, halo profiles as a function of
environment, and others) with an understanding of the implication of
systematic errors on the results. Similar investigations of systematics
will also be needed for weak lensing auto-correlation analyses, once 
the statistical errors become as small as they are for 
our SDSS galaxy-galaxy lensing analysis. 

\section*{Acknowledgments}
We gratefully acknowledge Edd Edmondson, Christian Wolf, and Lance
Miller for their work on the redshift distributions from COMBO-17 that
they allowed us to use.  We also acknowledge James Gunn,
Erin Sheldon, and David Johnston for useful discussion
regarding systematics in the  
lensing signal.  We thank the entire DEEP2 team for allowing us to use their
redshift data in the Groth strip, and are particularly thankful for
the help and feedback from Doug Finkbeiner, Alison Coil, and
Jeffrey Newman.  

R.M. is supported by an NSF Graduate Research Fellowship, and
C.H. by NASA grant NGT5-50383. U.S. supported by a fellowship from the
David and Lucile Packard Foundation,
NASA grants NAG5-1993, NASA NAG5-11489 and NSF grant CAREER-0132953.

   Funding for the creation and distribution of the SDSS Archive has
been provided by the Alfred P. Sloan Foundation, the Participating
Institutions, the National Aeronautics and Space Administration, the
National Science Foundation, the U.S. Department of Energy, the
Japanese Monbukagakusho, and the Max Planck Society. The SDSS Web site
is {\slshape http://www.sdss.org/}. 

    The SDSS is managed by the Astrophysical Research Consortium (ARC)
for the Participating Institutions. The Participating Institutions are
The University of Chicago, Fermilab, the Institute for Advanced Study,
the Japan Participation Group, the Johns Hopkins University, the Korean
Scientist Group, Los Alamos National Laboratory, the
Max-Planck-Institute for Astronomy (MPIA), the Max-Planck-Institute
for Astrophysics (MPA), New Mexico State University, University of
Pittsburgh, University of Portsmouth, Princeton University, the United States Naval Observatory,
and the University of Washington. 

Funding for the DEEP2 survey has been provided by NSF grant
AST-0071048 and AST-0071198.  Some of the data presented herein were
obtained at the W.M. Keck 
Observatory, which is operated as a scientific partnership among the
California Institute of Technology, the University of California and
the National Aeronautics and Space Administration. The Observatory was
made possible by the generous financial support of the W.M. Keck
Foundation. The DEEP2 team and Keck Observatory acknowledge the very
significant cultural role and reverence that the summit of Mauna Kea
has always had within the indigenous Hawaiian community and appreciate
the opportunity to conduct observations from this mountain. 

{}

\appendix

\section{$\chi^2$ for redshift distribution fits}
\label{app:chi2}

The redshift distribution fits were obtained by minimizing the ``$\chi^2$'' function
\begin{equation}
\chi^2 = -2\sum_{i=1}^{N_s} \ln p(z_i),
\label{eq:chi2-a}
\end{equation}
where $N_s$ is the number of spectra, and $p(z_i)=(\rmd P/\rmd
z)|_{z=z_i}$ is the normalized redshift distribution.   
Equation~(\ref{eq:chi2-a}) is a ``$\chi^2$'' in the sense that the
likelihood for a given model $p(z)$ is proportional to  
$\rme^{-\chi^2/2}$ in the case where the galaxy redshifts are
independent.  While minimizing $\chi^2$ is a reasonable way to get the  
redshift distribution, it is not trivial to test for a ``goodness of
fit'' since (i) even in the case of independent redshifts,  
Equation~(\ref{eq:chi2-a}) does not follow the standard $\chi^2$
distribution, and (ii) there is large scale structure (LSS) that  
introduces correlations between different redshifts, particularly in a
narrow survey such as DEEP2.  This Appendix is devoted to  
calculating the statistical properties of $\chi^2$ and the fit
parameters obtained via its minimization. 

\subsection{Mean and variance of $\chi^2$}

The mean of Eq.~(\ref{eq:chi2-a}) is
\begin{equation}
\mu_{\chi^2} \equiv \langle\chi^2\rangle = 2N_sS,
\label{eq:mu-chi2}
\end{equation}
where the ``entropy'' $S$ is the expectation value of $-\ln p(z)$ for a single galaxy, i.e.
\begin{equation}
S = -\int p(z)\ln p(z)\, dz.
\end{equation}
The entropy is a property of the redshift distribution only; e.g. for the $\Gamma$-distribution (Eq.~\ref{eq:gammadist}),
\begin{equation}
S = \alpha + \ln \Gamma(\alpha) - (\alpha-1)\psi(\alpha) + \ln z_s,
\end{equation}
where $\psi(\alpha) = \rmd[\ln\Gamma(\alpha)]/\rmd\alpha$ is the digamma function.  Note that $S$ can be positive, zero, or negative 
depending on the distribution; it is larger for wide distributions.

Equation~(\ref{eq:mu-chi2}) is the mean of a simple sum over galaxies and hence it is valid regardless of LSS.  The variance is more 
complicated; it is
\begin{equation}
\sigma^2_{\chi^2} \equiv \langle(\chi^2-\langle\chi^2\rangle)^2\rangle
= 4\sum_{i,j=1}^{N_s} \langle [\ln p(z_i)+S] [\ln p(z_j)+S] \rangle.
\end{equation}
The expectation value here contains {\em two} galaxy redshifts, and hence
depends on both the redshift distribution and the redshift correlation
function.  Defining the redshift correlation function $\xi_z$ by
\begin{equation}
\left. p(z_i,z_j)\right|_{i\neq j} = p(z_i)p(z_j)[1+\xi_z(z_i,z_j)],
\label{eq:pzizj}
\end{equation}
we find
\begin{equation}
\sigma^2_{\chi^2} = 4N_sS_{1,1} + 4N_s(N_s-1)S_2,
\end{equation}
where
\begin{equation}
S_{1,1} = \int p(z) [\ln p(z)+S]^2 \rmd z
\end{equation}
and
\begin{eqnarray}
S_2 \!\!\! &=& \!\!\! \int p(z_1) p(z_2) \xi_z(z_1,z_2) [\ln p(z_1)+S]
\nonumber \\ && \!\!\! \times [\ln p(z_2)+S] \rmd z_1 \rmd z_2.
\label{eq:s2}
\end{eqnarray}
The ``independent redshift'' contribution to $\sigma^2_{\chi^2}$ comes 
from the $S_{1,1}$ term, which does not involve the correlation 
function.  For the $\Gamma$-distribution,
\begin{equation}
S_{1,1} = (\alpha-1)^2\psi'(\alpha) - \alpha + 2.
\end{equation}

\subsection{Correlation function and $S_2$}

The evaluation of the clustering contribution $S_2$ to the variance of
$\chi^2$ is more complicated.  We begin by assuming that the
three-dimensional correlation function of the sources is a power law,
$\xi(r) = (r_0/r)^\gamma$.  Written in polar coordinates, this becomes
\begin{equation}
\xi(r_1,r_2,\theta)
\approx \frac{ r_0^\gamma }{ [(r_1-r_2)^2 + r_1^2\theta^2]^{\gamma/2} },
\end{equation}
where the approximation is valid for small $\theta$ (the relevant case 
here).  The correlation falls off at large $|r_1-r_2|$ as 
$|r_1-r_2|^{-\gamma}$, hence for $\gamma>1$ the integral is finite and 
$\xi$ is non-zero only for $r_1\approx r_2$.  We thus make a Limber 
approximation,
\begin{eqnarray}
\xi(r_1,r_2,\theta) &\approx &
r_0^\gamma\delta(r_1-r_2) \int_{-\infty}^\infty \frac{d\Delta r}{ (\Delta 
r^2 + r_1^2\theta^2)^{\gamma/2} }
\nonumber \\
&=& B\left({\gamma-1\over 2},{1\over 2}\right) r_0^\gamma 
(r_1\theta)^{1-\gamma}\delta(r_1-r_2).
\end{eqnarray}
The resulting correlation function $\xi_z(z_1,z_2)$ is obtained by
converting from comoving distance to redshift, and averaging over the
angular separations $\theta$ in the survey,
\begin{equation}
\xi_z(z_1,z_2) = A \langle(r_1\theta)^{1-\gamma}\rangle
   \left.{\rmd z\over \rmd r}\right|_{z_1} \delta(z_1-z_2),
\label{eq:xiz}
\end{equation}
where for the purposes of obtaining a rough estimate of the importance of
$S_2$ we assume $A=B({\gamma-1\over 2},{1\over 2})  r_0^\gamma$ to be a
constant.\footnote{Technically Eq.~(\ref{eq:xiz}) should contain an
additive constant to account for the integral constraint $\int
p(z_i,z_j)\rmd z_i\rmd z_j=1$ in Eq.~(\ref{eq:pzizj}).  An additive
constant in $\xi_z$ would not affect $S_2$ because its contribution in
Eq.~(\ref{eq:s2}) multiplies $\int p(z_1)p(z_2)[\ln p(z_1)+S][\ln
p(z_2)+S] \rmd z_1\rmd z_2=0$.} The result for $S_2$ is
\begin{equation}
S_2 = A \langle \theta^{1-\gamma}\rangle \int [p(z)]^2 r^{1-\gamma} H(z) 
[\ln p(z)+S]^2 \rmd z,
\label{eq:s2ref}
\end{equation}
where we have used $\rmd z/\rmd r=H(z)$.  We may find $A$ and $\gamma$ by 
comparing to the observed angular 2-point correlation function of the 
galaxies; the theoretical prediction is
\begin{eqnarray}
\omega(\theta) \!\!\! &=& \!\!\! \int \xi(r_1,r_2,\theta) {\rmd 
P\over\rmd r}(r_1){\rmd P\over\rmd r}(r_2) \rmd r_1\rmd r_2
\nonumber \\ &=& \!\!\!
A\theta^{1-\gamma} \int [p(z)]^2 r^{1-\gamma} H(z) \rmd z.
\label{eq:omega}
\end{eqnarray}
The angular correlation function in SDSS \citep{2002ApJ...579...42C} is
fit by a power law with $\gamma=1.7$.  The DEEP2/SDSS overlap region can
be approximated by a rectangle of dimensions $0.25\times 0.6\deg$, hence
averaged over all pairs of points in this region (obtained via the obvious
Monte Carlo procedure) is $\langle\theta^{1-\gamma}\rangle \approx
3.7\deg^{-0.7}$.

\subsection{Parameter uncertainties}

The uncertainty of the redshift distribution parameters is greater than
the naive result from $\Delta\chi^2$ contours because of source
clustering.  We begin our analysis of this effect by assuming that the
redshift distribution depends on a vector of $M$ parameters ${\bmath
a}=(a^1...a^M)$; in the case of the $\Gamma$-distribution, $M=2$ and
${\bmath a}=(z_s,\alpha)$.  We are then estimating the parameters ${\bmath
a}$ by minimizing $\chi^2({\bmath a})$.  We denote the true set of
parameters by ${\bmath a}_0$.

We begin by Taylor expanding $\chi^2$ around ${\bmath a}_0$,
\begin{equation}
\chi^2({\bmath a}) = \chi^2({\bmath a}_0)
   + {\bmath G}\cdot({\bmath a}-{\bmath a}_0)
   + ({\bmath a}-{\bmath a}_0)\cdot
   {\mathbfss K}({\bmath a}-{\bmath a}_0) + ...,
\end{equation}
where ${\bmath G}$ and ${\mathbfss K}$ are the derivatives of $\chi^2$ at 
${\bmath a}_0$.  The minimum value of ${\bmath a}$ in this 
approximation is then
\begin{equation}
\hat{\bmath a} = {\bmath a}_0 - {1\over 2}{\mathbfss K}^{-1}{\bmath G}.
\label{eq:delta-a}
\end{equation}
We can calculate ${\bmath G}$ and ${\mathbfss K}$ in terms of the 
measured source redshifts,
\begin{equation}
G_\mu = -2\sum_{i=1}^{N_s} \left.{\partial\over\partial a^\mu}\ln p(z_i|
   {\bmath a})
   \right|_{{\bmath a}_0}
\end{equation}
and
\begin{equation}
K_{\mu\nu} = -\sum_{i=1}^{N_s}
   \left.{\partial^2\over \partial a^\mu\partial a^\nu}\ln p(z_i|{\bmath 
   a})\right|_{{\bmath a}_0}.
\end{equation}
We now estimate the statistical properties of $\hat{\bmath a}$ using Eq.~(\ref{eq:delta-a}), approximating ${\mathbfss K}$ by its 
expectation value for simplicity:
\begin{eqnarray}
\langle K_{\mu\nu}\rangle &=& -N_s \left\langle
   {\partial^2\ln p(z_i)\over \partial a^\mu\partial a^\nu}\right\rangle_{{\bmath a}_0}
\nonumber \\
&=& N_s \left\langle {\partial\ln p(z_i)\over\partial a^\mu}
   {\partial\ln p(z_i)\over\partial a^\nu} \right\rangle_{{\bmath a}_0},
\label{eq:kmunu}
\end{eqnarray}
where we have suppressed the argument ${\bmath a}$ in $p(z_i)$ for
simplicity of notation.\footnote{The second equality in
Eq.~(\ref{eq:kmunu}) can be found by differentiating the relation
$\int p(z)dz=1$ twice to get $\langle [{\partial^2 p(z)/\partial a^\mu
\partial a^\nu}]/p(z) \rangle = 0$, and then re-writing this result in
terms of $\ln p(z)$.}

The expectation value of $G_\mu$ is easily verified to be $\langle
G_\mu\rangle = 0$, and so we have $\langle\hat{\bmath a}\rangle = {\bmath
a}_0$.  The covariance of ${\bmath G}$ is
\begin{eqnarray}
\langle G_\mu G_\nu \rangle \!\!\! &=& \!\!\! 4\sum_{i,j=1}^{N_s} \left\langle
   {\partial\ln p(z_i)\over\partial a^\mu}
   {\partial\ln p(z_j)\over\partial a^\nu}
   \right\rangle_{{\bmath a}_0}
\nonumber \\
&=& \!\!\! 4K_{\mu\nu} + 4\sum_{i\neq j} \left\langle
   {\partial\ln p(z_i)\over\partial a^\mu}
   {\partial\ln p(z_j)\over\partial a^\nu}
   \right\rangle_{{\bmath a}_0}.
\end{eqnarray}
(The $i=j$ terms become $K_{\mu\nu}$, following the argument of 
Eq.~\ref{eq:kmunu}.)  The second term can be evaluated using 
Eq.~(\ref{eq:pzizj}) and Eq.~(\ref{eq:xiz}) to yield
\begin{eqnarray}
\langle G_\mu G_\nu \rangle \!\!\! &=& \!\!\! 4K_{\mu\nu} + 
4N_s(N_s-1)A\langle 
   \theta^{1-\gamma}\rangle
\nonumber \\ && \times
   \int r^{1-\gamma} H(z) {\partial p\over \partial a^\mu}
   {\partial p\over \partial a^\nu} dz.
\end{eqnarray}
The covariance matrix of ${\bmath a}$ is then
\begin{eqnarray}
\langle\delta a^\mu \delta a^\nu\rangle \!\!\! &=& \!\!\!
 {1\over 4} [{\mathbfss K}^{-1}]^{\mu\lambda}
    \langle G_\lambda G_\rho \rangle
    [{\mathbfss K}^{-1}]^{\rho\nu}
\nonumber \\ &=& \!\!\!
    [{\mathbfss K}^{-1}]^{\mu\nu}
    +N_s(N_s-1)A\langle \theta^{1-\gamma}\rangle
    [{\mathbfss K}^{-1}]^{\mu\lambda}
\nonumber \\ && \times
    [{\mathbfss K}^{-1}]^{\rho\nu}
    \int r^{1-\gamma} H(z) {\partial p\over \partial a^\lambda}
    {\partial p\over \partial a^\rho} dz,
\label{eq:cov-a}
\end{eqnarray}
which is larger than the covariance matrix ${\mathbfss K}^{-1}$ 
that would be obtained via the usual $\Delta\chi^2=1$ prescription.

In the case of the $\Gamma$-distribution with parameters $(z_s,\alpha)$, 
we find
\begin{equation}
{\mathbfss K} = N_s \left( \begin{array}{cc}
\alpha z_s^{-2} & z_s^{-1} \\
z_s^{-1} & \psi'(\alpha) \end{array}\right).
\end{equation}

\section{Non-Gaussian Confidence Intervals for Correlated Variables}\label{app:confint}

In this section, we explain the method used to compute non-Gaussian
confidence intervals on the ratio of correlated variables.  First, we
consider the case of the ratio of two uncorrelated variables with
Gaussian noise, then generalize to the correlated case.  

When computing the error on the ratio of two variables
$R=y/x$, where $y$  and $x$ are Gaussian variables with standard
deviations $\sigma_y$ and
$\sigma_x$, the naive result $\sigma_R =
R\sqrt{\left(\left(\sigma_y/y\right)^2 +
  \left(\sigma_x/x\right)^2\right)}$ is unreliable in the case when
$x$ may be consistent with zero even at the many-$\sigma$ level.
Instead, the non-Gaussian confidence intervals should be used.  If we
want to find the confidence intervals on $R$ at the $z\sigma$ level
(i.e. $z=1$ would give the 68 per cent confidence interval),
then the result (\citealt{bliss1};
\citealt{bliss2}; \citealt{fieller}) is as follows:
\begin{equation}\label{E:gausserr}
\pm\mbox{ limit on }R = \frac{y \pm
  \frac{\sigma_y}{\sigma_x}\frac{z}{\sqrt{k^2-z^2}} x}{x \mp
  \frac{\sigma_x}{\sigma_y}\frac{z}{\sqrt{k^2-z^2}} y}
\end{equation}
where we define
\begin{equation}
k = \sqrt{\left(\frac{x}{\sigma_x}\right)^2 + \left(\frac{y}{\sigma_y}\right)^2}.
\end{equation}

If $y$ and $x$ are correlated, with nonzero correlation coefficient $\rho =
{\rm Cov}(x,y)/(\sigma_x\sigma_y)$, then Eq.~(\ref{E:gausserr}) is no
longer valid.  In order to get confidence intervals on
$R$, we do the following manipulations: we change variables so that we
are once again dealing with two uncorrelated variables, in this case
$w = y - m x$, where $m = \rho\sigma_y/\sigma_x$.
Here $w$ is clearly uncorrelated with $x$, and has error $\sigma_w^2 =
\sigma_y^2 (1-\rho^2)$.  Furthermore, $R = w/x + m$.  We can then use
the procedure for 
uncorrelated variables to compute confidence intervals on $w/x$, so
that the upper and lower limits on $y/x$ are related to those on $w/x$ 
by the addition of $m$. 

\section{Calibration bias from significance cut}
\label{app:sn5}

There is a contribution to the shear calibration bias originating from
selection effects from the signal-to-noise cut in our catalog.   
We can investigate the magnitude of this effect by the same methods
used in \S\ref{SSS:shearcalibration}.  {\sc Photo} convolves the 
observed image $I$ with the PSF $P$ to create a new image $J=I\otimes
P$, and finds pixels in $P$ above some threshold $J_c$, set to 
$\nu_t\sigma[J({\bmath x})]$ above the sky background, where the
signal-to-noise threshold is $\nu_t=5$.  This convention is chosen 
because, in the case where $P$ is symmetric under 180 degree rotation,
$J({\bmath x})$ represents the optimal statistic for searching 
for a point source at ${\bmath x}$.  Objects above $J_c$ are
designated as BINNED1 by {\sc Photo}; for detection, we require that
this 
flag be set in the $r$ and $i$ bands.  The threshold is
\begin{eqnarray}
J_c \!\!&=& \nu_t\sqrt{n\int [P({\bmath x})]^2 \rmd^2{\bmath x}}
\nonumber \\ &=& \left\{\begin{array}{lcl}
0.47 \nu_t n^{1/2}\theta_{FWHM}^{-1} & & {\rm (Gaussian)} \\
0.57 \nu_t n^{1/2}\theta_{FWHM}^{-1} & & {\rm (Kolmogorov)}
\end{array}\right.,
\end{eqnarray}
where $n$ is the noise variance per pixel and $\theta_{FWHM}$ is the
full-width at half maximum of the PSF.  We have shown both the  
case of the Gaussian PSF and the Kolmogorov turbulence-induced PSF
$\ln\tilde P(k)\propto k^{5/3}$.  For typical parameters,  
$J_c=0.13$~nmgy~arcsec$^{-2}$ in $r$ band and 0.22~nmgy~arcsec$^{-2}$
in $i$ band. 

The value of $J({\bmath 0})$, where we have translated the object's
centroid to the origin ${\bmath 0}$, is given by 
\begin{equation}
J({\bmath 0}) = \int I({\bmath x}) P({\bmath x}) \rmd^2{\bf x}
= \int f({\bmath x}) K({\bmath x}) \rmd^2{\bf x},
\label{eq:j0}
\end{equation}
where $K=P\otimes P$ and we have assumed $P({\bmath x})=P(-{\bmath
  x})$.  The simplest way to evaluate $J$ is in the case of a  
circular Gaussian PSF and elliptical Gaussian galaxy, for which
\begin{equation}
J({\bmath 0}) = \frac{F}{2\pi}\left[ \det \left( {\mathbfss
    M}^{(f)}+2{\mathbfss M}^{(P)} \right) \right]^{-1/2}, 
\end{equation}
where $F$ is the galaxy flux.  Under a shear leaving $\sigma^{(f)}$
constant, this varies by 
\begin{equation}
\frac{\delta J({\bmath 0})}{J({\bmath 0})} = -\frac{2(1-R_2){\mathbfss
    e}\cdot\bgamma}{4R_2^{-1}-4+R_2-e^2}. 
\end{equation}
The resulting shear selection bias can then be computed by the same
method as was used in \S\ref{SSS:shearcalibration}.  It yields 
\begin{equation}
\frac{\delta\gamma}{\gamma} =
-\left\langle\frac{(1-R_2)e^2}{(4R_2^{-1}-4+R_2-e^2)\cal
  R}\right\rangle 
\left.{\rmd N(j)\over\rmd\ln j}\right|_{j=J_c},
\end{equation}
where $N(j)$ is the (weighted) fraction of galaxies with $J({\bmath
  0})<j$.  The coefficient of $\rmd N(j)/\rmd\ln j$ has a mean value  
(averaged over the $R_2$ distribution) of $-0.024$, $-0.028$, and
  $-0.026$ for the $r<21$, $r>21$, and LRG samples respectively in  
$r$ band, and $-0.023$, $-0.026$, and $-0.024$ in $i$ band.
  Conservative values of $\rmd N(j)/\rmd\ln j$ (obtained by maximizing
  over $j$, since we have not calculated the threshold exactly)  are
  1.40, 2.37, and 
  1.42 respectively in $r$ band, and 1.23, 2.01, and 1.46 in $i$
  band.  We conservatively estimate the calibration bias due to the
  signal to noise cut using the $r$ band values, and obtain $-0.034$,
  $-0.066$, and $-0.037$ for $r<21$, $r>21$, and LRG samples respectively.

\end{document}